САНКТ-ПЕТЕРБУРГСКИЙ ГОСУДАРСТВЕННЫЙ ТЕХНИЧЕСКИЙ УНИВЕРСИТЕТ

На правах рукописи

ГАРСИЯ ФУЭНТЕС Франсиско Игнасио

# МЕТОДИКА ОПТИМИЗАЦИИ СИСТЕМЫ ДЕТЕКТИРОВАНИЯ КОРПУСКУЛЯРНОГО ИЗЛУЧЕНИЯ ПЛАЗМЫ ДЛЯ РАБОТЫ В УСЛОВИЯХ ИНТЕНСИВНОГО n-γ ИЗЛУЧЕНИЯ

Специальность 01.04.16 – Физика атомного ядра

Д И С С Е Р Т А Ц И Я

На соискание ученой степени
Кандидат физико-математических наук

Научный руководитель:
доцент,
кандидат физ.-мат. наук
С.С. Козловский

Санкт-Петербург
2 0 0 0 г.

# О Г Л А В Л Е Н И Е

# ВВЕДЕНИЕ

Прогресс в области термоядерного синтеза базируется как на теоретических исследованиях, так и на экспериментальных данных, полученных с помощью диагностики параметров плазмы на существующих термоядерных установках. Одним из самых информативных методов диагностики плазмы является метод корпускулярной диагностики, который дает возможность одновременно производить массовый и энергетический анализ атомов, выходящих из плазмы, с помощью анализаторов нейтральных частиц.

В настоящее время в плазменных разрядах крупных термоядерных установок интегральный поток нейтронного и гамма излучения достигает величины ~ $10^{18}$ $сек^{-1}$. При проектировании термоядерного реактора нового поколения ITER уровень n-γ излучения плазмы будет еще выше. Такие потоки нейтронного и гамма излучения плазмы создают серьезную проблему для выделения полезного сигнала над уровнем фоновых сигналов. Решение этой проблемы может быть осуществлено двумя способами:

- Установка защиты от проникающего излучения с целью ослабления интенсивности фоновых сигналов до приемлемого уровня.
- уменьшение чувствительности детекторов к фоновому нейтронному и гамма излучению.

В первом варианте велики затраты на сооружение защитного бокса, в ряде случаев задача не может быть решена в полной мере из-за ограниченного пространства, выделенного на соответствующее диагностическое оборудование.

В связи с этим в настоящей работе была поставлена цель создания детектирующей системы для регистрации корпускулярного излучения, работоспособной в условиях интенсивного нейтронного и гамма излучения плазмы.

Основные результаты были получены в период с 1995 по 1999 гг. и изложены в работах [1-6]. В течение этого времени на кафедре Экспериментальной Ядерной Физики СП-бГТУ были разработаны системы детектирования для анализаторов нейтральных частиц GEMMA-2M и для изотопного сепаратора ISEP. Оба этих анализаторов в настоящее время эксплуатируются на термоядерной установке JET (Англия). Приборы такого типа также были ранее установлены на токамаках TFTR (США) и JT-60U (Япония).

Диссертационная работа состоит из четырех глав, введения и заключения.

В первой главе рассмотрены источники фонового нейтронного и гамма излучения. Проведен анализ радиационной стойкости детекторов и компонентов, используемых в системах детектирования анализаторов, кроме того, представлены опубликованные результаты по чувствительности различных детекторов к фоновому нейтронному и гамма излучению.

Во второй главе описан экспериментальный стенд, на котором были проведены измерения амплитудных распределений исследуемых детекторов к фоновому нейтронному и гамма излучению. Кроме того, приводится методика измерения чувствительности детекторов к фоновому излучению. Описаны способы обработки полученных данных и изложена методика напыления сцинтилляторов CsI(Tl).

В третьей главе приведены основные результаты измерения чувствительности исследуемых детекторов к фоновому нейтронному и гамма излучению. Кроме того, изложена модель формирования амплитудного распределения сигналов на выходе детектора под действием фонового излучения.

Четвертая глава посвящена краткому описанию конструкции детектирующей системы анализаторов GEMMA-2M и ISEP. В ней показаны результаты оптимизации детектирующей системы анализатора GEMMA-2M.

Также приведены первые результаты измерения корпускулярных потоков с помощью анализатора ISEP на термоядерной установке JET.

# ГЛАВА I. АНАЛИЗ ФАКТОРОВ, ОПРЕДЕЛЯЮЩИХ ВЫБОР ДЕТЕКТОРА ДЛЯ КОРПУСКУЛЯРНОЙ ДИАГНОСТИКИ ПЛАЗМЫ В УСЛОВИЯХ ИНТЕНСИВНОГО n-$\gamma$ ИЗЛУЧЕНИЯ

Стремительное развитие исследований высокотемпературной плазмы требует быстрых темпов совершенствования уже существующих методов ее диагностики. Одним из самых информативных является метод корпускулярной диагностики, который позволяет восстановить энергетическое распределение ионов в плазме / 1 – 5 /. Однако рост мощности плазменных установок привел к существенным проблемам при регистрации частиц в условиях интенсивного фонового излучения / 6 /. В большинстве систем регистрации корпускулярного излучения используются магнитные и электрические диспергирующие системы, при этом, детектирующая система, как правило, находится вне прямой видимости плазмы. Поэтому основную проблему представляет проникающее через защиту и материалы конструкции фоновое излучение, т.е. потоки нейтронов и гамма квантов высокой энергии. В связи с этим и была поставлена задача по разработке методики оптимизации системы регистрации корпускулярного излучения в условиях интенсивного фонового нейтронного и гамма излучения.

Рассмотрим источники проникающего фонового излучения на существующих термоядерных установках. Первичными источниками излучения являются термоядерные реакции, идущие в плазме, и тормозное излучение быстрых электронов. Потоки нейтронов высокой энергии возникают в основном по следующим каналам реакций:

$$D + D \rightarrow {}^{3}He\ (0.82\ \text{МэВ}) + n\ (2.45\ \text{МэВ})$$

$$D + T \rightarrow \alpha\ (3.56\ \text{МэВ}) + n\ (14.03\ \text{МэВ})$$

Видно, что в зависимости от типа плазменного разряда энергия вылетающих нейтронов сильно различается. При рассеянии нейтронов,

вылетающих из плазмы, конструкционными элементами установки происходит изменение первичного энергетического спектра нейтронов. Преобразование энергетического спектра происходит с обогащением медленной компонентой, т.е. количество тепловых и медленных нейтронов становится больше, а количество первичных нейтронов рождающихся в процессе термоядерного горения уменьшается. В результате поток фонового нейтронного излучения, падающего на детектор, имеет относительное широкое энергетическое распределение.

Второй компонентой фонового излучения являются гамма кванты, которые возникают как в процессе тормозного излучения быстрых электронов в плазме, так и в процессах взаимодействия нейтронов с элементами конструкции установки (реакции n,n' и n,γ). Как показывают измерения, выполненные на крупных термоядерных установках, спектр фонового гамма излучения имеет также широкое энергетическое распределение со средней энергией ~ 1 МэВ.

В таблице 1.1 приведены потоки нейтронов и гамма квантов в различных термоядерных установках, таких как JT-60U / 7 /, JET, TFTR / 8 /, ITER / 9 /.

Таблица 1.1

| Поток нейтронов ($\frac{1}{см^2 \cdot сек}$) | | | |
|---|---|---|---|
| JT-60U | JET | TFTR | ITER |
| $5.4 \cdot 10^{10}$ | $1.4 \cdot 10^{12}$ | $3.0 \cdot 10^{13}$ | $2.5 \cdot 10^{14}$ |
| Поток гамма квантов ($\frac{1}{см^2 \cdot сек}$) | | | |
| JT-60U | JET | TFTR | ITER |
| $3.5 \cdot 10^{10}$ | $0.9 \cdot 10^{12}$ | $2.0 \cdot 10^{13}$ | $1.6 \cdot 10^{14}$ |

Все приведенные выше величины даны для потоков излучения вблизи первой стенки вакуумной камеры. При этом видно, что потоки нейтронов и гамма квантов сравнимы между собой для каждой отдельной установки.

Как следует из этих данных, крупные термоядерные установки (TFTR, JT-60U, JET), характеризуются высоким уровнем нейтронного и гамма излучения. Потоки нейтронного и гамма излучения зависят от местоположения детектирующей системы и, как правило, на несколько порядков ниже в месте расположения детекторов, чем на первой стенке вакуумной камеры.

Таким образом, детектирующая система анализатора нейтральных частиц должна оставаться работоспособной в течение всего срока эксплуатации при облучении нейтронами и гамма квантами с интенсивностью ~ $10^8$ – $10^{10}$ $\frac{1}{\text{см}^2 \cdot \text{сек}}$. В задаче по разработке методики оптимизации детекторов для регистрации корпускулярного излучения плазмы при интенсивном фоновом нейтронном и гамма излучении можно выделить два наиболее важных аспекта. С одной стороны, радиационная стойкость детектирующей системы, которая определяет срок службы детектора и компонентов системы, используемых при регистрации корпускулярного излучения, а с другой – чувствительность к регистрации фонового излучения, которая определяет способность детектирующей системы к разделению полезных и фоновых событий.

## 1. Радиационная стойкость детекторов используемых для регистрации корпускулярного излучения плазмы

За последние сорок лет получены обширные данные по радиационной стойкости различных детекторов и их компонентов к воздействию нейтронов и гамма квантов. Перед тем, как приступить к анализу радиационной стойкости различных детекторов, мы разделим эту тематику на две группы. Одна группа представляет результаты, полученные при исследовании радиационной стойкости по отношению к гамма квантам, а другая – по отношению к нейтронам.

Далее представим результаты по радиационной стойкости детекторов и компонентов детектирующей системы под действием гамма квантов. При облучении сцинтилляционных детекторов большими дозами гамма квантов деградация детекторов связана в основном с потерями прозрачности входного окна ФЭУ и кристалла / 10 - 12 /. Отсюда следует, что радиационный ресурс детектора определяется дозой, при которой становится заметным изменение пропускающей способности кристаллов и входного окна ФЭУ, а также однородности кристаллов, которая влияет на величину энергетического разрешения детектора.

В работе / 13 / представлены результаты, полученные при облучении ФЭУ-85 и ФЭУ-84 гамма квантами $^{137}Cs$ и $^{60}Co$. Заметная потеря прозрачности входных окон марки С095-2 (ФЭУ-85) и С050-2 (ФЭУ-84) наблюдалась при дозах ~ 1 Мрад (2.9 $10^{15}$ $\frac{\gamma}{\text{см}^2}$) / 14 /. Также было проведено исследование воздействия облучения гамма квантами от источника $^{60}Co$ на динодную систему ФЭУ. Это исследование показало, что при дозах ~ 1 Мрад (2.2 $10^{15}$ $\frac{\gamma}{\text{см}^2}$) никаких изменений усилительной способности динодной системы не наблюдалось в пределах 5% точности.

Исследование основных параметров сцинтилляционных кристаллов в зависимости от величины поглощенной дозы является ключевым вопросом при анализе их радиационной стойкости. Эффекты наблюдаемые, при облучении сцинтилляционных кристаллов как показано в анализируемых работах, в основном связаны с потерей прозрачности к собственному излучению и с увеличением фосфоресцентного послесвечения.

В работе / 15 / представлены результаты изменения световыхода и пропускающей способности различных сцинтилляционных кристаллов. На рис.

1.1 представлены сравнительные кривые нормализованного световыхода для сцинтилляционных кристаллов CsI(Tl) изготовленных по разным технологиям. Необходимо отметить, что в качестве образцов были использованы массивные кристаллы.

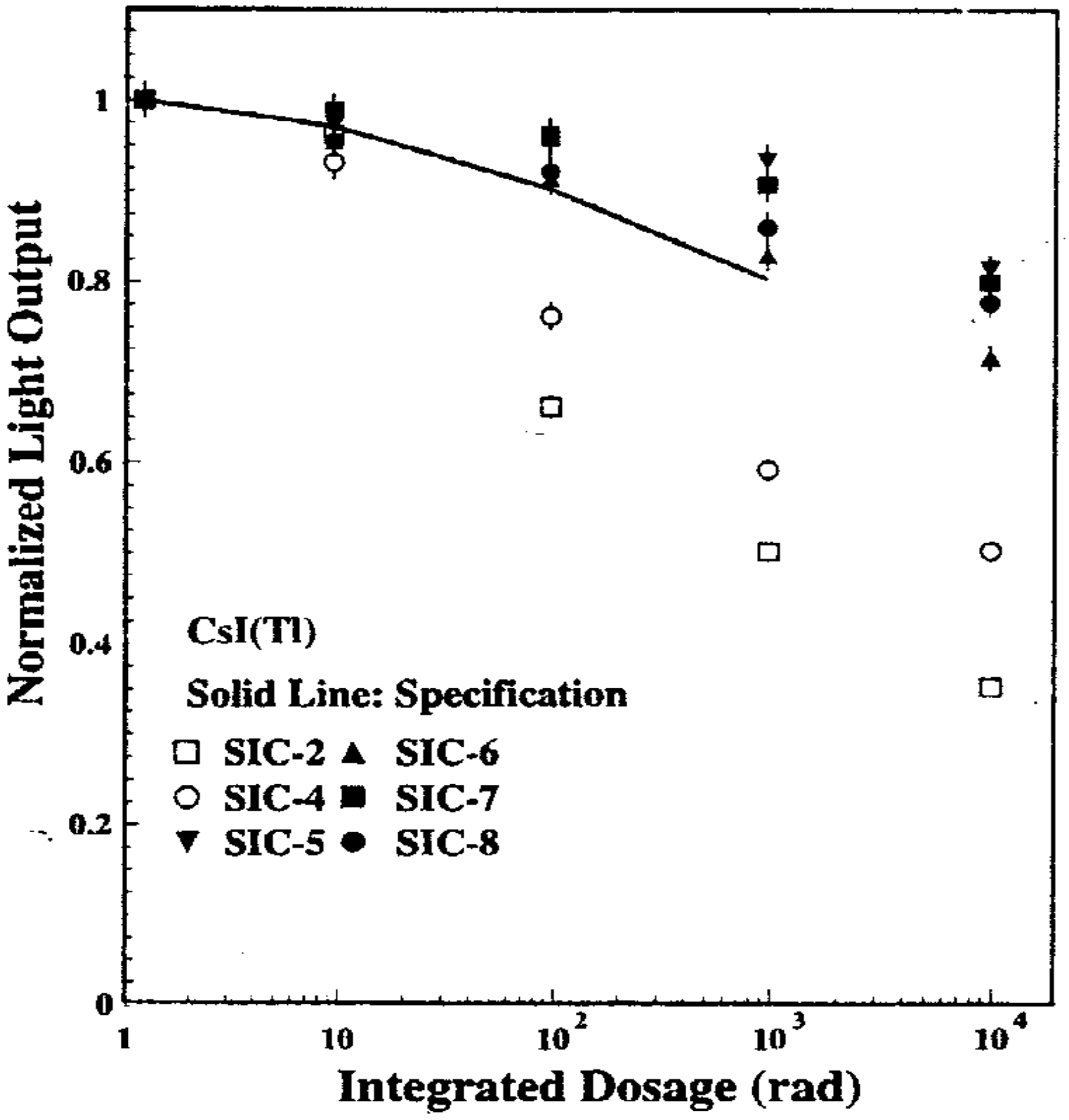

Рис 1.1 Изменение относительного световыхода различных сцинтилляторов при разных дозах облучения

Из рис. 1.1 видно, что независимо от типа сцинтилляционного кристалла световыход уменьшается при увеличении интегральной дозы облучения. Для образцов марки SIC-2 и SIC-4 наблюдается большее изменение световыхода, нежели для кристаллов SIC-5, SIC-6, SIC-7, SIC-8. Отсюда следует, что в сцинтилляционных детекторах, работающих в условиях интенсивного фонового излучения целесообразно, применять кристаллы типа SIC-5, SIC-6, SIC-7, SIC-8. Эти кристаллы были изготовлены по технологии, которая сводит к минимуму концентрацию примеси кислорода / 16 /. Кристаллы марки BGRI / 17 / изготовлены в Beijing Glass Research Institute, SIC / 18 / в Shanghai Institute of Ceramics а BTCP / 19 / изготовлены в Bogorodisk Techno-Chemical plant.

На рисунках 1.2 – 1.4 из этой же работы / 15 / представлены результаты измерения пропускающей способности сцинтилляционных кристаллов CsI(Tl), $PbWO_4$, $BaF_2$ в зависимости от длины волн для разных доз облучения.

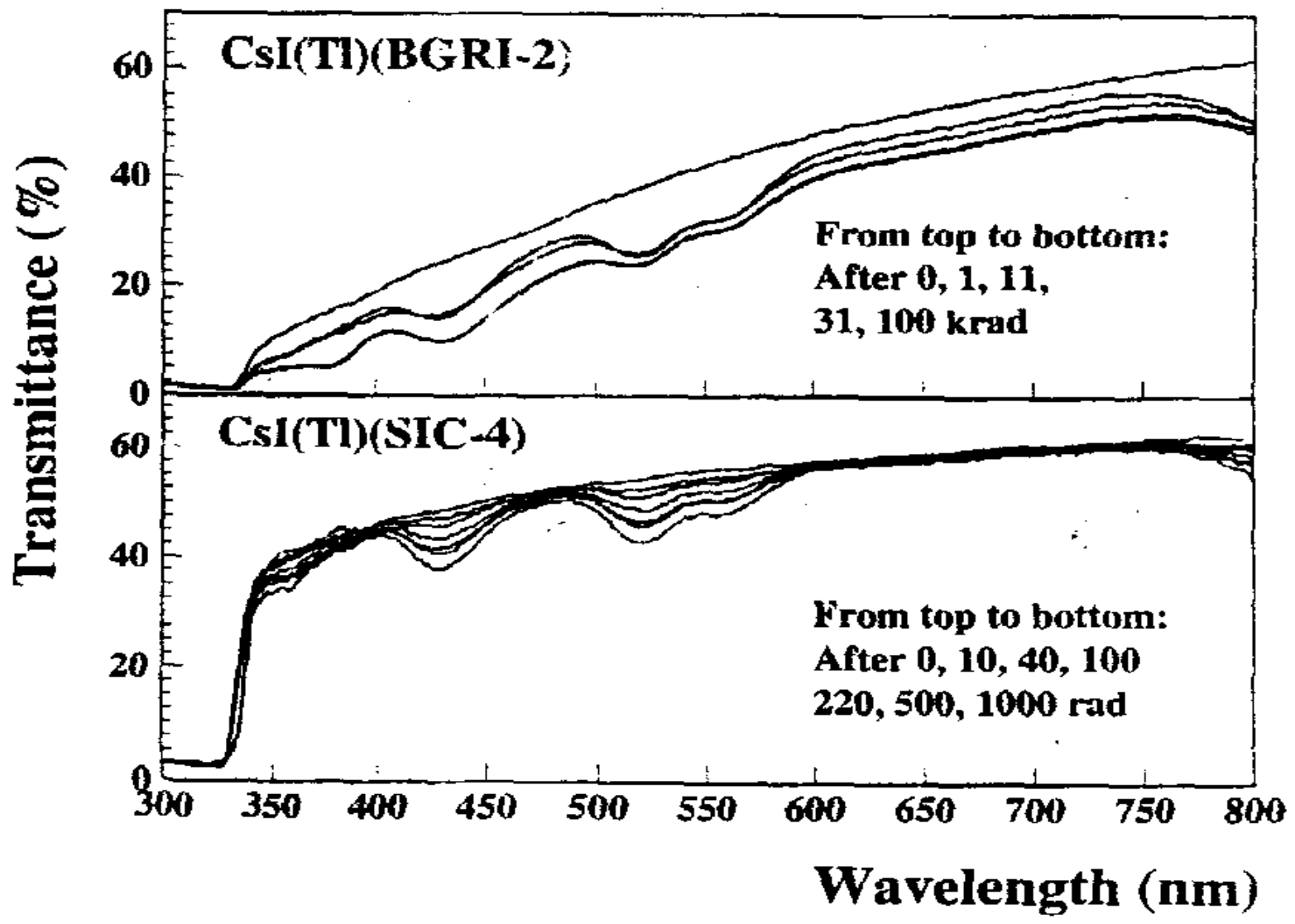

Рис 1.2 Изменение пропускающей способности сцинтиллятора CsI(Tl) в зависимости от длины волны для различных доз облучения

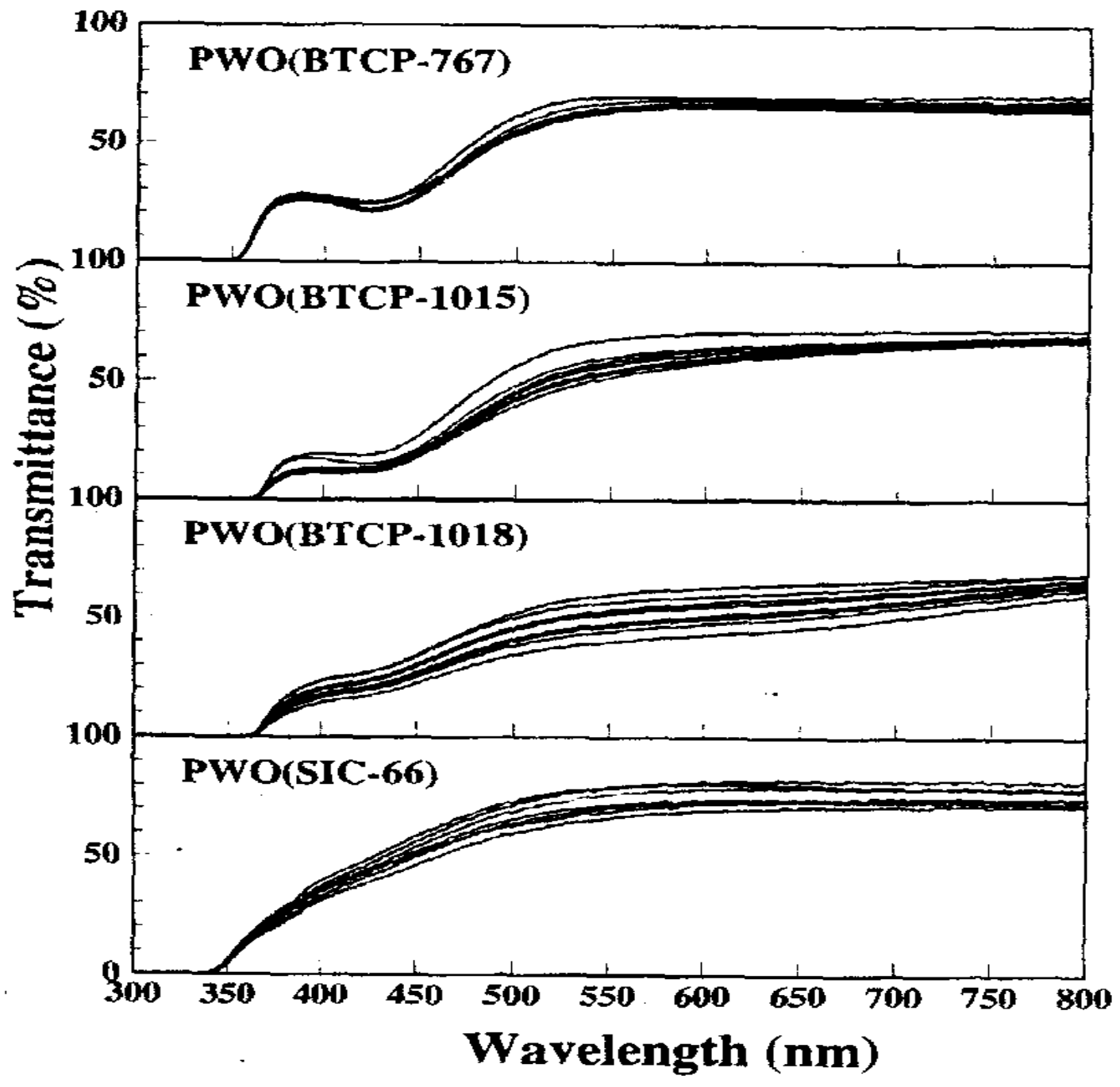

Рис 1.3 Изменение пропускающей способности сцинтиллятора $PbWO_4$ в зависимости от длины волны для различных доз облучения

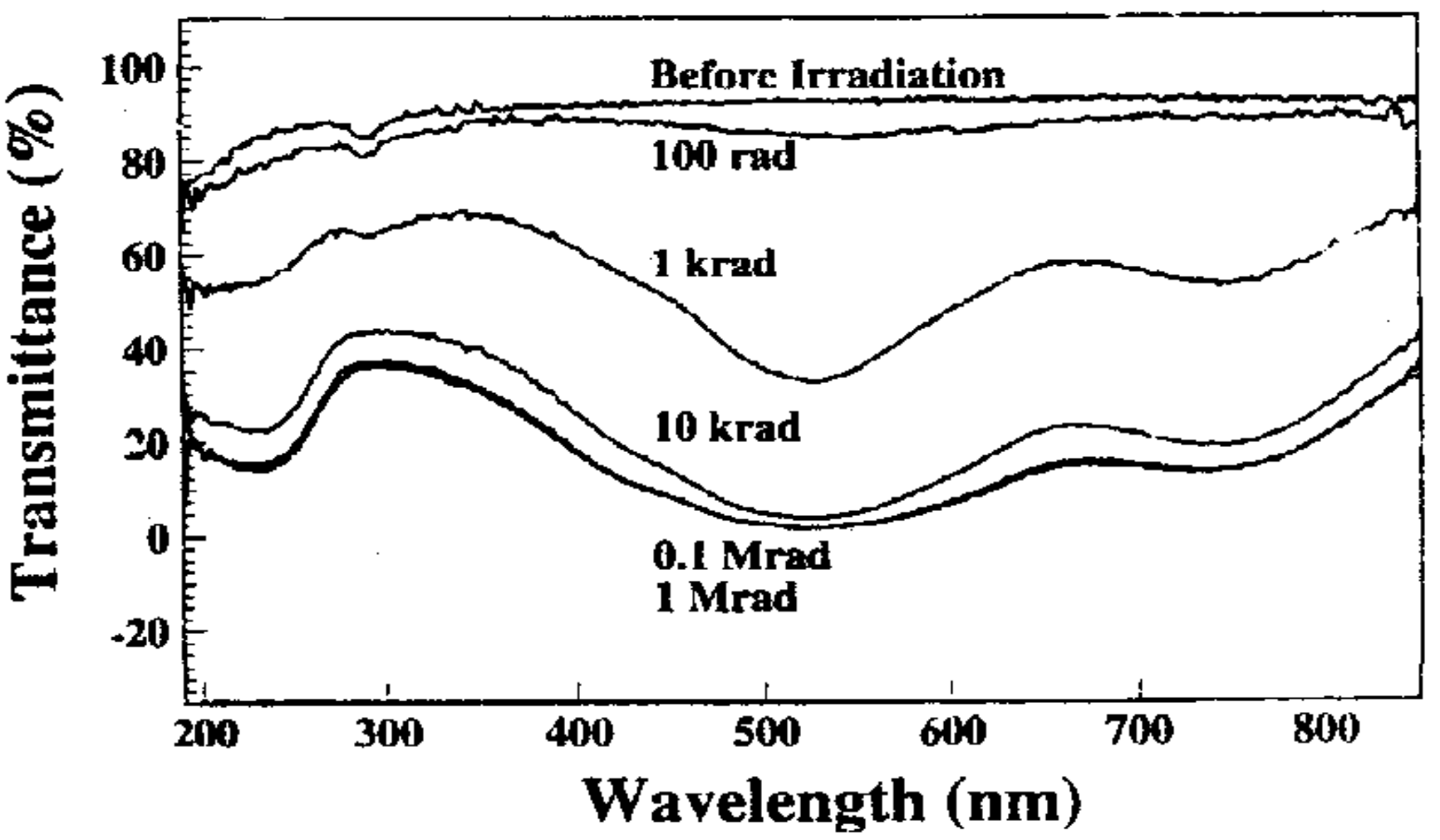

Рис 1.4 Изменение пропускающей способности сцинтиллятора $BaF_2$ в зависимости от длины волны для различных доз облучения

Из рисунков 1.2 – 1.4 видно, что пропускающая способность различных сцинтилляционных кристаллов при облучении заметно изменяется в полосе длин волн собственного излучения. Так при дозе облучения ~ 10 Крад ($2.2\cdot10^{13}$ $\frac{\gamma}{см^2}$), наблюдается, уменьшение пропускающей способности кристаллов: CsI(Tl) (λ ~ 550 нм) примерно на 20%, $PbWO_4$ (λ ~ 460 нм) на 10% - 20% и $BaF_2$ (λ ~ 300 нм) на 60%. Кроме того, следует отметить, что изменение пропускающей способности как следует из рис. 1.4 – 1.5 зависит от технологии изготовления конкретного кристалла / 20, 21 /. Надо отметить, что все эти измерения проводились на массивных образцах сцинтилляционных кристаллов длиной около 30 см.

В таблице 1.2 приведенной в работе / 15 /, показаны эффекты, к которым приводит облучение сцинтилляционных кристаллов большими дозами гамма излучения.

Таблица 1.2

| Кристалл | CsI(Tl) | CsI | $BaF_2$ | BGO | PWO |
|---|---|---|---|---|---|
| Образование центров поглощения | Да | Да | Да | Да | Да |
| Фосфоресцентное послесвечение | Да | Да | Да | Да | Да |
| Изменение механизма сцинтилляции | Нет | Нет | Нет | Нет | Нет |

Из таблицы видно, что механизм сцинтилляции для всех исследуемых кристаллов не испытал ни каких изменений после облучения, и напротив, образование центров поглощения собственного света и фосфоресцентное послесвечение появились во всех образцах / 15, 16, 22, 23 /. Процесс образования центров поглощения собственного света, который приводит к уменьшению амплитуды сигнала на выходе детектора, представляется в нашем случае не существенным. Исходя, из данных представленных выше следует, что при облучении сцинтилляционных кристаллов CsI(Tl) толщиной меньше 1 см пропускающая способность практически не изменяется для исследованных доз облучения. А в нашем случае толщины используемых кристаллов еще меньше и находятся в диапазоне от 1 до 10 мкм. Поэтому для таких толщин кристалла CsI(Tl) изменением пропускающей способности при облучении очевидно можно пренебречь. Надо отметить, что в нашем случае более значительным является процесс фосфоресцентного послесвечения под действием фонового излучения плазмы, так как он приводит к увеличению уровня шума.

Воздействие нейтронов на детекторы и их компоненты при больших интегральных дозах облучения также приводит к деградации их основных параметров / 24 – 26 /. Воздействие нейтронов на сцинтилляционные детекторы, приводит к уменьшению амплитуды сигнала и возрастанию уровня шумов. Это происходит вследствие уменьшения прозрачности стекла входного окна ФЭУ, как и в случае облучения, гамма квантами. Кроме того, облучение нейтронами приводит к деградации эмиссионной способности фотокатода и динодной системы. Воздействие нейтронов на сцинтиллятор приводит к уменьшению его прозрачности к собственному излучению и как следствие к уменьшению амплитуды сигналов. Для фотодиода эффекты, вызываемые радиационным воздействием нейтронов сказываются в уменьшении квантовой эффективности и возрастании напряжения пробоя / 27, 28 /. Нейтроны воздействуют на полупроводниковые детекторы путем генерацией дефектов в кристалле, в результате чего возрастает ток утечки и ухудшается энергетическое разрешение / 29 – 34 /.

На рисунке 1.6 приведены результаты, полученные в работе / 15 / по радиационной стойкости входных окон ФЭУ. Здесь показано изменение пропускающей способности стекла входного окна ФЭУ в зависимости от марки стекла и от длины волны излучения для различных интегральных потоков нейтронов.

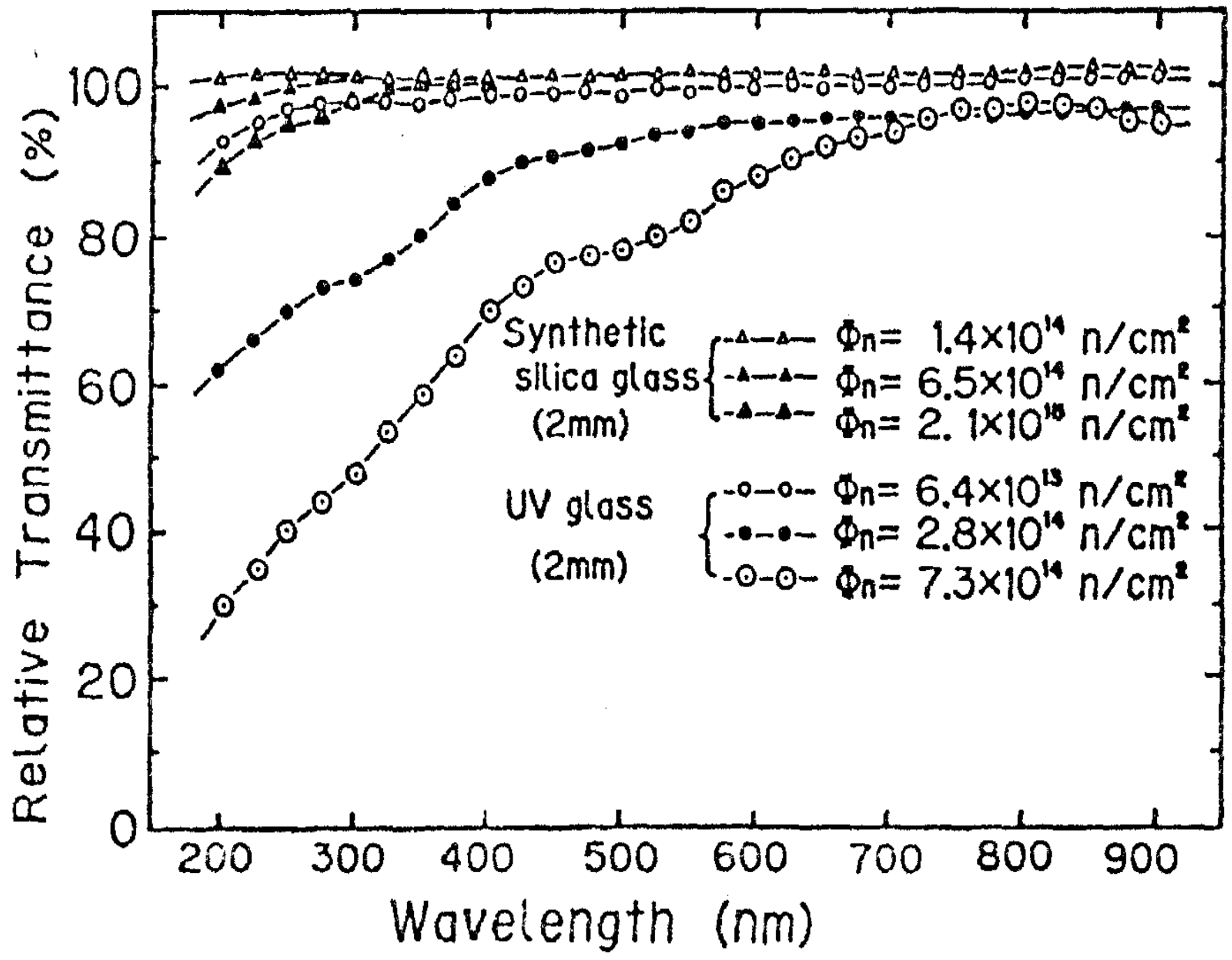

Рис 1.6 Изменение спектральной чувствительности стекла при облучении быстрыми нейтронами

Видно что, вне зависимости от сорта стекла входного окна ФЭУ наибольшая деградация наблюдается в области коротких длин волн / 35 /. В интересующей нас области длин волн ($\lambda \sim 550$ нм, т.е. то значение длин волн, которое соответствует собственному излучению сцинтилляционного кристалла CsI(Tl)) пропускающая способность, для потока $\sim 6 \cdot 10^{14}\ \frac{n}{см^2}$ падает на 20 % для увиолевого стекла входного окна ФЭУ, а для кварцевого стекла практически не изменяется / 36 /. Резюмируя вышесказанное, можно отметить, что выбор ФЭУ с входным окном из кварцевого стекла является целесообразным для применения в детектирующих системах анализаторов подвергающихся интенсивному фоновому излучению / 37 /.

Рассмотрим данные, полученные в работе / 13 / по радиационной стойкости различных детекторов при облучении нейтронами с энергией 14 МэВ В таблице 1.3 показана величина потоков быстрых нейтронов, при которой начинается деградация различных параметров детекторов и их компонентов.

Таблица 1.3

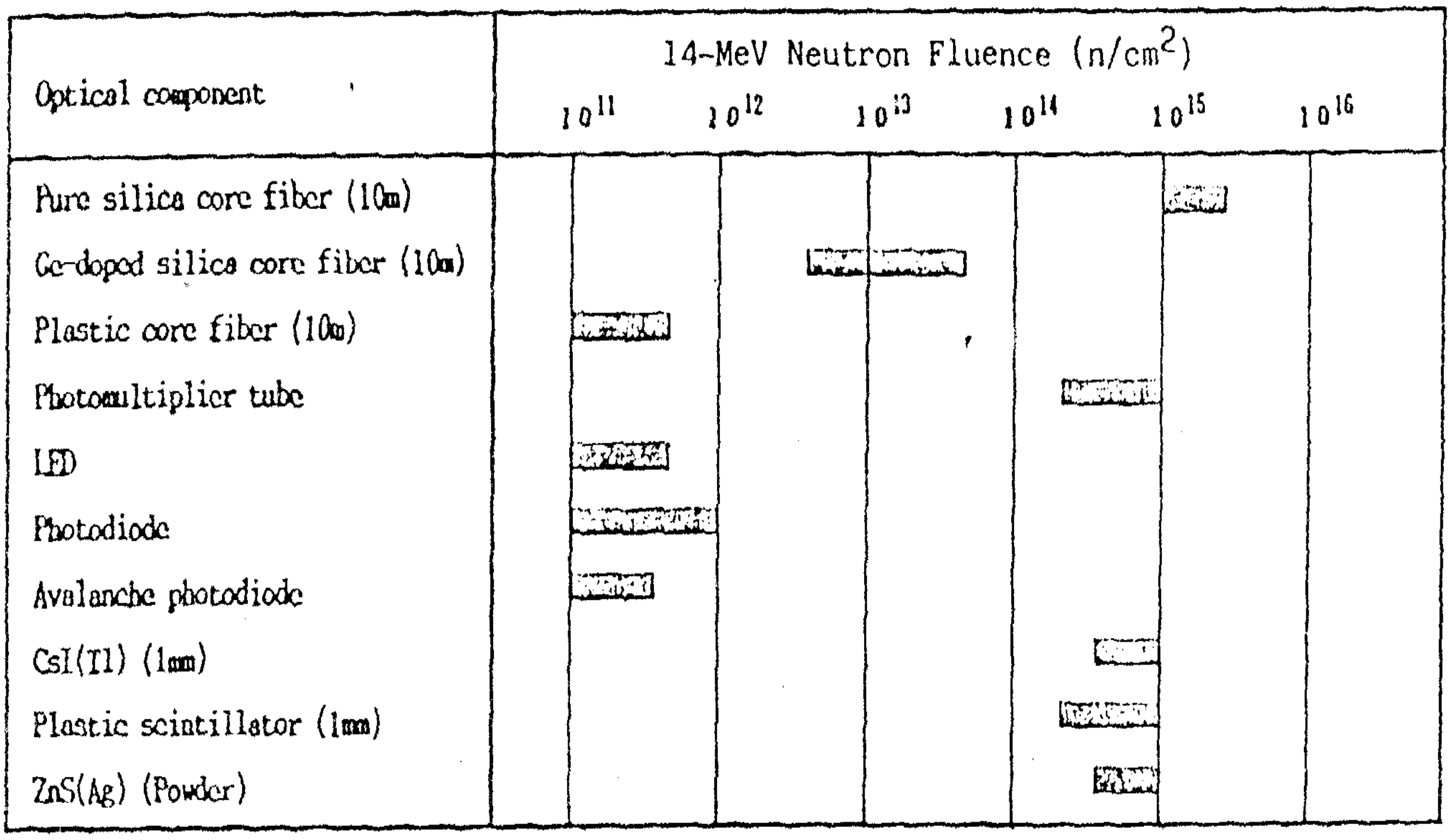

Из таблицы видно, что как для ФЭУ, так и для сцинтилляционного кристалла CsI(Tl) заметное изменение параметров начинается при потоке нейтронов ~ $10^{14}\frac{n}{см^2}$, а радиационная деградация начинает сказываться при потоке $6\cdot10^{14}\frac{n}{см^2}$, таким образом, радиационная стойкость сцинтилляционного кристалла CsI(Tl) сравнима с радиационной стойкостью ФЭУ. Следует отметить, что для фотодиода, лавинного фотодиода и полупроводникового / 38 – 40 / детектора радиационная деградация параметров происходит при существенных меньших потоках ~ $10^{11}$ - $10^{12}$ $\frac{n}{см^2}$ / 41 /.

В работе / 42 / представлены результаты исследования параметров различных типов детекторов, облучаемых нейтронами с энергией 14 МэВ в зависимости от величины интегрального потока нейтронов. На рис.1.1 представлены данные, приведенные в этой работе.

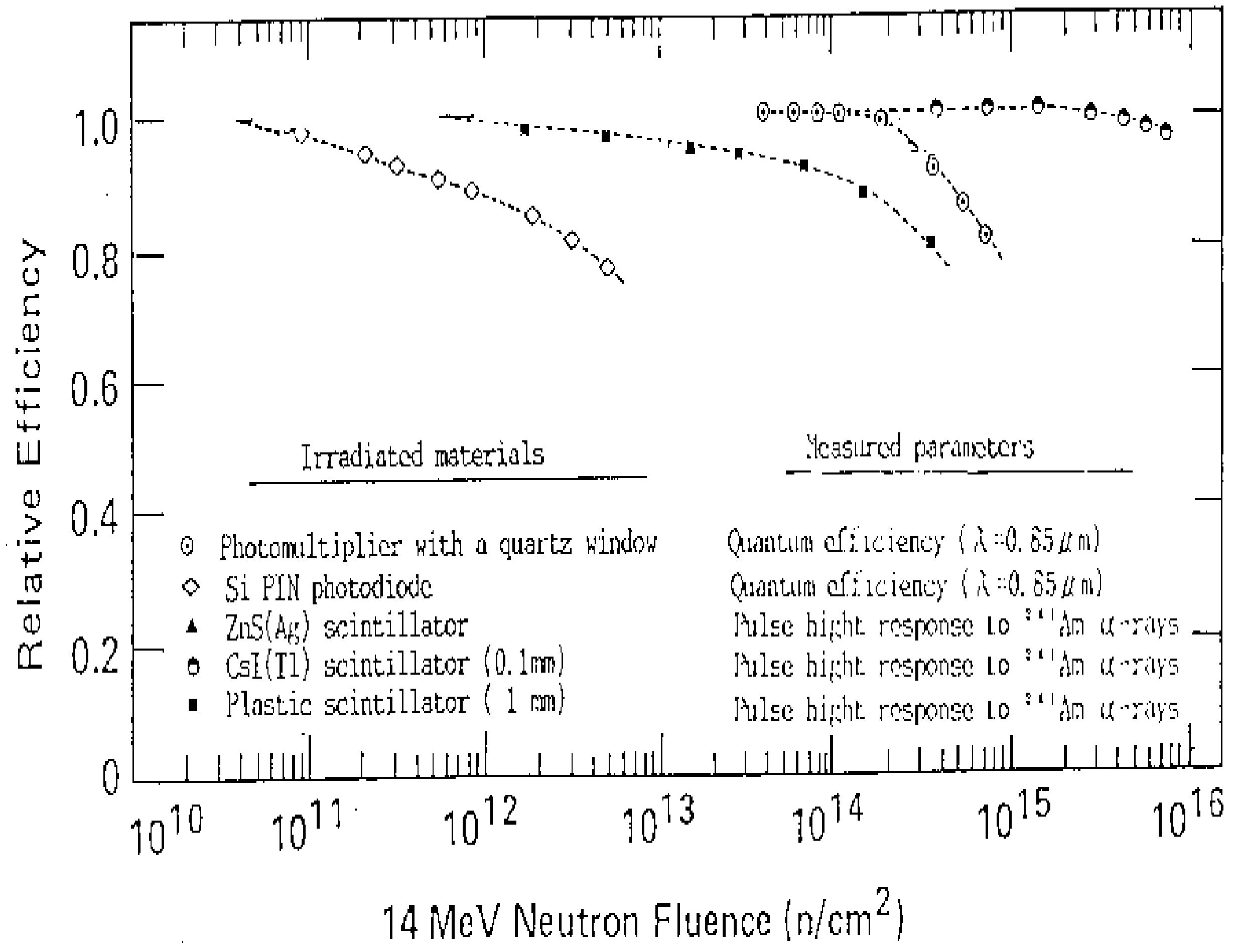

Рис. 1.5 Зависимость изменения относительной эффективности различных детекторов от дозы облучения нейтронами с энергией 14 МэВ

Из рис. 1.5 видно, что квантовая эффективность ФЭУ при длине волны $\lambda$ ~ 0.85 мкм деградирует на 20% при потоке около $10^{15}$ $\frac{n}{см^2}$. Для сцинтилляционного кристалла CsI(Tl) толщиной 0.1 мм заметное ухудшение энергетического разрешения при регистрации альфа частиц от источника $^{241}Am$ наблюдается при потоке нейтронов ~ $10^{16}$ $\frac{n}{см^2}$ / 43, 44 /. Отметим, что кремниевый фотодиод имеющий наименьшую радиационную стойкость из всех сравниваемых детекторов, уменьшает энергетическое разрешение при регистрации альфа частиц от источника $^{241}Am$ на 20 % при потоке ~ $10^{12}$ $\frac{n}{см^2}$ / 45 – 49, 30 /.

Исходя, из представленных выше данных видно, что одним из самых радиационно-стойких детекторов при облучении нейтронами с энергией 14 МэВ является сцинтилляционный детектор, изготовленный на основе кристалла CsI(Tl).

Величина требуемой радиационной стойкости детекторов, используемых в настоящее время на существующих крупных термоядерных установках, зависит от места установки таких детекторов.

При установке детектирующей системы вблизи первой стенки токамака интегральный поток нейтронов и гамма излучения за один импульс для установки JET (см. таблицу 1.1) не превышает $10^{13}$ $\frac{частиц}{см^2}$ и даже для сцинтилляционных детекторов без защиты полное число мощных импульсов, приводящих к заметной деградации их параметров, может составить от нескольких десятков до сотни. При установке детектирующей системы на расстояние от 5 – 10 м (обычное положение анализаторов) от первой стенки радиационный ресурс существенно повышается (так как величина полного потока нейтронов и гамма квантов падает на 2 – 4 порядках). Таким образом, даже полупроводниковые детекторы / 50, 51 / и фотодиоды могут использоваться для регистрации корпускулярного и оптического излучения плазмы. Существенно более высокие интегральные потоки нейтронов и гамма квантов ожидаются на установке ITER, где реактор проектируется для работы в непрерывном режиме.

## 2. Чувствительность к фоновому нейтронному и гамма излучению

Проанализируем теперь данные измерений чувствительности к проникающему фоновому излучению плазмы для детекторов используемых в системах корпускулярной диагностики.

В таблице 1.4 представлены экспериментальные данные, полученные в работе / 36 / по чувствительности к нейтронному потоку с энергией 14 МэВ для ФЭУ (фотоэлектронного умножителя), МКП (микроканальной пластины), КЭУ (канально-электронного умножителя), ЭУ (электронного умножителя).

Из данных представленных в таблице видно, что чувствительность всех детекторов к фоновому излучению различается между собой, при этом, различие достигает величины нескольких порядков. Электронный умножитель обнаруживает наименьшую чувствительность к потоку нейтронов с энергией 14 МэВ, а фотоэлектронный умножитель – наибольшую. Данные приведенные, в этой таблице отвечают, использованию этих детекторов в режиме пороговой регистрации (когда регистрируются события с амплитудой превосходящей некоторую заданную величину). При этом, как видно из приведенных выше амплитудных распределений разделение, фоновых и полезных сигналов практически не возможно.

Таблица 1.4

Response of some diagnostic detectors to 14-MeV neutrons

| Detector | Pulse hight distribution | Detection efficiency $(n/cm^2)^{-1}$ |
|---|---|---|
| Photomultiplier (R647-04) | 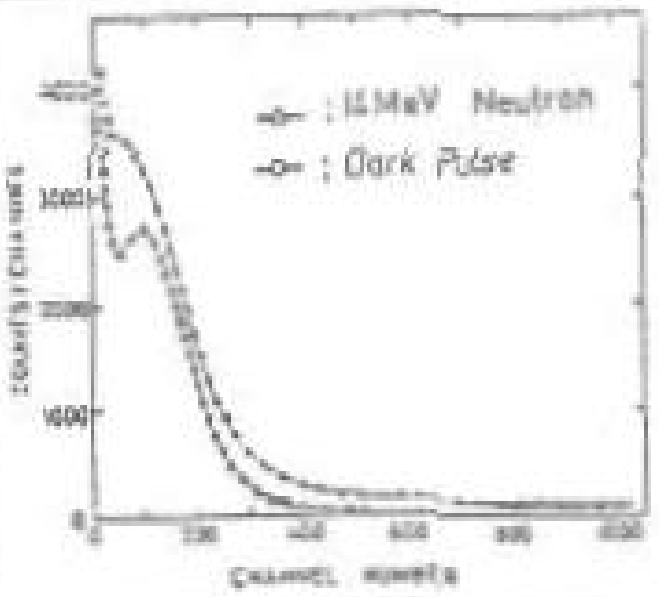  | $2\times10^{-2}$ |
| Microchannel plate (F1551) | 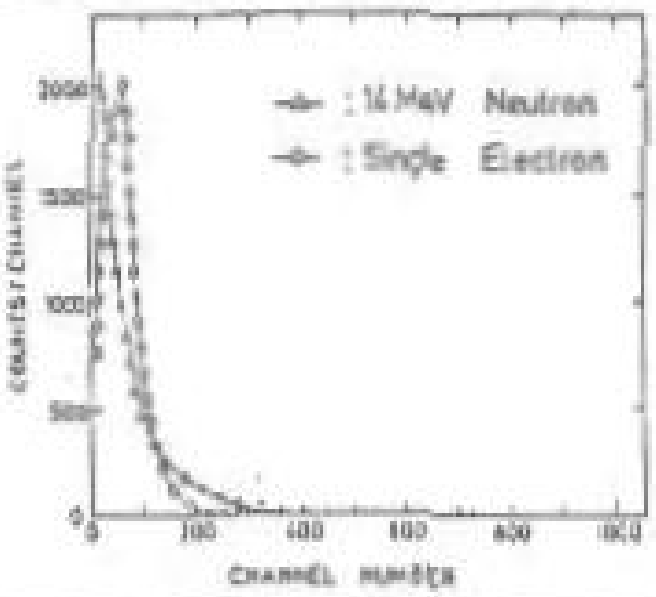  | $1\times10^{-3}$ |
| Ceratron (EMS-6081B) | 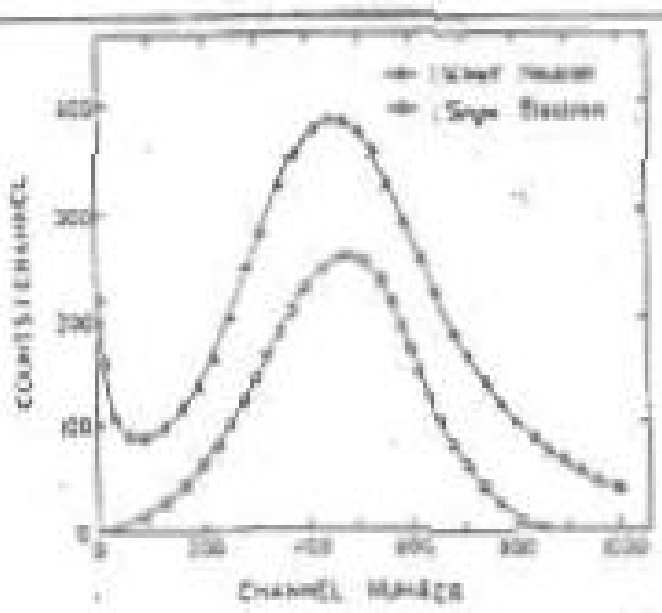  | $2\times10^{-3}$ |
| Electron multiplier (R515) | 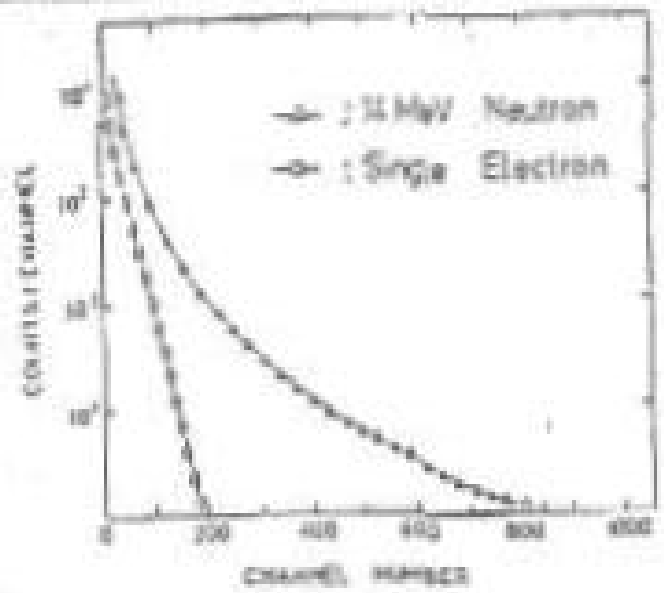  | $3\times10^{-4}$ |

Однако при использовании ФЭУ в составе сцинтилляционного детектора можно реализовать амплитудное разделение сигналов возникающих при регистрации корпускулярного и фонового излучения и тем самым уменьшить чувствительность к фоновому излучению.

Исходя из представленных выше характеристик различных детекторов используемых в корпускулярной диагностике можно заключить, что более перспективным для регистрации корпускулярного излучения при условиях интенсивного нейтронного и гамма фона, характерного для крупных современных термоядерных установок является детектор, который позволяет реализовать амплитудный анализ сигналов, возникающих при регистрации фонового и корпускулярного излучения плазмы. К такому классу детекторов относятся полупроводниковый и сцинтилляционный детекторы / 52 – 57 /. Однако последний обладает существенно более высокой радиационной стойкостью и, кроме того, использование сцинтилляционных кристаллов CsI(Tl) позволяет легко получить сцинтилляторы оптимизированные по толщине.

# ГЛАВА II. ОПИСАНИЕ ЭКСПЕРИМЕНТАЛЬНОЙ УСТАНОВКИ И МЕТОДИКИ ИЗМЕРЕНИЙ

## 1. Методика напыления сцинтилляционных кристаллов

Из соображений приведенных выше в качестве сцинтиллятора, был выбран CsI(Tl). Для получения оптимальных по толщине сцинтилляционных кристаллов была использована методика вакуумного напыления CsI(Tl) на подложку из кварцевого стекла. Анализ результатов измерений чувствительности детекторов к фоновому излучению показал, что оптимальная толщина сцинтилляционных кристаллов должна быть близка к пробегу регистрируемых частиц. В нашем случае это соответствует диапазону толщин кристаллов в пределах от одного до десятков микрон.

Блок-схема установки для напыления кристаллов, созданная на базе универсального вакуумного поста «ВУП-2К» приведена на рис. 2.1. Распыление CsI(Tl) осуществлялось из вольфрамового нагревателя, установленного непосредственно под печкой, на которой крепилась подложка. Температура подложки контролировалась в процессе напыления с помощью термопары. Скорость распыления регулировалась величиной тока, пропускаемого через нагреватель. Вакуум в процессе напыления был не хуже, чем $5 \cdot 10^{-5}$ мм рт. ст.

С целью получения хорошей адгезии паров CsI(Tl) к подложке поверхность последней перед напылением тщательно очищалась с помощью толуола, затем подложка просушивалась чистой батистовой тканью и после откачки до рабочего вакуума прогревалась в течение 35 минут при температуре близкой к рабочей.

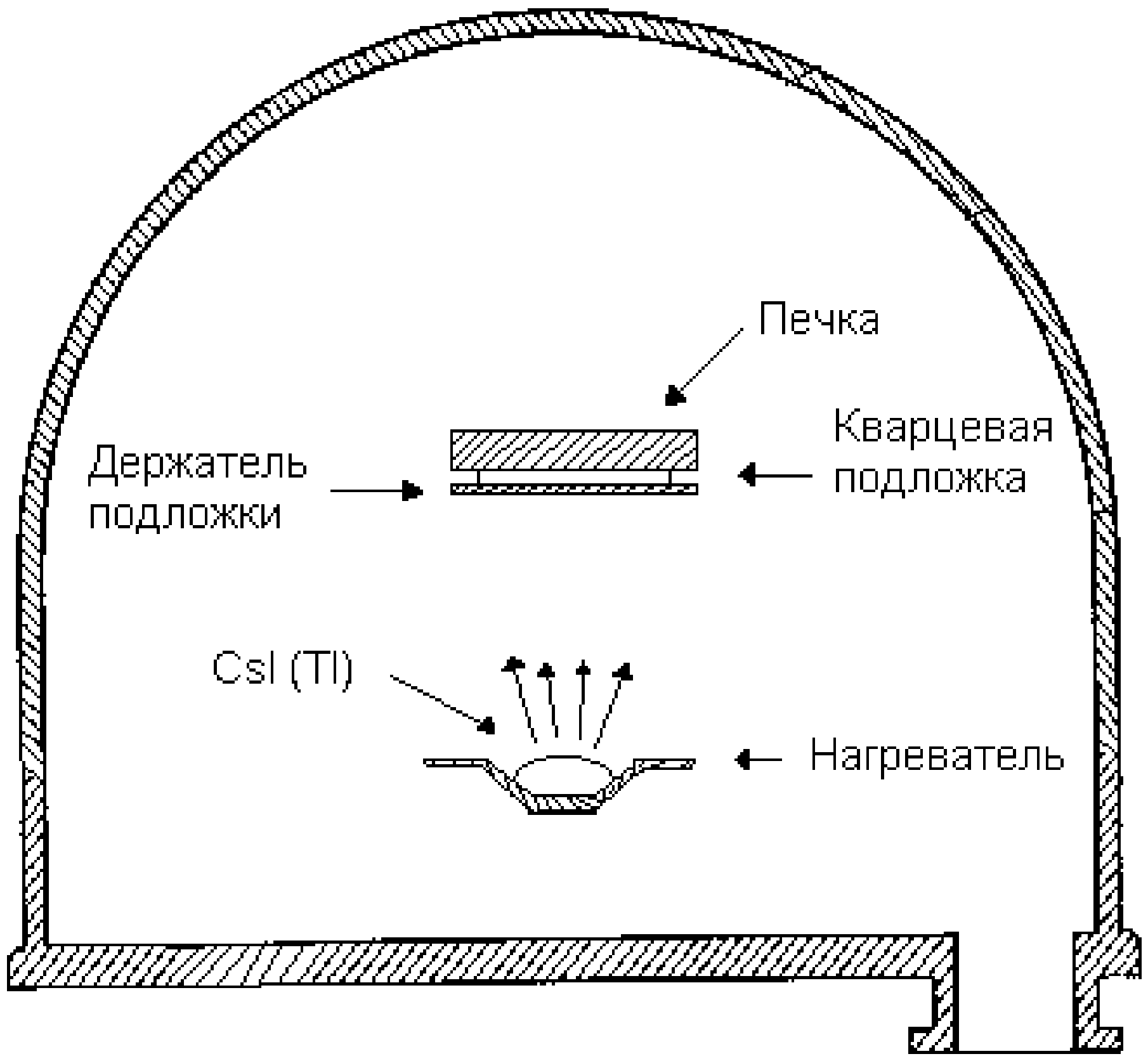

Рис. 2.1- Блок-схема установки для напыления кристаллов CsI(Tl)

Как было отмечено выше в качестве исходного материала для напыления, использовались сцинтилляционные кристаллы CsI(Tl). Известно, что сцинтилляционные свойства этих кристаллов в существенной мере определяются процентным содержанием Tl / 58 /. Поэтому основной проблемой при напылении кристаллов на подложку является сохранение их сцинтилляционных свойств. Предварительные исследования показали, что световыход напыленного сцинтиллятора (а следовательно, и процентное содержание Tl) зависит от температуры подложки во время напыления и скорости напыления. Поэтому была проведена серия экспериментов по оптимизации световыхода напыленного кристалла. В результате было получено, что оптимальная температура подложки должна быть 135°С и скорость напыления – приблизительно 0.5 мкм/мин.

При этих условиях напыления световыход полученных кристаллов оказался даже несколько выше, чем у массивного образца, который использовался для напыления. Полученные результаты иллюстрируются рис. 2.2, на котором приведены амплитудные спектры гамма квантов ($E_\gamma$ = 5.9 КэВ) для массивного и напыленного образцов.

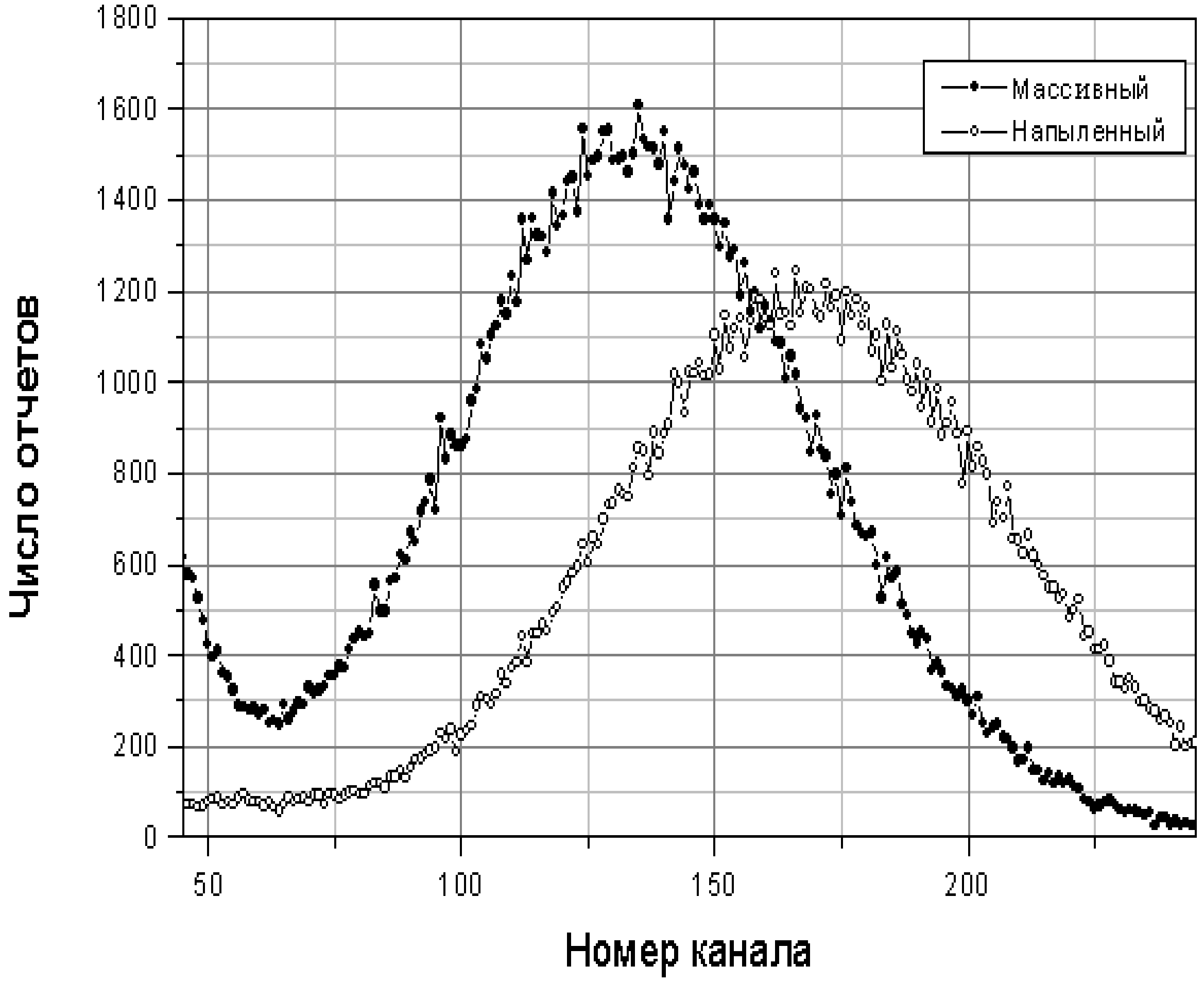

Рис. 2.2- Амплитудные спектры гамма квантов ($E_\gamma$ = 5.9 КэВ) для массивного и напыленного образцов

Так как при напылении необходимо было получать кристаллы определенной толщины, то была проведена калибровка напылительной установки. Предварительно масса распыляемого кристалла определялась исходя из заданной геометрии и требуемой толщины напыленного на подложку кристалла. Измерение толщины полученных при напылении кристаллов проводилось двумя методами.

Первый метод заключался в определении толщины кристалла, напыленного на подложку, по поглощению коллимированного пучка гамма квантов с энергией $E_\gamma$ = 5.9 КэВ от источника $^{55}Fe$. Вначале, с помощью массивного кристалла, установленного на ФЭУ измерялась полная интенсивность коллимированного пучка $I_0$. Затем, на тот же ФЭУ устанавливался напыленный кристалл, и измерялась интенсивность поглощенных в нем гамма квантов (I'). После этого, толщина кристалла ($x$) определялась по известному выражению:

$$I' = I_o\left(1 - e^{-\mu \cdot x}\right) \qquad (2.1)$$

где: μ- коэффициент линейного ослабления гамма квантов с энергией 5.9 КэВ в CsI(Tl), x – толщина исследуемого кристалла CsI(Tl).

Второй способ заключался в определении толщины кристалла по величине энергии $\Delta E$, выделяющейся в кристалле при прохождении через него альфа частиц с энергией $E_0$. По энергии $E_1 = E_0 - \Delta E$ определялся остаточный пробег $R_1$ и далее находилась толщина кристалла по методике предложенной в работе / 59 /.

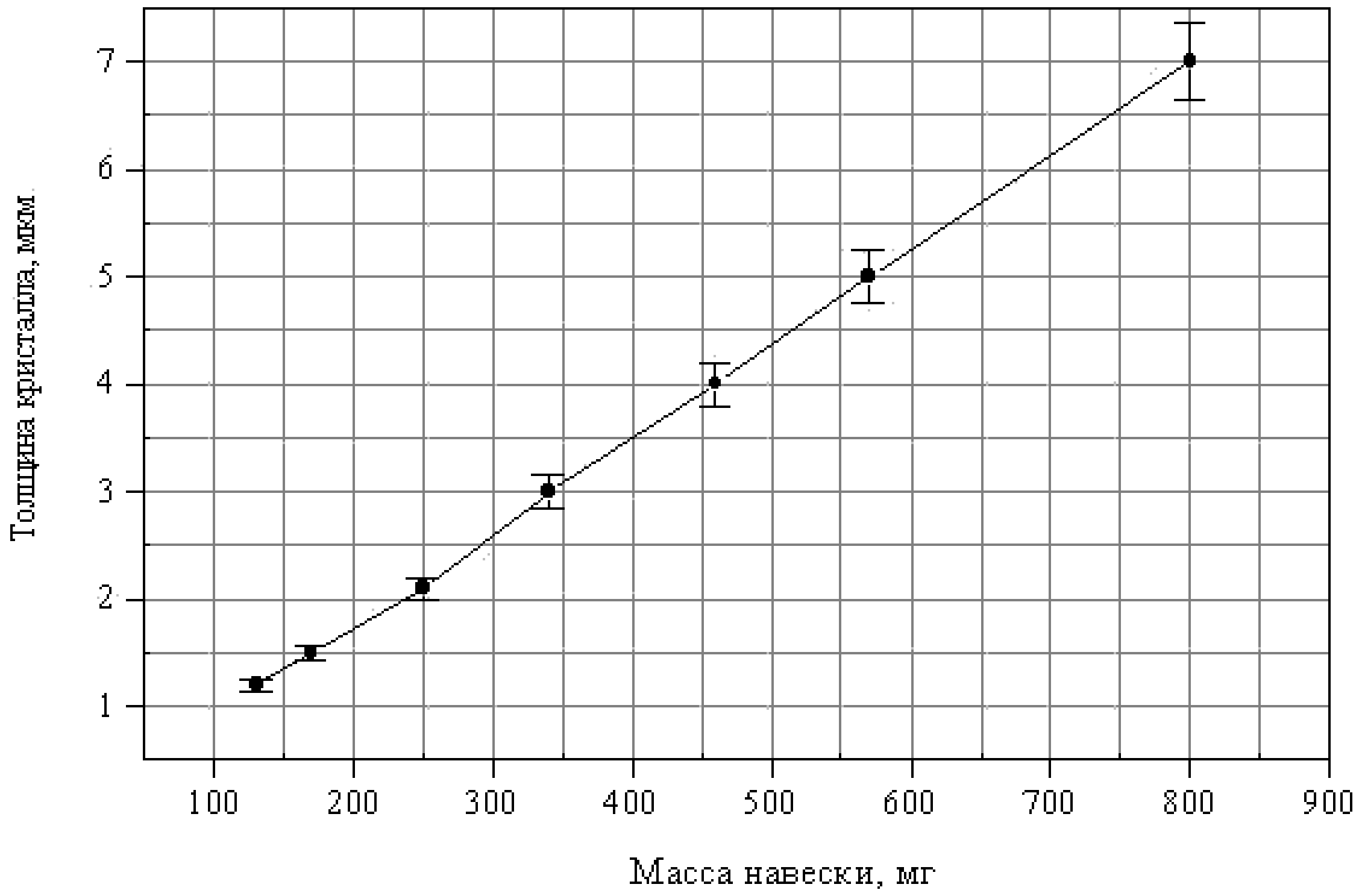

Рис. 2.3- Калибровочная кривая зависимости толщины напыленного кристалла от массы распыляемого CsI(Tl)

На рис. 2.3 показаны результаты проведенной калибровки. Из рисунка видно, что зависимость толщины напыленного кристалла от массы навески близка к линейной. Величина погрешности, приведенная на графике, включает в основном погрешность неконтролируемых процессов, которые происходят при распылении массивного кристалла путем сублимации и образования сцинтилляционного кристалла на подложке.

Эта методика позволила нам изготовить свыше 60 кристаллов оптимизированных по толщине и имеющих хорошие сцинтилляционные характеристики. Все эти сцинтилляторы были использованы в детектирующих системах анализаторов, установленных на крупнейших в мире токамаках.

## 2. Методика измерения фоновых характеристик детекторов

С целью исследования детекторов в условиях схожих с теми, которые встречаются в реальных термоядерных установках, был сконструирован специальный стенд. Этот стенд использовался для измерения чувствительности детекторов (сцинтиллятор и фотоэлектронный умножитель) к фоновому нейтронному и гамма излучению. Схема стенда представлена на рис. 2.4.

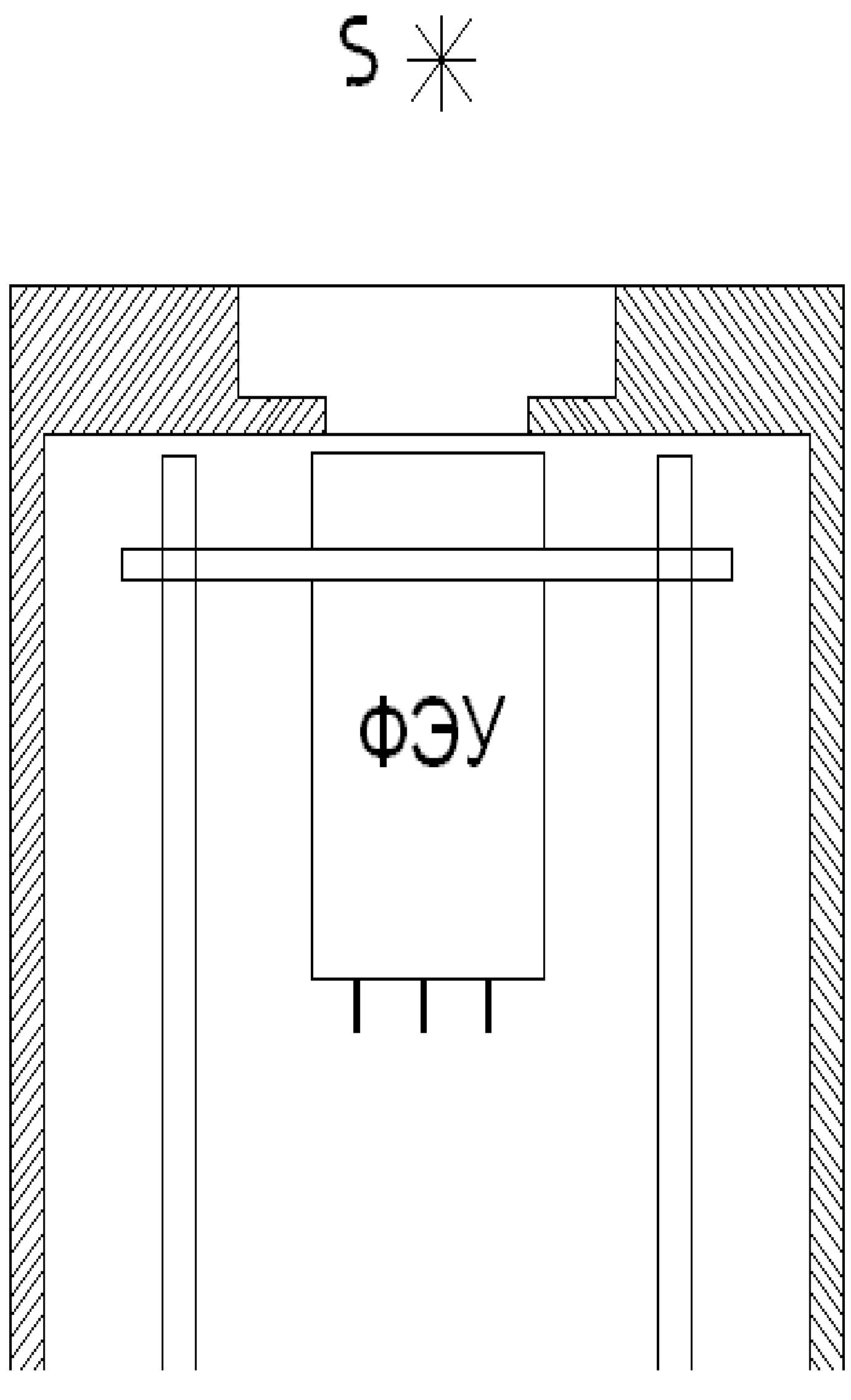

Рис. 2.4- Схема экспериментального стенда для исследования фоновых характеристик сцинтилляторов и фотоэлектронных умножителей

Детектор устанавливался в железной камере с толщиной стенок ~ 20 мм. В верхней части камеры было предусмотрено отверстие для установки различных источников облучения. При облучении детекторов источниками, имитирующими фоновое излучение, это отверстие закрывалось крышкой. При этом чувствительная плоскость детектора располагалась практически вплотную к крышке. В такой геометрии исключалось образование вторичных частиц в реакциях (n,p), (n,$\alpha$) в слое воздуха между детектором и крышкой камеры / 60, 61 /.

С целью исследования фоновых характеристик других типов детекторов, таких как: микроканальная пластина и канальный электронный умножитель в условиях, имитирующих реальные термоядерные установки, был сконструирован фланец, который стыковался с вакуумной камерой электронографа. Блок-схема этого фланца представлена на рисунке 2.5.

Детектор устанавливался на подвижной подставке, которая перемещалась вдоль горизонтальной плоскости, давая возможность установить детектор под различные источники облучения. Таким образом, сконструированный фланец был использован для исследования фоновых характеристик, как микроканальной пластины, так и канального электронного умножителя.

Определение чувствительности детектора к фоновому излучению является одним из ключевых вопросов при разработке детектирующей системы способной регистрировать альфа частицы, протоны, дейтроны и тритоны в условиях интенсивного фонового излучения / 62 – 65 /. Поэтому детально рассмотрим все фоновые характеристики, которые были измерены для каждого отдельного детектора.

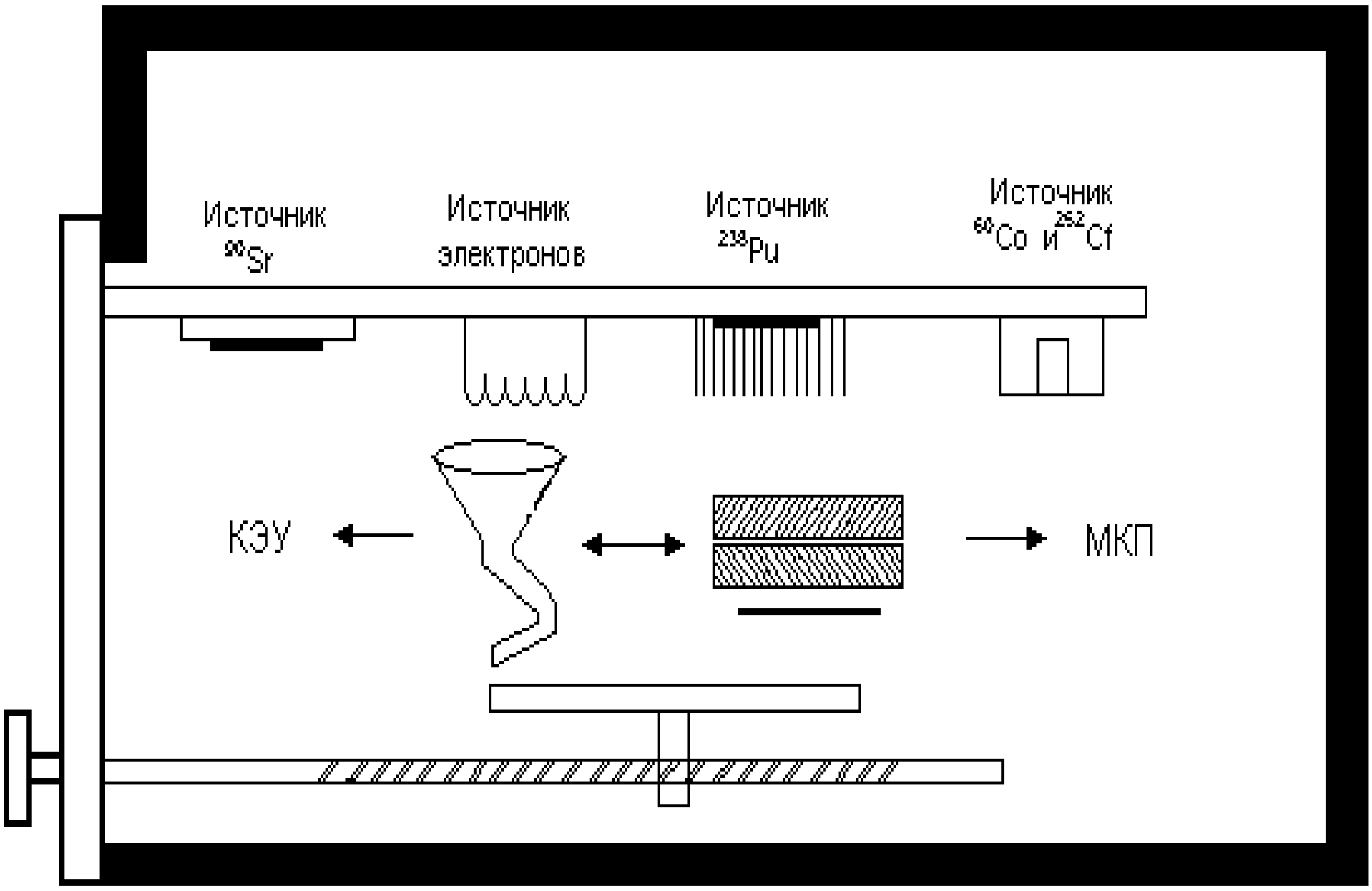

Рис 2.5- Блок-схема фланца для измерения фоновых характеристик детекторов типа МКП и КЭУ.

Для определения чувствительности детектора, первоначально измерялось амплитудное распределение сигналов, возникающих в детекторе под действием радиоактивного источника, имитирующего фоновое излучение плазмы.

Для измерения амплитудного распределения использовалась стандартная измерительная система. Электрические сигналы с выхода детектора поступали в ЗЧПУ (зарядо-чувствительный предусилитель), где формировались и усиливались. Затем сигналы поступали на спектрометрический усилитель с гауссовским формированием сигналов (со временем нарастания $\tau_н$ от 1 мкс. до 3 мкс.) и далее на вход АЦП (аналого-цифровой преобразователь). Использовался АЦП с буферной памятью в стандарте КАМАК. Блок-схема измерительного тракта представлена на рис. 2.7.

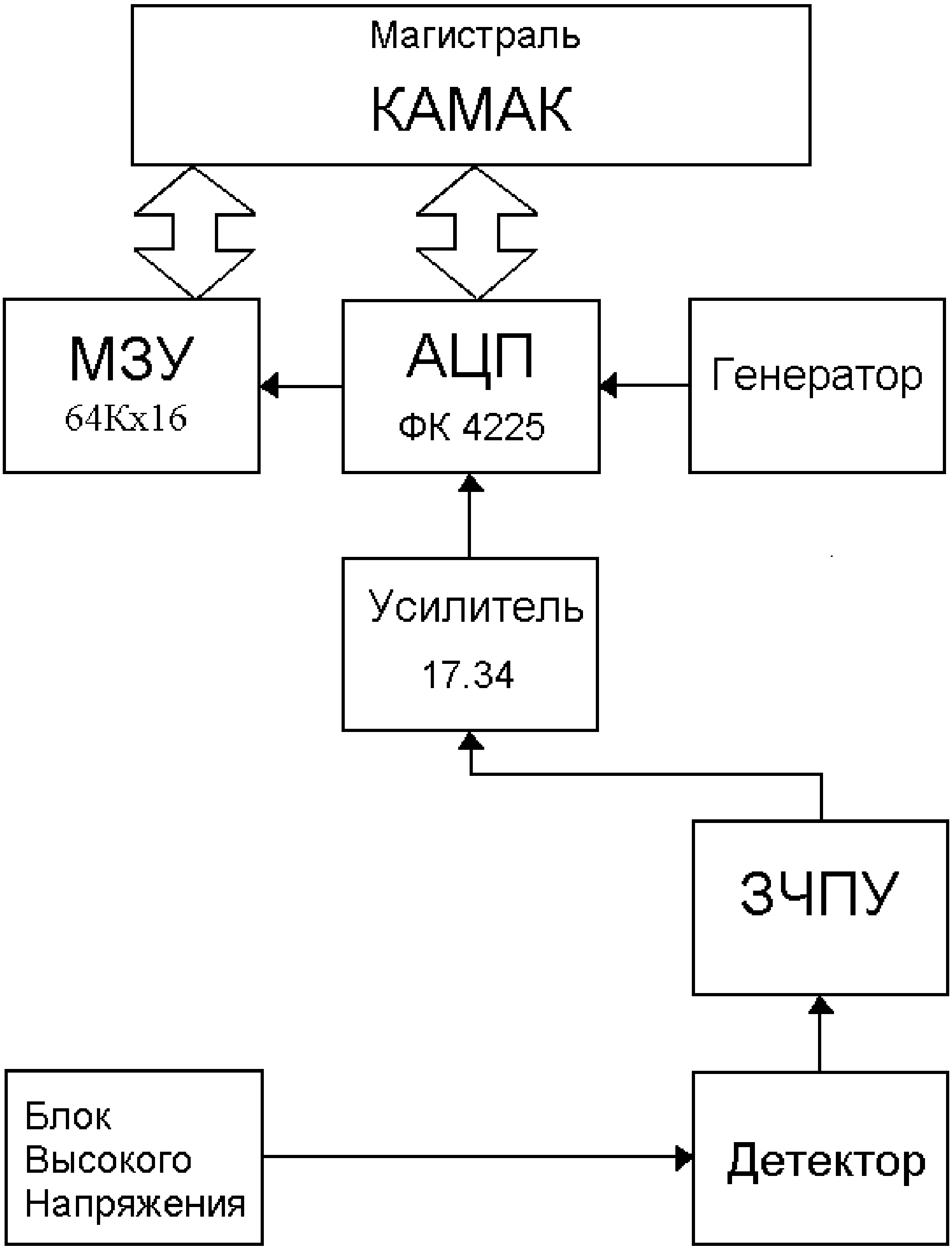

Рис. 2.7 - Блок-схема измерительного тракта для измерения амплитудного распределения фонового излучения

В качестве примера на рисунке 2.6 приведен амплитудный спектр, полученный таким образом для ФЭУ.

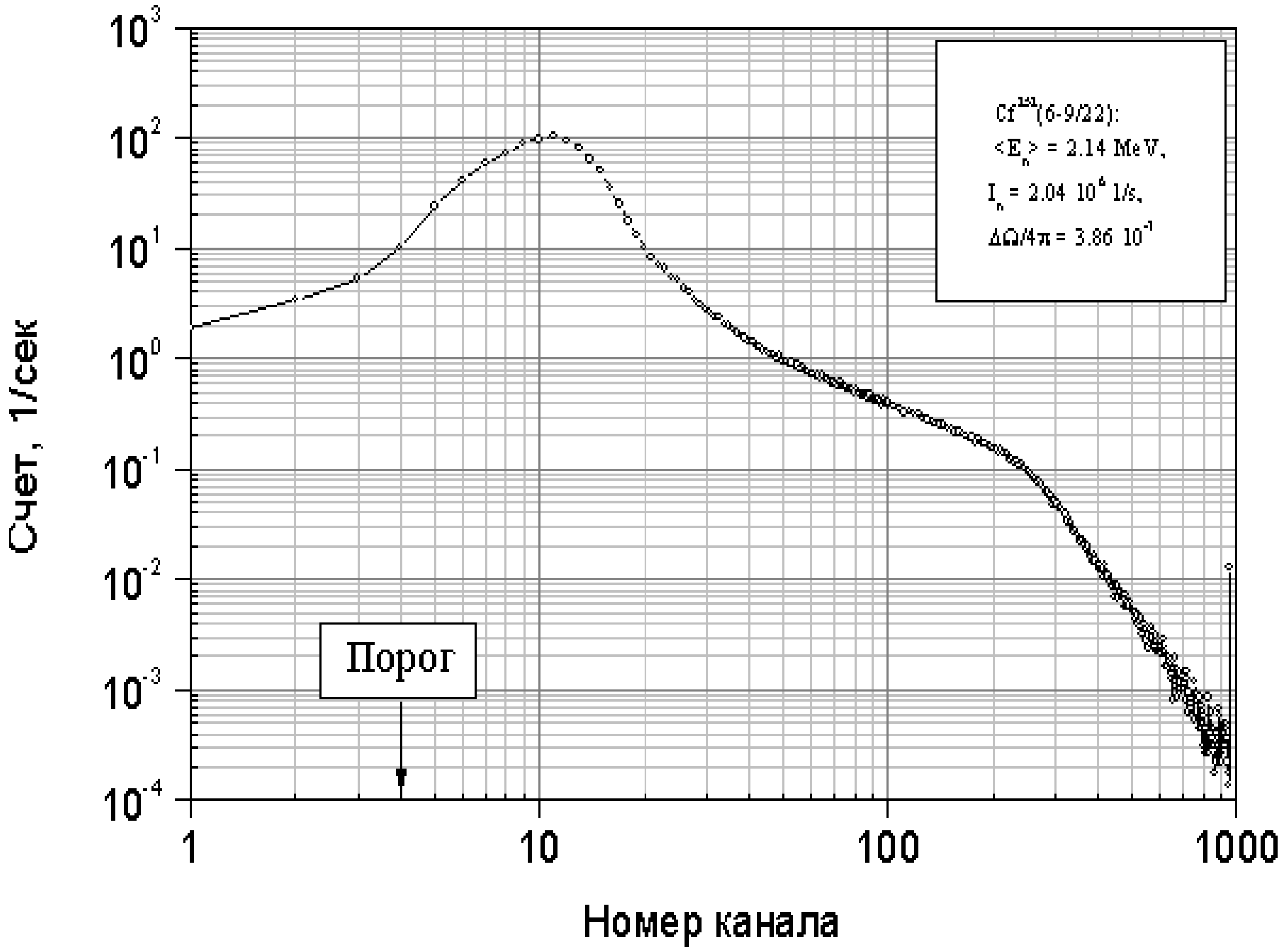

Рис 2.6- Амплитудное распределение при облучении ФЭУ источником $^{252}$Cf.

Для сравнения детекторов мы пользовались величинами дифференциальной и интегральной чувствительностей детекторов к фоновому нейтронному и гамма излучению. Рассмотрим определения каждой из этих величин.

Дифференциальная чувствительность детекторов к фоновому излучению $\frac{dS}{dU}$ определялась по следующему выражению:

$$\frac{dS}{dU} = \frac{1}{I_{П} \cdot t} \cdot \frac{dn}{dU} \qquad (2.2)$$

где: $\frac{dn}{dU}$ - амплитудное распределение сигналов, возникающих при регистрации фонового нейтронного и гамма излучения, $I_{П}$ - полная интенсивность падающих на детектор нейтронов и/или гамма квантов, $t$ - время измерения.

Величина $I_{П}$ - полная интенсивность нейтронов и/или гамма квантов, падающих на чувствительную поверхность детектора в заданной геометрии, рассчитывалась из соотношения:

$$I_{П} = I_0 \cdot \frac{d\Omega}{4\pi} \qquad (2.3)$$

где: $I_0$ - интенсивность частиц вылетевших в телесный угол $4\pi$. Величина $I_0$ определялась активностью данного радиоактивного источника на день измерения.

Относительный телесный угол $\frac{d\Omega}{4\pi}$ был рассчитан по соотношению.

$$\frac{d\Omega}{4\pi} = \frac{1}{2}\left(1 - \frac{h}{\sqrt{h^2 + r^2}}\right) \qquad (2.4)$$

где: $h$ - расстояние от источника до плоскости чувствительной поверхности детектора, $r$ - радиус чувствительной поверхности детектора.

Данное выражение получено для точечного источника, что в нашем случае было оправдано, в связи с тем, что размеры активной области источника были достаточные малы по отношению к  расстоянию до детектора.

Полная чувствительность детектора $S_t(U)$ определялась как:

$$S_t(U) = \int_0^{\infty} \frac{dS}{dU} \cdot dU \qquad (2.5)$$

где: $\frac{dS}{dU}$ - Дифференциальная чувствительность детекторов к фоновому излучению. Интегрирование проводилось от нулевого порога до максимального значения порога. Кроме того, мы использовали интегральную чувствительность детектора при заданном пороге регистрации $S(U_П)$. Расчет проводился по следующему выражению:

$$S(U_П) = \frac{1}{I_П \cdot t} \int_{U_п}^{\infty} \frac{dn}{dU} \cdot dU \qquad (2.6)$$

где: $\frac{dn}{dU}$ - амплитудное распределение сигналов, возникающих при регистрации фонового нейтронного и гамма излучения. Интегрирование проводилось от заданного порога $U_п$ до максимального значения порога.

Дифференциальная и интегральная чувствительности представляются в виде зависимости от амплитуды сигнала, выраженной либо в числе, регистрируемых фотоэлектронов, либо в энергетической шкале. Для представления в фотоэлектронах проводилось определение амплитуды, одноэлектронного пика и по этим значениям пересчитывалась вся шкала. Для пересчета в энергетическую шкалу был использован источник альфа частиц $^{238}Pu$ с максимальной энергией альфа частиц ($E_\alpha$= 5.5 МэВ). Энергия падающих частиц варьировалась путем изменения толщины воздушного слоя между источником и детектором, при этом расстояние определялось с точностью до 0.1 мм. В качестве второго источника излучения использовался светодиод в импульсном режиме, длительность вспышки которого соответствовала времени высвечивания сцинтиллятора. Эти источники устанавливались в отверстие в верхней части камеры при снятой крышке.

Полученные таким образом зависимости дифференциальной и интегральной чувствительностей из амплитудного распределения, приведенного на рис. 2.6, представлены ниже на рис. 2.8.

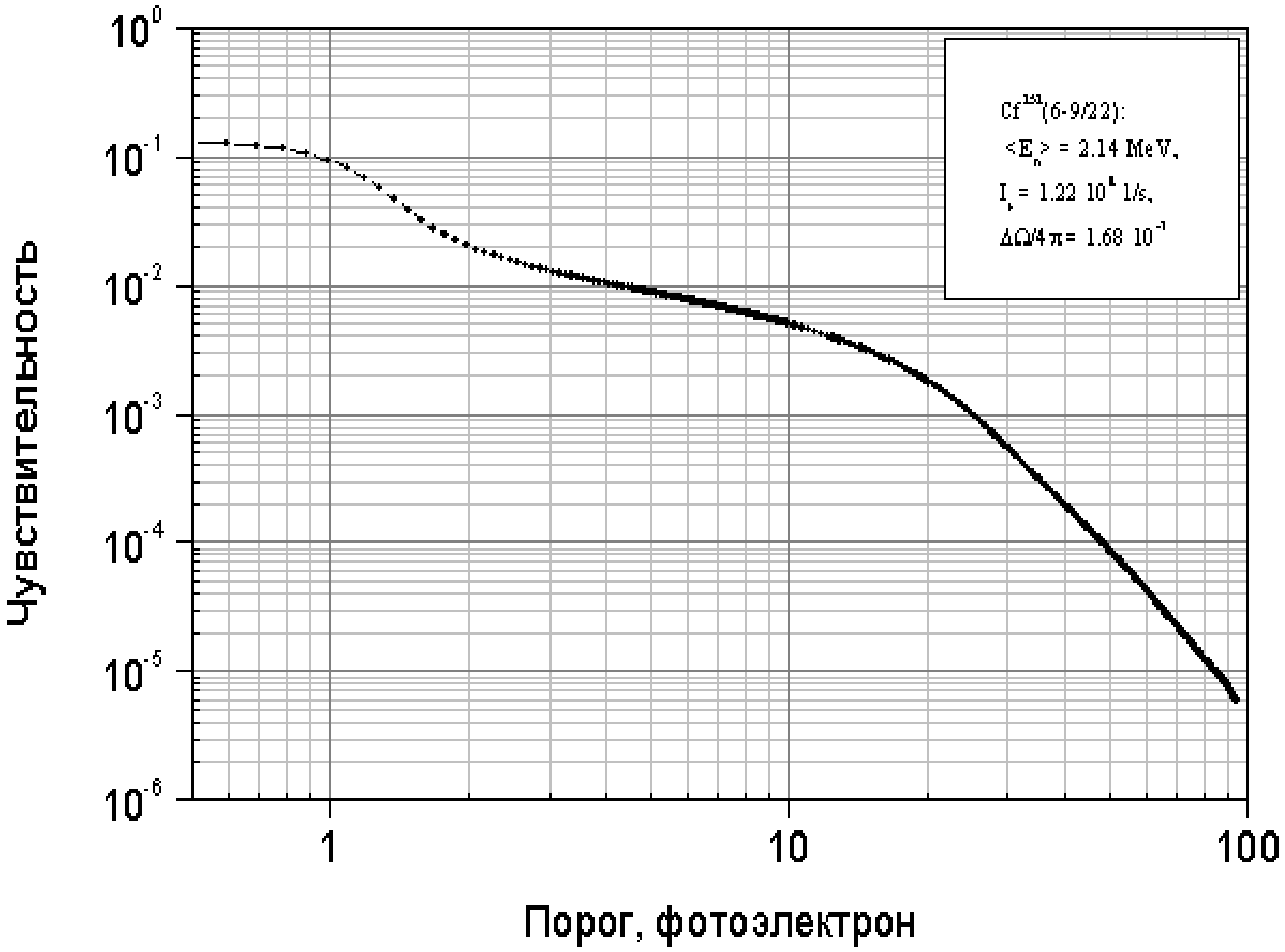

Рис 2.8- Зависимости интегральной чувствительности к фоновому нейтронному и гамма излучению $^{252}Cf$ от порога регистрации.

Определенные таким образом, дифференциальную и интегральную чувствительности детекторов к фоновому излучению мы использовали при оптимизации детекторов для регистрации корпускулярного излучения плазмы.

## 3. Имитация фонового излучения плазмы

Известно, что в зависимости от типа плазменного разряда энергетические спектры фонового излучения различаются между собой. В D-D плазме энергия нейтронов попадающих на детектирующую систему лежит в диапазоне от тепловых до 2.45 МэВ, а в D-T плазме от тепловых до 14 МэВ, при этом, средняя энергия гамма квантов составляет около 1 МэВ.

Основанием для использования излучения источника $^{252}Cf$, для имитации фонового излучения плазмы являются результаты представленные в работе / 66 /, где показаны сравнительные кривые амплитудных распределений фоновых сигналов, возникающих при облучении детекторов источником $^{252}Cf$ и фоновым излучением плазмы.

На рис. 2.9 представлены результаты измерения фоновых амплитудных распределений в реальных условиях работы детекторов в термоядерной установке JT-60U и результаты измерения при облучении источником $^{252}Cf$.

Из рис. 2.9 видно, что амплитудные распределения сигналов, возникающих при облучении детекторов, находящихся в разных каналах, достаточно близки к результатам полученным при облучении источником $^{252}Cf$. Надо отметить, фоновая кривая от источника $^{252}Cf$ была получена с детектором, имеющим такие же характеристики, как и детектор, находящийся в третьем канале. Поэтому использование излучения от источника $^{252}Cf$ является достаточно корректным приближением для исследования фоновых характеристик детектирующей системы.

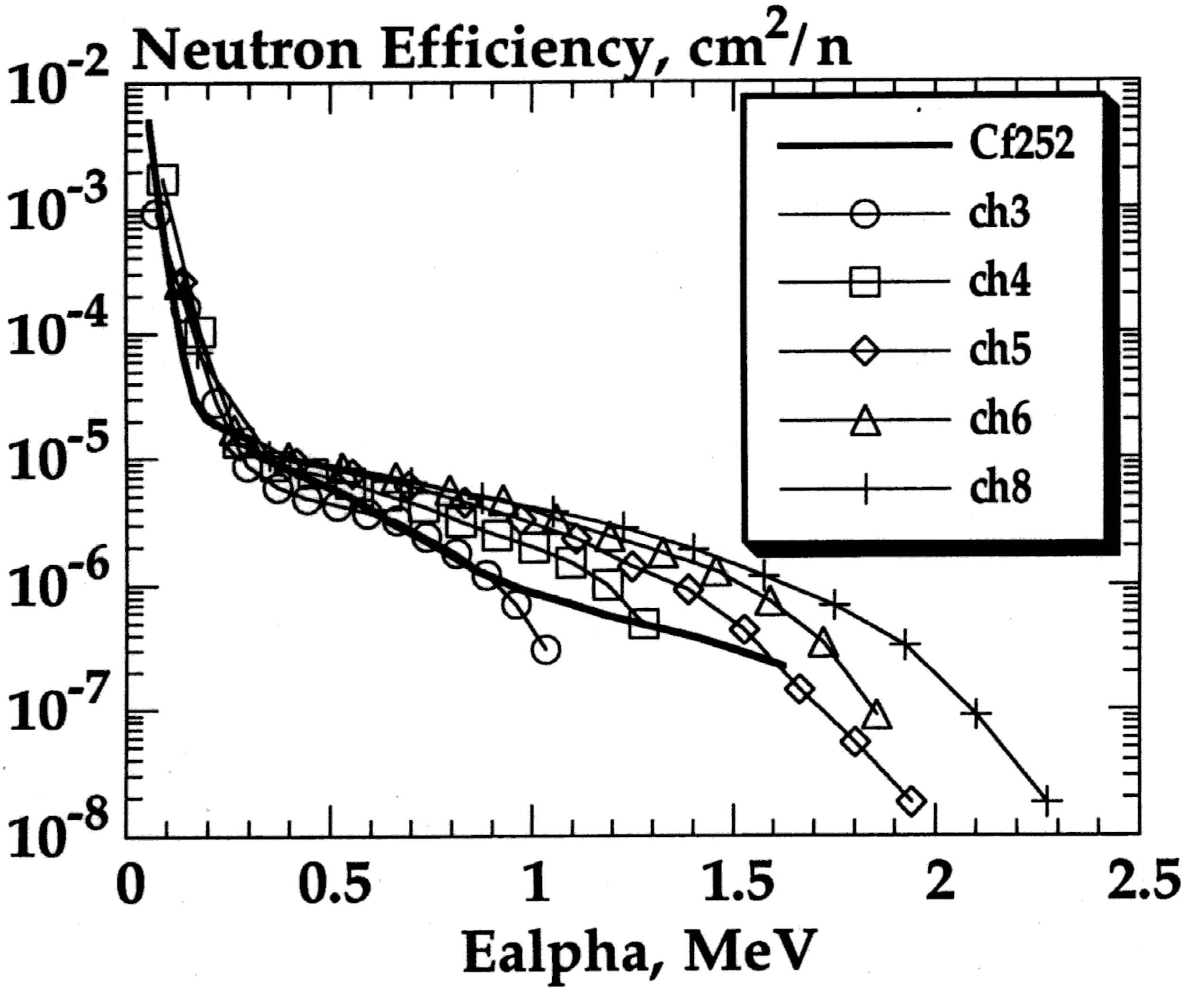

Рис. 2.9 Сравнительные кривые чувствительности сцинтиллятором детектором к фоновому излучению от источника и от каналов 3его – 8ого на установке JT-60U

На рис. 2.10, показан спектр излучения $^{252}Cf$, который имеет спектр нейтронов деления от 0.025 эВ до 10 МэВ с средней энергией $\langle E \rangle_n$ = 2.14 МэВ / 67 / и, кроме того широкий спектр гамма излучения до 10 МэВ с средней энергией $\langle E \rangle_\gamma$ = 850 КэВ.

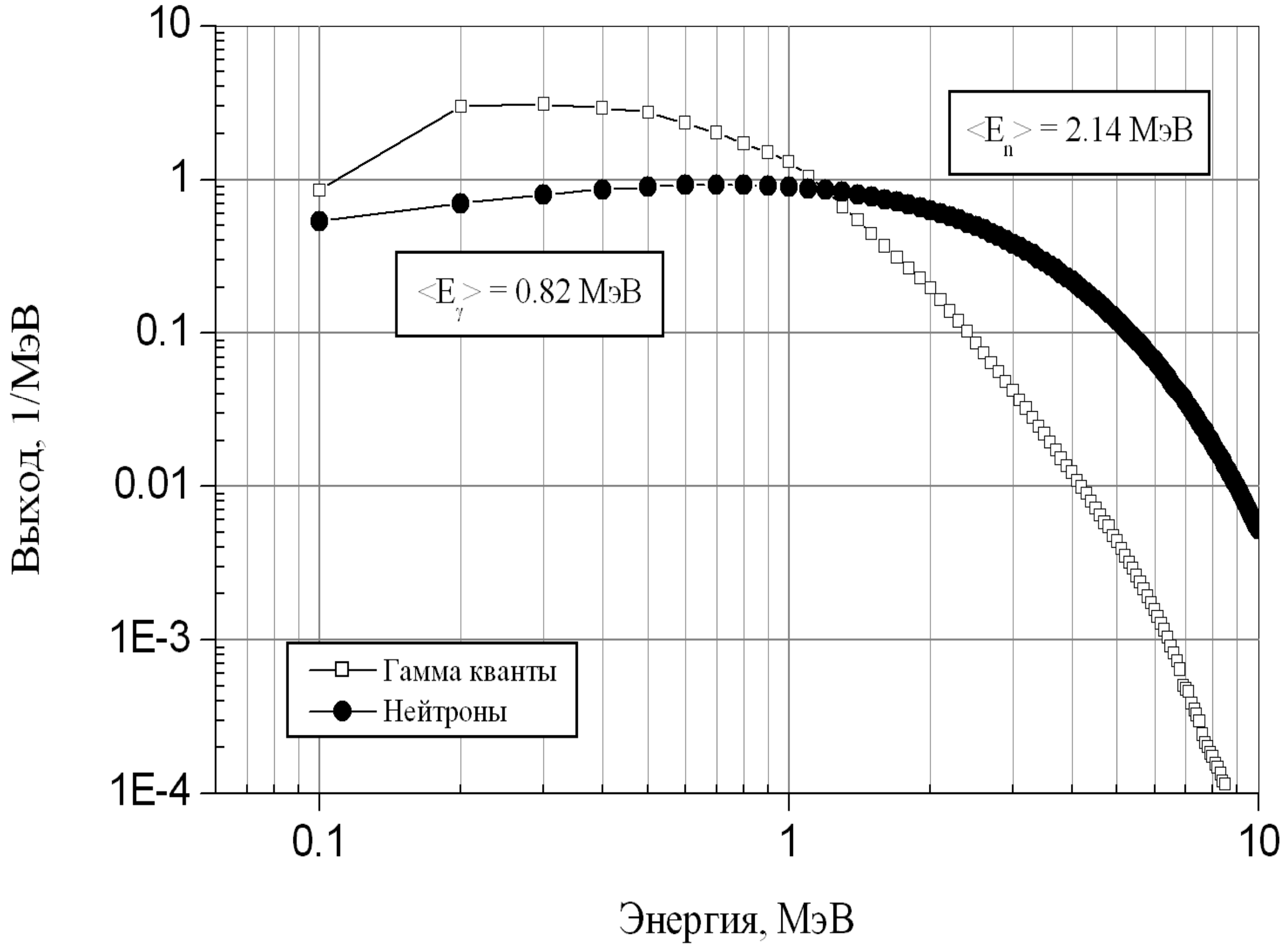

Рис. 2.10 - Спектр излучения радиоактивного источника $^{252}Cf$

Источники $^{60}Co$, $^{137}Cs$ и $^{90}Sr$ использовались для определения вклада от гамма излучения и электронов в общую чувствительность детектора к фоновому излучению. Приведем параметры выше перечисленных источников:

- $^{60}Co$, $E_\gamma$ = 1173 КэВ и $E_\gamma$ =1332 КэВ,
- $^{137}Cs$, $E_\gamma$ = 662 КэВ,
- $^{90}Sr$, $\langle E \rangle_e$ = 566 КэВ,

4. Типы и основные параметры детекторов, используемых для измерения их фоновых характеристик

Известно, что в корпускулярной диагностике плазмы используются такие детекторы как: сцинтилляционные детекторы, микроканальные пластины, канальные электронные умножители, полупроводниковые и другие / 68 /. Поэтому при решении поставленной задачи об оптимизации детекторов для регистрации корпускулярного излучения плазмы мы исследовали фоновые характеристики для различных типов детекторов.

В таблице 2.1 представлены основные параметры тех детекторов, которые были исследованы в этой работе.

Таблица 2.1

| Тип детектора | Максимальный ток на аноде, $I_{анод\,max}$, A | Средний коэффициент усиления, $G_У$ | Примечания |
|---|---|---|---|
| МКП N° 04572 | $10^{-6}$ | $10^5 \div 10^7$ | Шевронная сборка, $d_К$=10 мкм |
| КЭУ – 6 8710 | $10^{-6}$ | $10^6 \div 10^8$ | Спиральный, $d_{раструб}$= 8 мм, $к_{анал}$= 1 мм |
| ФЭУ – 85 | $5{\cdot}10^{-5}$ | $1.6{\cdot}10^5$ | Входное окно - боросиликатное, $d_{фк}$= 25 мм |
| ФЭУ R3998-02 | $10^{-4}$ | $1.3{\cdot}10^6$ | Входное окно – боросиликатное, $d_{фк}$= 25 мм |
| ФЭУ R5900U-00-L16 | $1.6{\cdot}10^{-4}$ | $2{\cdot}10^6$ | Входное окно – боросиликатное, многоанодный, $S_{фк}$= 12.8x16 мм$^2$ |
| ФЭУ 5600U | $1.6{\cdot}10^{-4}$ | $1.3{\cdot}10^6$ | Входное окно – боросиликатное, $d_{фк}$= 8 мм |
| ФЭУ 5600U-06 | $1.6{\cdot}10^{-4}$ | $1.3{\cdot}10^6$ | Входное окно – кварцевое, $d_{фк}$= 8 мм |

# ГЛАВА III. ИССЛЕДОВАНИЕ ВОЗДЕЙСТВИЯ ФОНОВОГО n-γ ИЗЛУЧЕНИЯ НА ДЕТЕКТОРЫ, ИСПОЛЬЗУЕМЫЕ В КОРПУСКУЛЯРНОЙ ДИАГНОСТИКЕ ПЛАЗМЫ

Из представленных выше экспериментальных данных по чувствительности различных детекторов, используемых в корпускулярной диагностике видно, что опубликованная информация по этой тематике не достаточна для проведения полноценного анализа при разработке новых детекторов корпускулярного излучения плазмы, предназначенных для работы на крупнейших термоядерных установках мира, таких как: JT-60, TFTR, JET и др. В связи с этим мы провели детальное исследование чувствительности различных типов детекторов к фоновому излучению плазмы / 62 – 65, 69, 70 /. В качестве детекторов, используемых в детектирующих системах корпускулярной диагностики плазмы, обычно применяют микроканальные пластины (МКП) / 71 – 75 /, канальные электронные умножители (КЭУ) /68, 76 / и детекторы на основе фотоэлектронных умножителей (ФЭУ). Имитация фонового излучения плазмы осуществлялась радиоактивным источником $^{252}Cf$.

## 1. Результаты измерения чувствительности МКП к фоновому нейтронному и гамма излучению.

Результаты измерения чувствительности МКП к фоновому излучению в зависимости от величины порога регистрации приведены на рисунке 3.1. Величина порога регистрации выражена в числе электронов, генерируемых на аноде МКП. На этом же рисунке приведена и зависимость эффективности регистрации альфа частиц с энергией ($E_\alpha$= 1.4 МэВ).

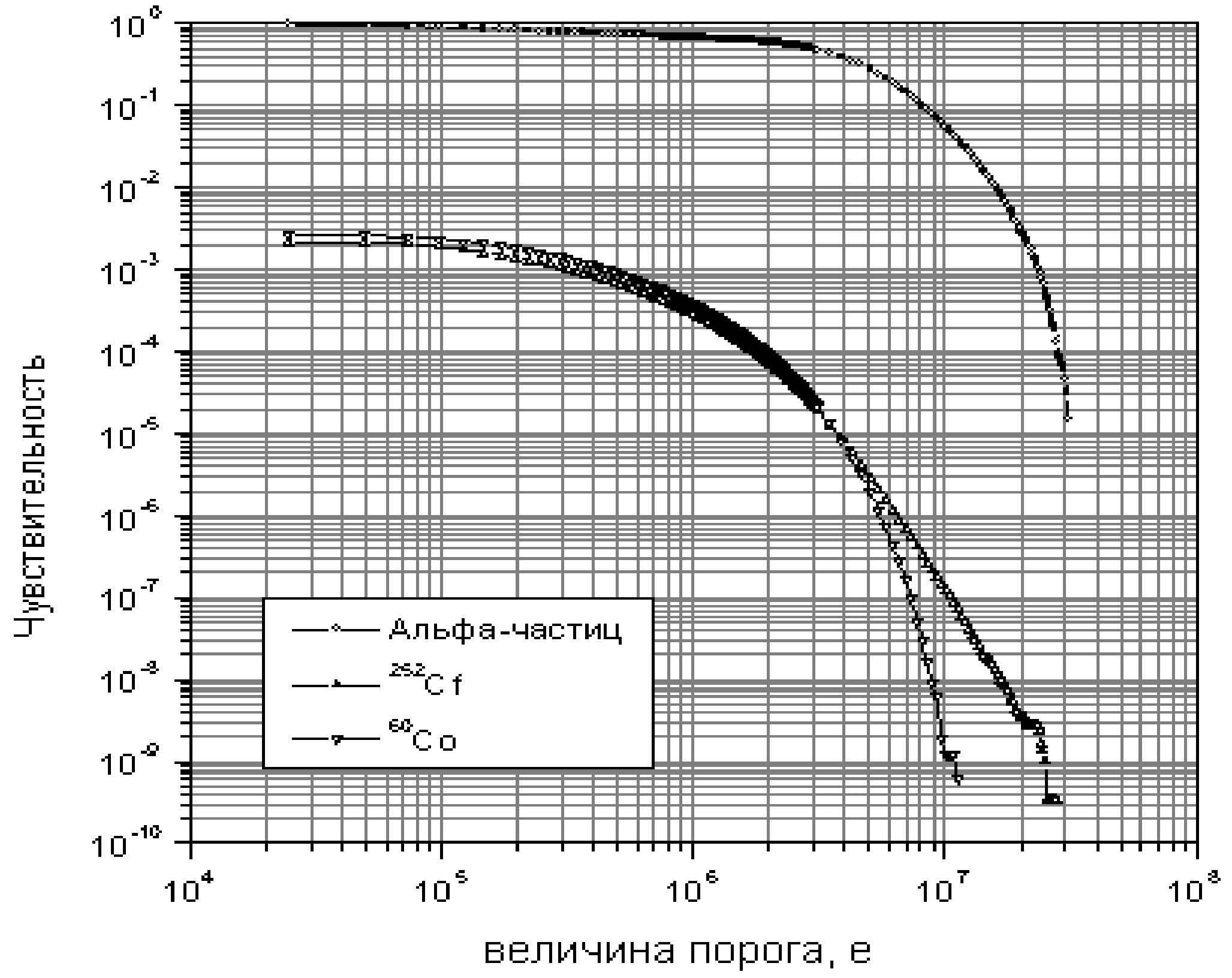

Рис. 3.1 – Зависимость чувствительности МКП
от порога регистрации для фонового излучения и альфа частиц.

Из рисунка 3.1 видно, что зависимости чувствительности МКП от порога регистрации для $^{60}Co$ (<E$_\gamma$>= 1.25 МэВ) и $^{252}Cf$ (<E$_\gamma$>= 850 КэВ и < E$_n$>= 2.14 МэВ) в области малых порогов идентичны, заметное расхождение наблюдается только в области больших порогов регистрации. По всей видимости, основной вклад в чувствительность детектора к фоновому излучению плазмы дает гамма излучение, а именно электроны, генерируемые вследствие взаимодействия гамма квантов с элементами конструкции детектора и окружающей его камеры. Оценка показывает, что доля таких электронов, отнесенная к потоку падающих на детектор гамма квантов, составляет для указанного диапазона энергии величину ~ $6 \cdot 10^{-3}$ / 77 /, а эффективность их регистрации МКП достаточно высока.

Наблюдаемое небольшое различие в полной чувствительности детектора (при нулевом пороге) к излучению $^{60}Co$ – $2.9 \cdot 10^{-3}$ и $^{252}Cf$ – $2.0 \cdot 10^{-3}$ объясняется тем, что с ростом средней энергии гамма квантов растет и коэффициент выхода электронов. Различие кривых в области больших порогов регистрации вероятно можно объяснить вкладом гамма квантов более высоких энергий, испускаемых источником $^{252}Cf$ а также, генерируемых при взаимодействии потока нейтронов с элементами конструкции камеры.

Типичный поток фонового излучения **Ф**, в месте расположения детектирующей системы, характерный для крупных термоядерных установок (JET, JT-60U, TFTR), составляет величину от $10^8$ до $10^{10}$ 1/см$^2$·сек / 8, 78 /, как для нейтронов так и для гамма квантов (при средней энергии гамма квантов ~ 1 МэВ). При чувствительности МКП к фоновому излучению **S** ~ $10^{-3}$, мы получим интенсивность фоновых импульсов на выходе МКП от $10^5$ до $10^7$ имп./сек ($n = \Phi \cdot S$). Известно, что линейность работы МКП обеспечивается при условии:

$$\bar{I}_A << I_Д, \quad (1)$$

где: $\bar{I}_A$ - Средний ток на аноде, $I_Д$ - ток распределенного делителя.

Оценим предельную интенсивность фоновых сигналов исходя из условия (1), при этом:

$$n_{max} << \frac{I_Д}{\bar{q}}, \quad (2)$$

где $\bar{q}$ – средняя величина заряда, снимаемого с анода детектора (МКП, КЭУ или ФЭУ) при регистрации единичного сигнала.

При типичном токе питания, протекающем через МКП $I_Д$= 1 мкА и при средней величине снимаемого заряда $\bar{q}=5.1\cdot10^5$ е (расчет выполнен по спектру сигналов на аноде МКП), мы получим предельную интенсивность фоновых импульсов $n_{max} << 1.2\cdot10^7$ имп/сек, т.е. $n_{max} \sim 10^6$ имп/сек.

Казалось бы, что можно уменьшить чувствительность детектора, к фоновому излучению увеличив порог регистрации до величины ~ $2{\cdot}10^6$ е. При этом эффективность регистрации альфа частиц близка к 100% / 79 – 81 /, а чувствительность детектора к фоновому излучению падает примерно на порядок до $10^{-4}$. Однако практически это можно реализовать при очень небольшой фоновой загрузке. В реальном случае приходится устанавливать порог регистрации существенно меньше вследствие заметного уменьшения коэффициента усиления МКП под действием фоновых сигналов, при этом реальная чувствительность детектора составляет ~ $10^{-3}$. Таким образом детектор на основе МКП оказывается работоспособным если поток фонового излучения в месте его установки не превышает $10^9$ 1/см$^2$·сек.

2. Результаты измерения чувствительности КЭУ к фоновому нейтронному и гамма излучению.

Детекторы на основе КЭУ также широко применяются в детектирующих системах, используемых в устройствах корпускулярной диагностики плазмы. В связи с этим были проведены измерения чувствительности КЭУ к фоновому нейтронному и гамма излучению плазмы.

Результаты измерения зависимости чувствительности КЭУ к фоновому излучению от величины порога регистрации приведены на рис. 3.2. Там же приведена и кривая эффективности регистрации альфа частиц ($E_\alpha$= 600 КэВ).

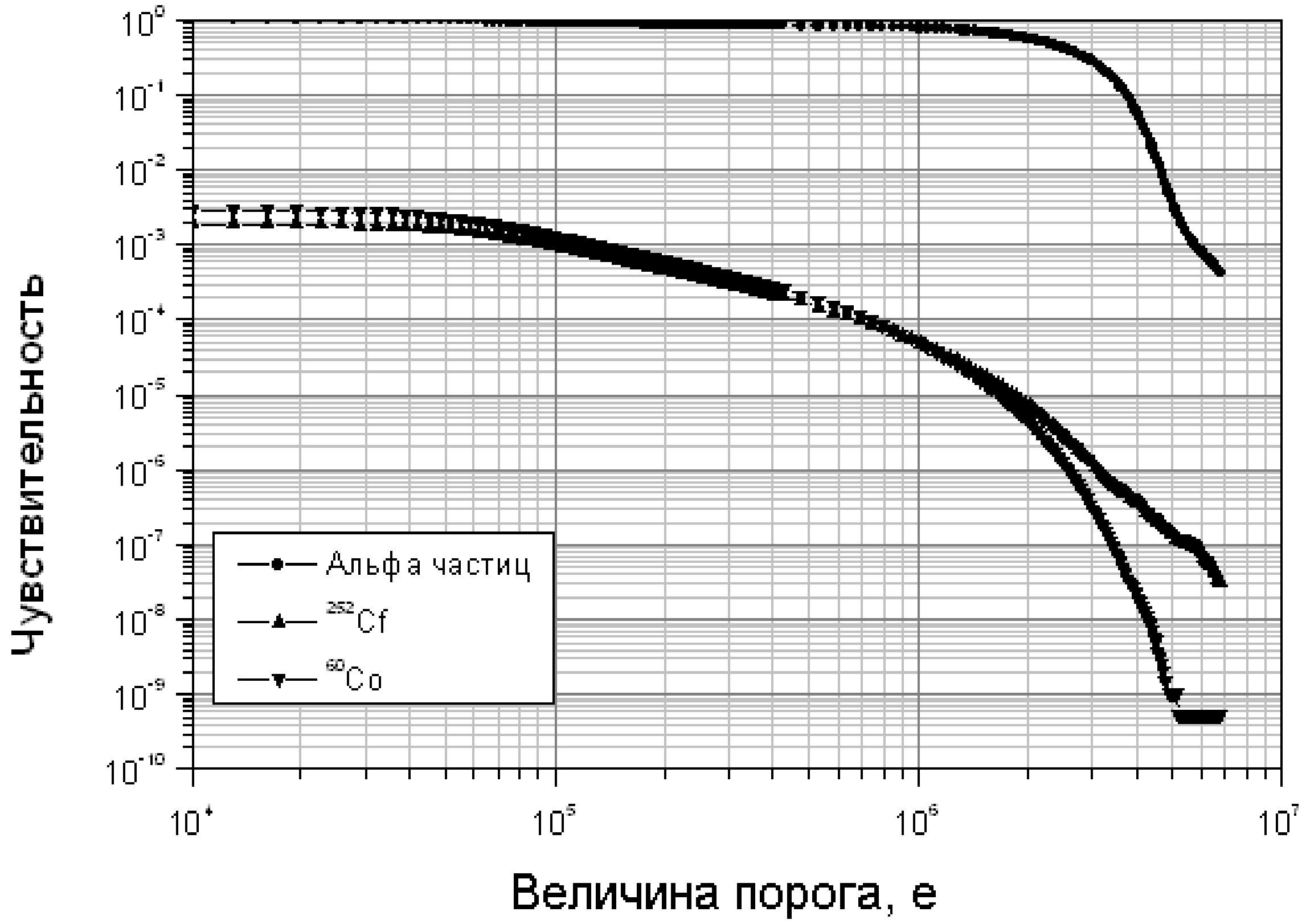

Рис. 3.2 – Зависимость чувствительности КЭУ от порога регистрации для фонового излучения и альфа частиц.

Из рисунка 3.2 видно, что значения полной чувствительности КЭУ к фоновому излучению (при нулевым пороге регистрации) для источников $^{60}$Co и $^{252}$Cf сравнимы с теми, которые были получены для МКП, а именно чувствительность детектора к фоновому излучению $^{60}$Co – $3.5\cdot10^{-3}$ и $^{252}$Cf – $1.9\cdot10^{-3}$.

По всей видимости, характер взаимодействия фонового излучения с КЭУ, такой же, как и для МКП. Отметим, что и величины чувствительности для КЭУ и МКП весьма близки.

Оценим предельную интенсивность фоновых сигналов исходя из условия (1) для КЭУ. При токе делителя $I_д$= 1 мкА и при средней величине фонового заряда на выходе КЭУ $\bar{q}=1.8\cdot10^5$ е (расчет выполнен по спектру сигналов на аноде КЭУ) мы получим предельную интенсивность фоновых импульсов $n_{max}$

<< $3.4 \cdot 10^7$ имп/сек, т.е. $n_{max} \sim 10^6$ имп/сек / 82, 83 /. Следует отметить, что такой высокий уровень загрузки, может быть, достигнут при работе КЭУ с относительно небольшим усилением. Таким образом детектор на основе КЭУ также оказывается работоспособным, если поток фонового излучения в месте его установки не превышает $10^9$ $1/см^2 \cdot сек$, а его чувствительность к фоновому излучению составляет величину близкую $10^{-3}$.

## 3. Результаты измерения чувствительности фотоэлектронного умножителя к фоновому n-γ излучению.

Детекторы на основе ФЭУ, применяются в детектирующих системах, используемых в устройствах как корпускулярной, так и оптической диагностики плазмы. Также как и МКП и КЭУ для ФЭУ были проведены измерения чувствительности к фоновому нейтронному и гамма излучению. В качестве примера на рис. 3.3 приведены результаты измерений для ФЭУ фирмы Hamamatsu R3998-02. На этом рисунке представлены измеренные интегральные чувствительности ФЭУ в зависимости от величины порога регистрации, выраженного в числе фотоэлектронов, испускаемых фотокатодом ФЭУ.

Из рисунка 3.3 видно, что зависимости чувствительности ФЭУ от порога регистрации при облучении $^{60}Co$ и $^{252}Cf$ не сильно отличаются между собой. Заметное расхождение, как и для детекторов рассмотренных выше, наблюдается лишь в области средних и больших порогов регистрации (>10 фэ). Следует отметить существенно более высокую полную чувствительность ФЭУ к фоновому излучению по сравнению с МКП и КЭУ, а именно $\sim 6 \cdot 10^{-2}$ против $3 \cdot 10^{-3}$. Несмотря на то, что чувствительность ФЭУ к фоновому излучению заметно выше, чем у КЭУ и МКП предельная загрузка, при которой можно эксплуатировать ФЭУ может быть даже больше, чем для КЭУ и МКП. Это связано с тем, что ток делителя задается внешней цепью и поэтому в качестве предельного значения тока нагрузки надо использовать $I_{a\ макс}$ (из паспортных данных). Оценим предельную величину загрузки фоновыми сигналами для ФЭУ. При типичном значении $I_{a\ макс} \approx$ 100 мкА и при средней величине снимаемого заряда $\bar{q}=4.0\cdot10^6$ e (расчет выполнен по спектру сигналов на аноде ФЭУ) мы получим предельную интенсивность фоновых импульсов $n_{max} << 10^{10}$ имп/сек т.е. $n_{max} \sim 10^9$ имп/сек. Таким образом, детектор на основе ФЭУ оказывается работоспособным если поток фонового излучения в месте его установки не превышает $10^{11}$ 1/см$^2$·сек, при этом его чувствительность к фоновому излучению составляет величину $4 \div 6\cdot10^{-2}$.

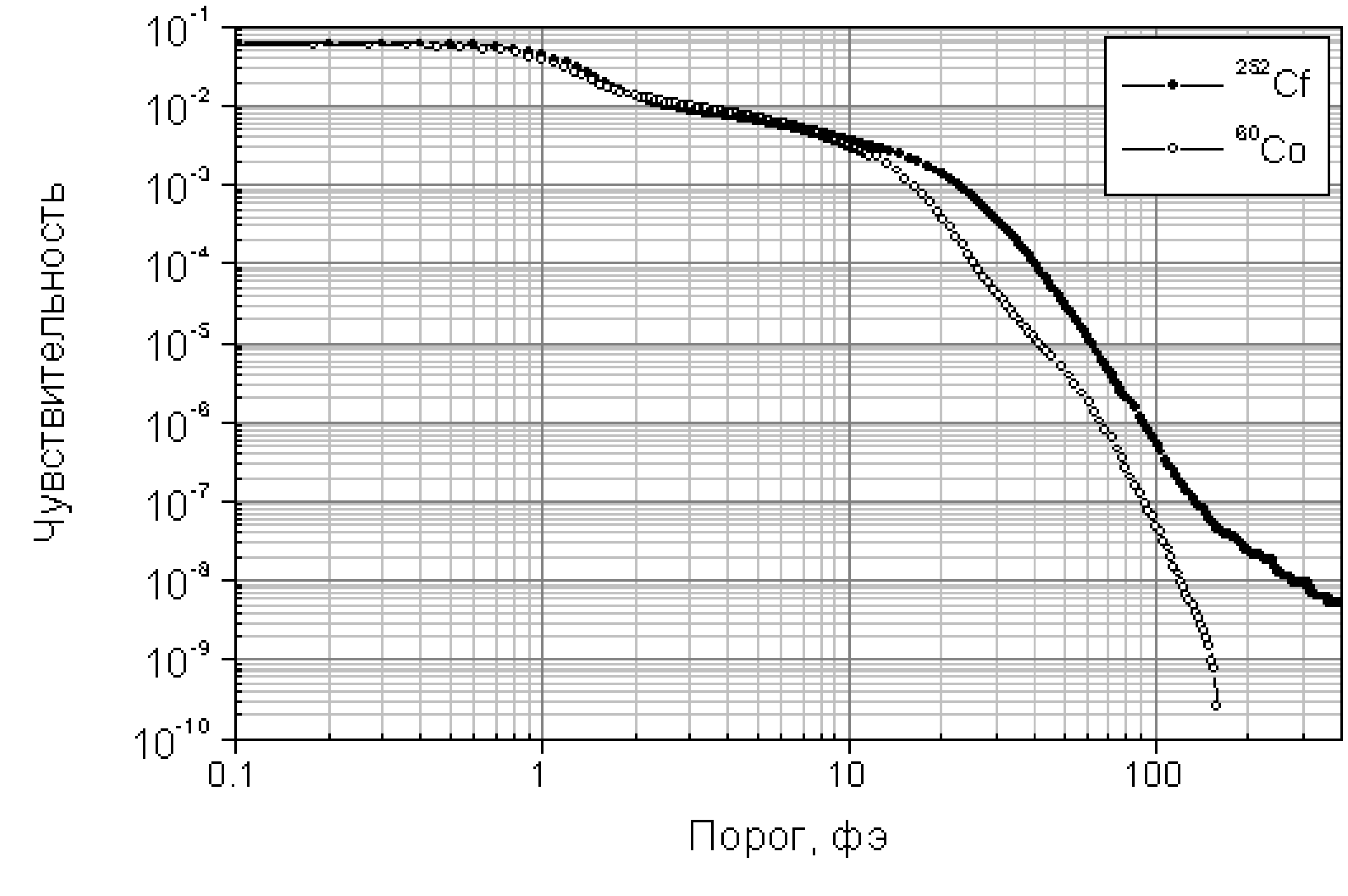

Рис. 3.3- Зависимость чувствительности ФЭУ (R3998-02) к фоновому излучению от порога регистрации.

Аналогичные измерения были выполнены и для других типов ФЭУ. На рисунках 3.4 – 3.7 представлены результаты, измерения чувствительности к фоновому излучению для различных фотоэлектронных умножителей.

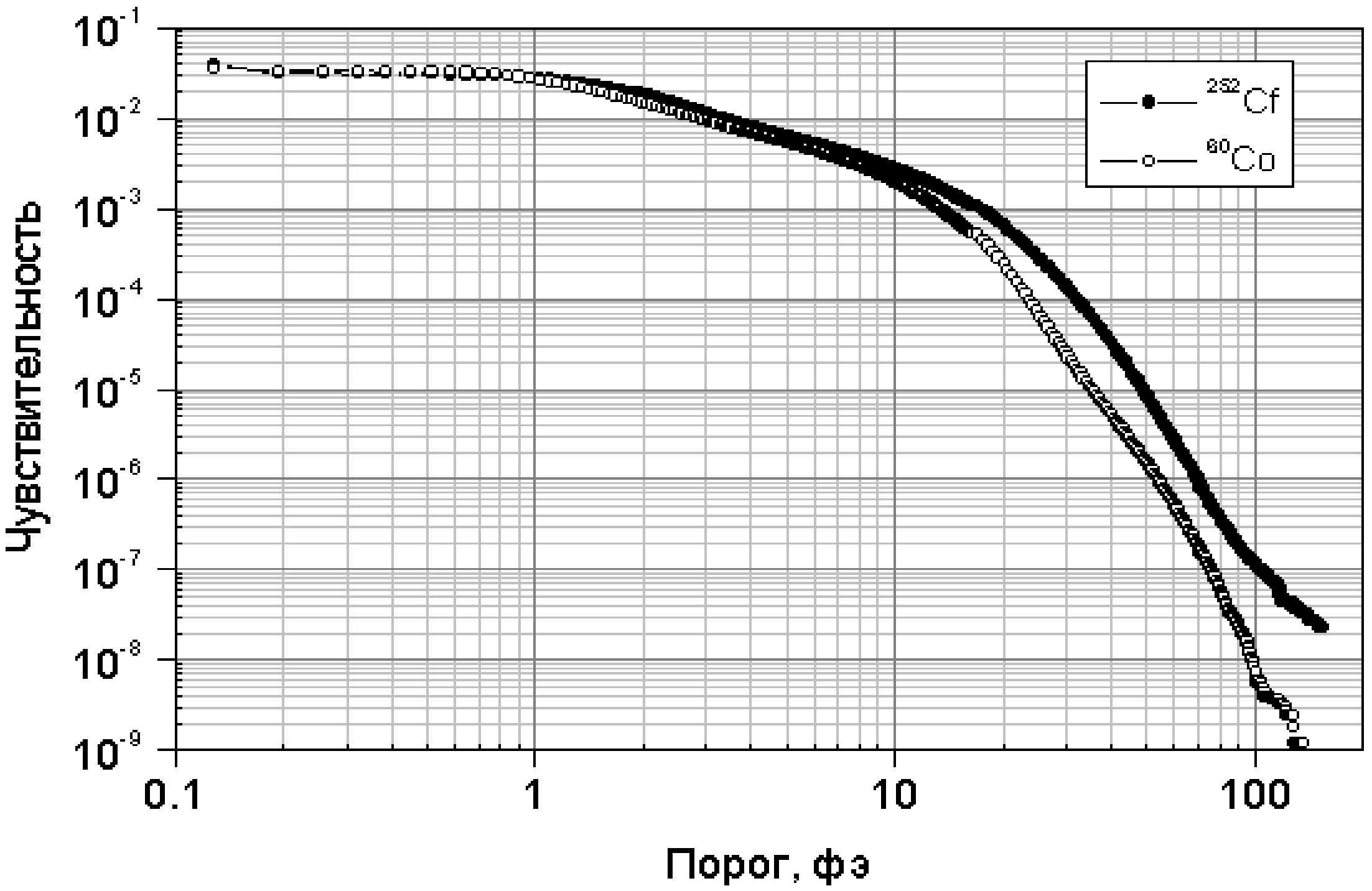

Рис. 3.4- Интегральная чувствительность ФЭУ-85 к фоновому излучению в зависимости от порога регистрации.

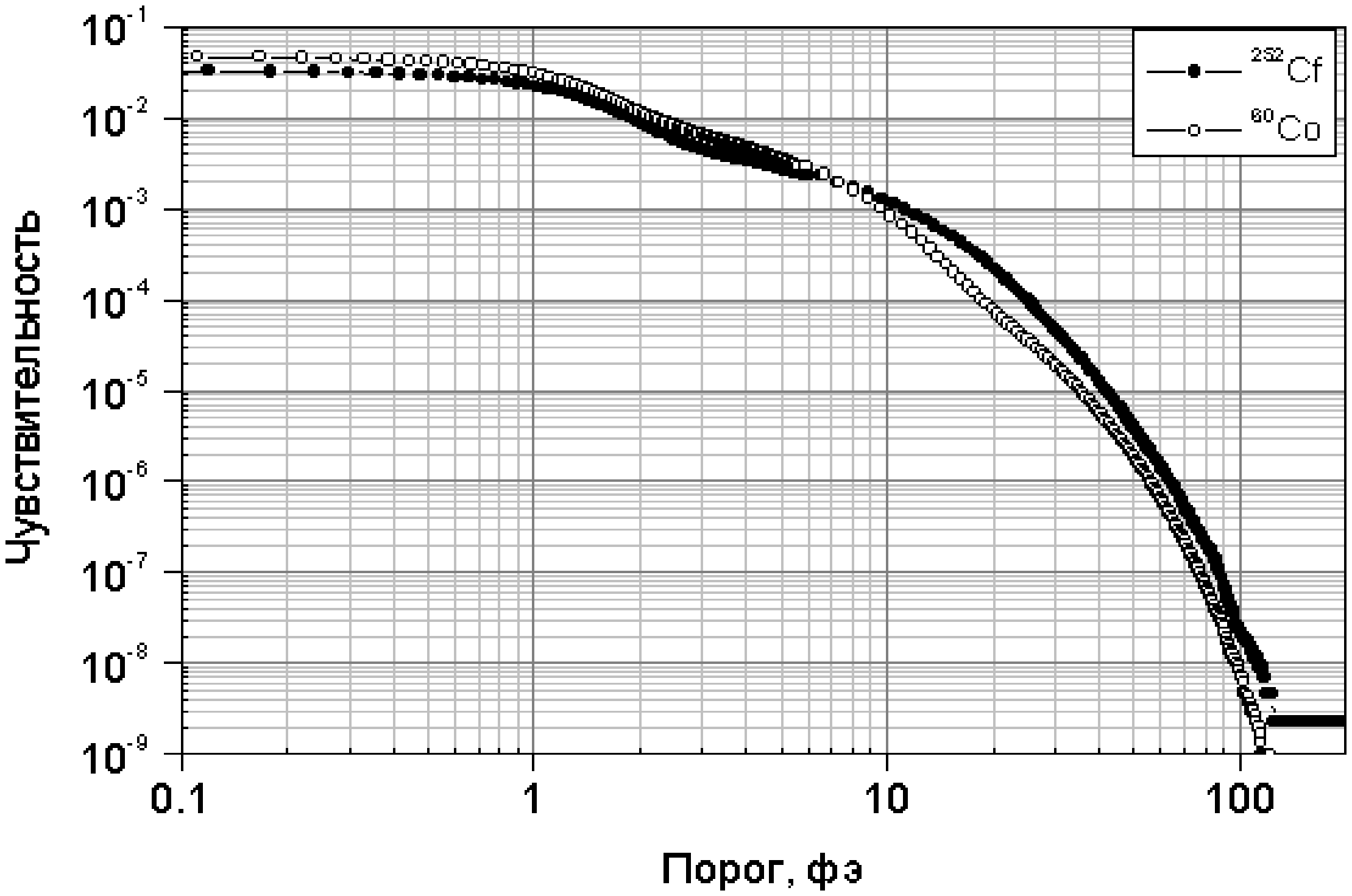

Рис. 3.5- Зависимость интегральной чувствительности ФЭУ (R5900U-00-L16) от порога регистрации.

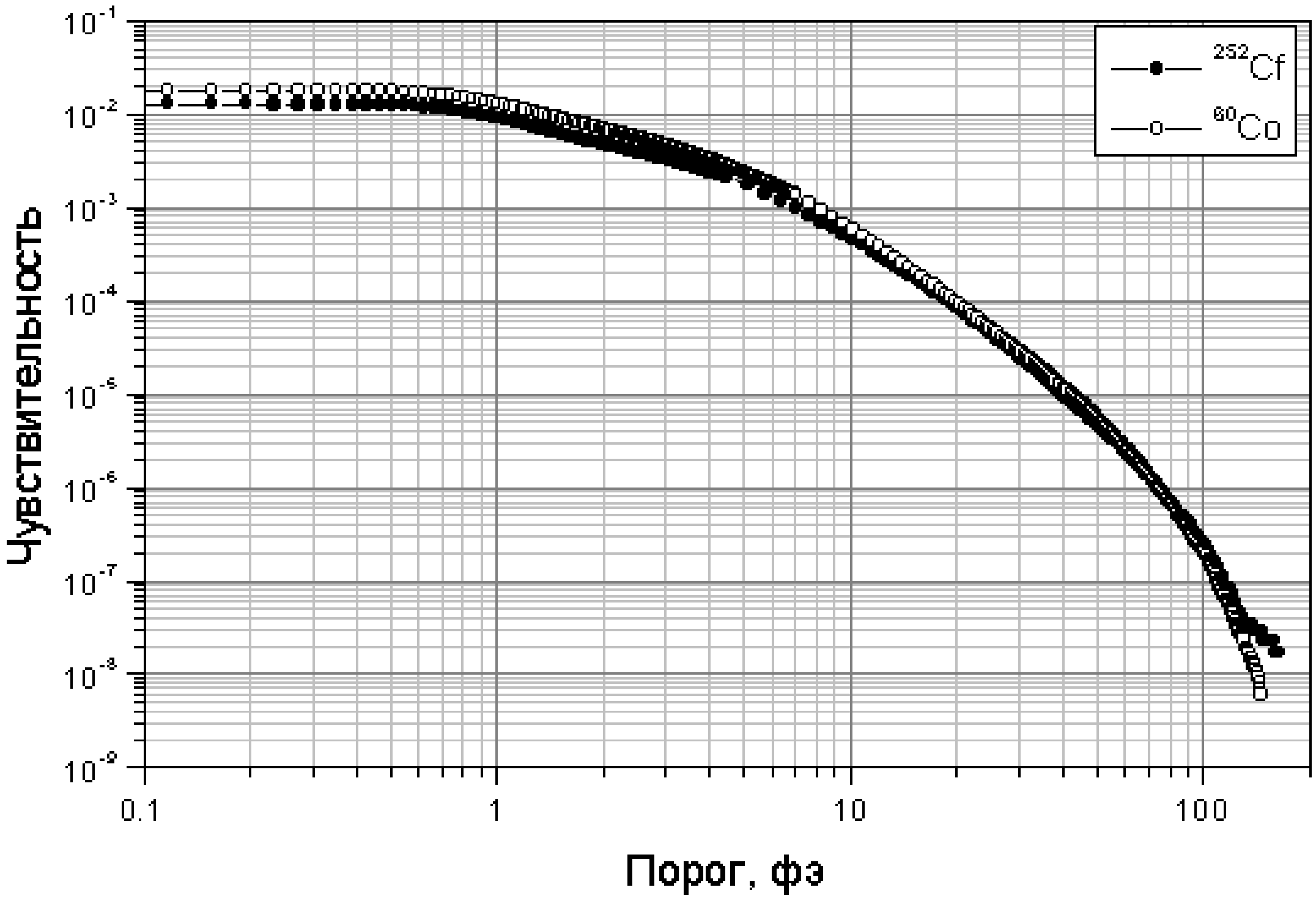

Рис. 3.6- Зависимость интегральной чувствительности ФЭУ (R5600U) от порога регистрации.

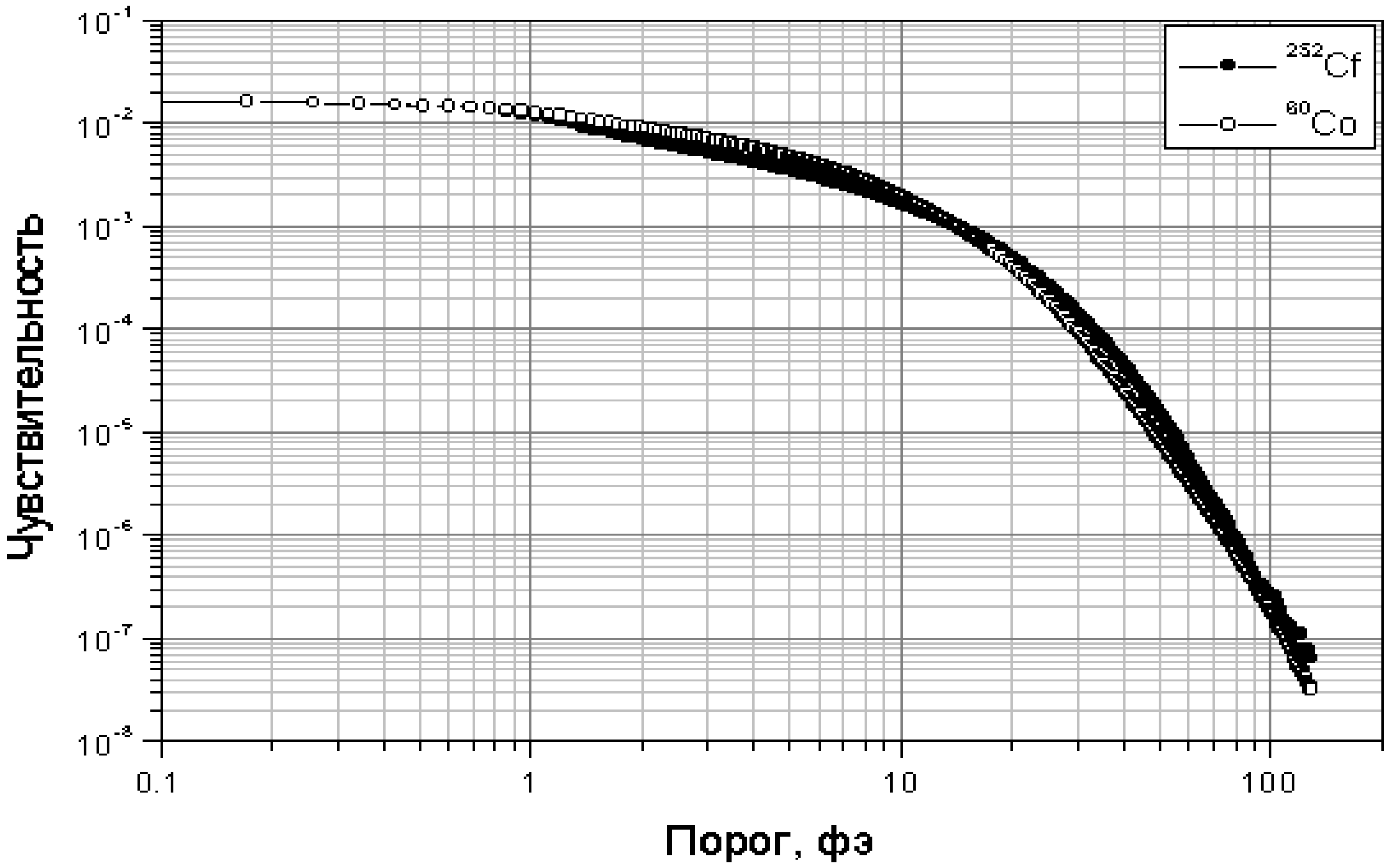

Рис. 3.7- Зависимость интегральной чувствительности ФЭУ (R5600U-06) от порога регистрации.

Все эти измерения показали, что характер наблюдаемой зависимости чувствительности различных типов ФЭУ от порога регистрации идентичен, кроме того, близки и значения полной чувствительности. Ниже приведена таблица 3.1, которая содержит значения полной чувствительности для исследованных выше детекторов.

Таблица 3.1

| Тип ФЭУ | Полная чувствительность, S | | Размеры Фотокатода | Толщина Входного окна, мм |
|---|---|---|---|---|
| | $^{60}$Co | $^{252}$Cf | | |
| R3998-02 | $5.6 \cdot 10^{-2}$ | $5.9 \cdot 10^{-2}$ | ∅ 25 мм | 2 |
| ФЭУ-85 | $3.8 \cdot 10^{-2}$ | $3.2 \cdot 10^{-2}$ | ∅ 25 мм | 1.5 |
| R5900U-00-L16 | $4.6 \cdot 10^{-2}$ | $3.4 \cdot 10^{-2}$ | □ 16x16 мм | 1.5 |
| R5600U | $1.7 \cdot 10^{-2}$ | $1.3 \cdot 10^{-2}$ | ∅ 8 мм | 0.5 |
| R5600U-06 | $1.4 \cdot 10^{-2}$ | $1.3 \cdot 10^{-2}$ | ∅ 8 мм | 0.5 |

Исходя из данных, представлены в этой таблице / 84 /, можно заключить, что величина полной чувствительности ФЭУ к фоновому излучению коррелирует с их размерами, а именно с толщиной входного окна (миниатюрные ФЭУ имеют более тонкое входное окно).

Таким образом, для всех исследованных нами детекторов (ФЭУ, МКП и КЭУ) общим является быстрый спад чувствительности к фоновому излучению при увеличении порога регистрации. Однако полная чувствительность для ФЭУ заметно выше, чем для КЭУ и МКП. При этом основное отличие между ФЭУ, МКП и КЭУ заключается в том, что в ФЭУ имеется дополнительный элемент – фотокатодная камера. Поэтому логичным является эксперимент по измерению фоновой чувствительности ФЭУ при отключенной фотокатодной камере, т.е. при подаче запирающего напряжения на фотокатод. Результаты этого измерения приведены на рис. 3.8. Где представлены кривые интегральной чувствительности ФЭУ к фоновому гамма излучению при запирании фотокатода и при обычной подаче напряжения на фотокатод / 85 /.

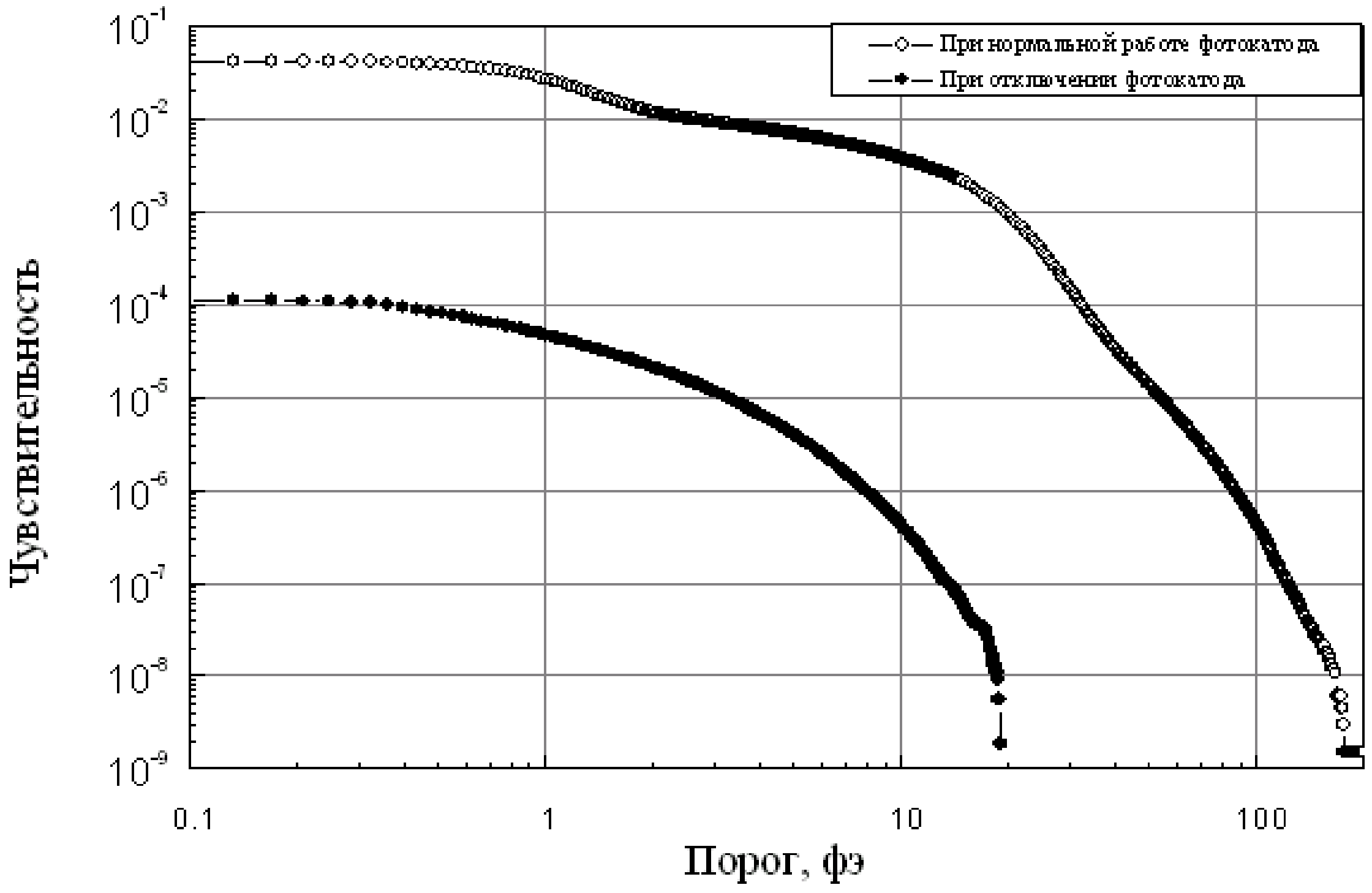

Рис. 3.8- Зависимость интегральной чувствительности ФЭУ к фоновому излучению от порога при отключении и при нормальном работе фотокатода.

Из рисунка 3.8 видно, что при подаче запирающего напряжения на фотокатод, полная чувствительность ФЭУ упала до $10^{-4}$, и характер зависимости стал весьма близок к зависимости, наблюдаемой для КЭУ и МКП. При этом измеренная величина чувствительности, возможно, получилась, несколько заниженной, Это связанно с тем, что полная интенсивность, падающих гамма квантов рассчитывалась для площади чувствительной поверхности фотокатода. Однако при запирании фотокатода реальная чувствительная область определялась динодной системой. Сравнивая эти кривые можно констатировать, что определяющий вклад в чувствительность ФЭУ к фоновому излучению дает фотокатодная камера.

4. Анализ возможных источников фоновых сигналов для ФЭУ.

Так как определяющий вклад в чувствительность ФЭУ к фоновому излучению (т.е. в генерацию фоновых сигналов на выходе детектора) дает фотокатодная камера, то рассмотрим более подробно процессы взаимодействия фонового излучения с этим элементом ФЭУ.

Типичная фотокатодная камера состоит из стеклянного цилиндра со стенками миллиметровой толщины, на внутреннюю поверхность которого напылен фотокатод. Фотокатод представляет собой тонкую пленку (толщиной ~ 1000 Å)/ 86 /, обладающую высокой эффективностью преобразования света в фотоэлектроны. Так же следует отметить, что такая пленка имеет и высокий коэффициент вторичной электронной эмиссии / 87, 88 /.

Известно, что все стекла люминесцируют под действием излучения, т.е. поглощение энергии в стекле приводит к генерации фотонов, которые могут достичь фотокатода и вызвать эмиссию фотоэлектронов / 89 /. Таким образом, облучение входного окна ФЭУ может вызвать сигналы на выходе детектора. Оценим вклад в образование фоновых сигналов различных процессов, происходящих при взаимодействии излучения с входным окном ФЭУ. Для этого, вероятность взаимодействия определим по следующему выражению:

$$\upsilon = \sigma n_a t, \quad (3)$$

где: $\upsilon$ – среднее число взаимодействий, $\sigma$ – сечение взаимодействия для данного вида излучения, $n_a$ – концентрация атомов на см$^3$, t – толщина рассматриваемого элемента конструкции.

Оценка показывает, что для гамма квантов с энергией $<E_\gamma>=1.25$ МэВ от источника $^{60}Co$, полное сечение взаимодействия $\sigma \approx 2$ барн и при толщине входного стекла фотокатодной камеры t = 1 мм мы получим $\upsilon = 1.7 \cdot 10^{-2}$ / 90 /. Т.е. значение достаточно близкое к полученному в эксперименте (см. таблицу 3.1).

Известно, что при облучении гамма квантами возникает эмиссия электронов, выходящих из элементов конструкции детектора и окружающей камеры. Эти электроны, поглощаясь во входном стекле в свою очередь, могут вызывать фоновые сигналы. Зависимость выхода быстрых электронов от энергии гамма квантов, представленная в работе / 69 /, меняется от $10^{-3}$ до $10^{-2}$. Для энергии гамма квантов $<E_\gamma>=1.25$ МэВ, выход быстрых электронов составляет величину 1.1 $10^{-2}$. Эти электроны поглощаются в стекле фотокатодной камеры с вероятностью близкой к 100%. Таким образом, их вклад в образование фоновых сигналов сравним с вкладом вносимым прямым взаимодействием гамма квантов с входным окном.

Еще одним источником фоновых сигналов может быть эмиссия электронов из фотокатода под действием быстрых электронов и тяжелых заряженных частиц, выходящих из входного окна и других элементов конструкции ФЭУ и проходящих через слой фотокатода.

Рассмотрим вероятность выхода тяжелых заряженных частиц из стекла под действием нейтронов. Для этого рассчитаем среднее число взаимодействий по этому каналу реакции отнесенное к количеству падающих нейтронов. В нашем случае, исходя из того, что глубина выхода тяжелых частиц из входного окна ФЭУ ~ 10 мкм, можно рассматривать взаимодействие нейтронов со стеклом в приближении однократного взаимодействия. Используя данные о сечениях взаимодействия нейтронов с различными элементами таблицы Менделеева ENDF / 23, 60, 61 / получим среднее сечение взаимодействия нейтронов, испускающих $^{252}Cf$ по реакции $(n,\alpha)$ с боросиликатным стеклом $\sigma$ =

11 мб, а затем по формуле (3) определим среднее число событий по каналу реакции (n,α), в расчете на 1 падающий нейтрон (толщина стекла t = 1см) ν = $9.0 \cdot 10^{-5}$. С учетом того, что глубина выхода тяжелых заряженных частиц из стекла ~ 10 мкм, то выход альфа частиц из стекла под действием нейтронного излучения от источника $^{252}$Cf не превышает $9.0\ 10^{-8}$. Выход других тяжелых заряженных частиц из боросиликатного стекла еще менее вероятны.

Дополнительный вклад фонового излучения в образовании сигналов на выходе ФЭУ может быть обусловлен также генерацией гамма квантов под действием нейтронов. С помощью формулы (3) было рассчитано число взаимодействий нейтронов по каналу реакции (n,n') с последующим выходом гамма квантов из стенок камеры, окружающей детектор. Для этого, из данных ENDF вычислялось среднее сечение взаимодействия нейтронов ($^{252}$Cf) реакции (n,n') с железом, величина этого сечения составила 650 мб. В результате расчета было получено, что вероятность генерации гамма квантов под действием нейтронов $^{252}$Cf равна $1.3 \cdot 10^{-1}$ / 25 /. С учетом того, что эти гамма кванты изотропно генерируется по всему объему камеры, а также того, что вероятность их взаимодействия со входным стеклом (толщиной t = 1 мм) составляет ~ $10^{-2}$, мы получим, что вклад этих гамма квантов в генерацию фоновых сигналов не превышает $10^{-3}$.

Таким образом, из всех рассмотренных выше процессов, взаимодействие гамма квантов со стеклом фотокатодной камеры дает самый значительный вклад. Следующими за ним идут процессы поглощения быстрых электронов, генерируемых в результате взаимодействия гамма квантов со стенками камеры, окружающей детектор и элементами самого детектора, а также поглощение продуктов ядерных реакций, инициируемых нейтронами.

Для иллюстрации вклада входного окна ФЭУ были проведены измерения его чувствительности в зависимости от толщины, добавочного стекла, которое было установлено на входном окне. Результаты этих измерений приведены на рис. 3.9, 3.10.

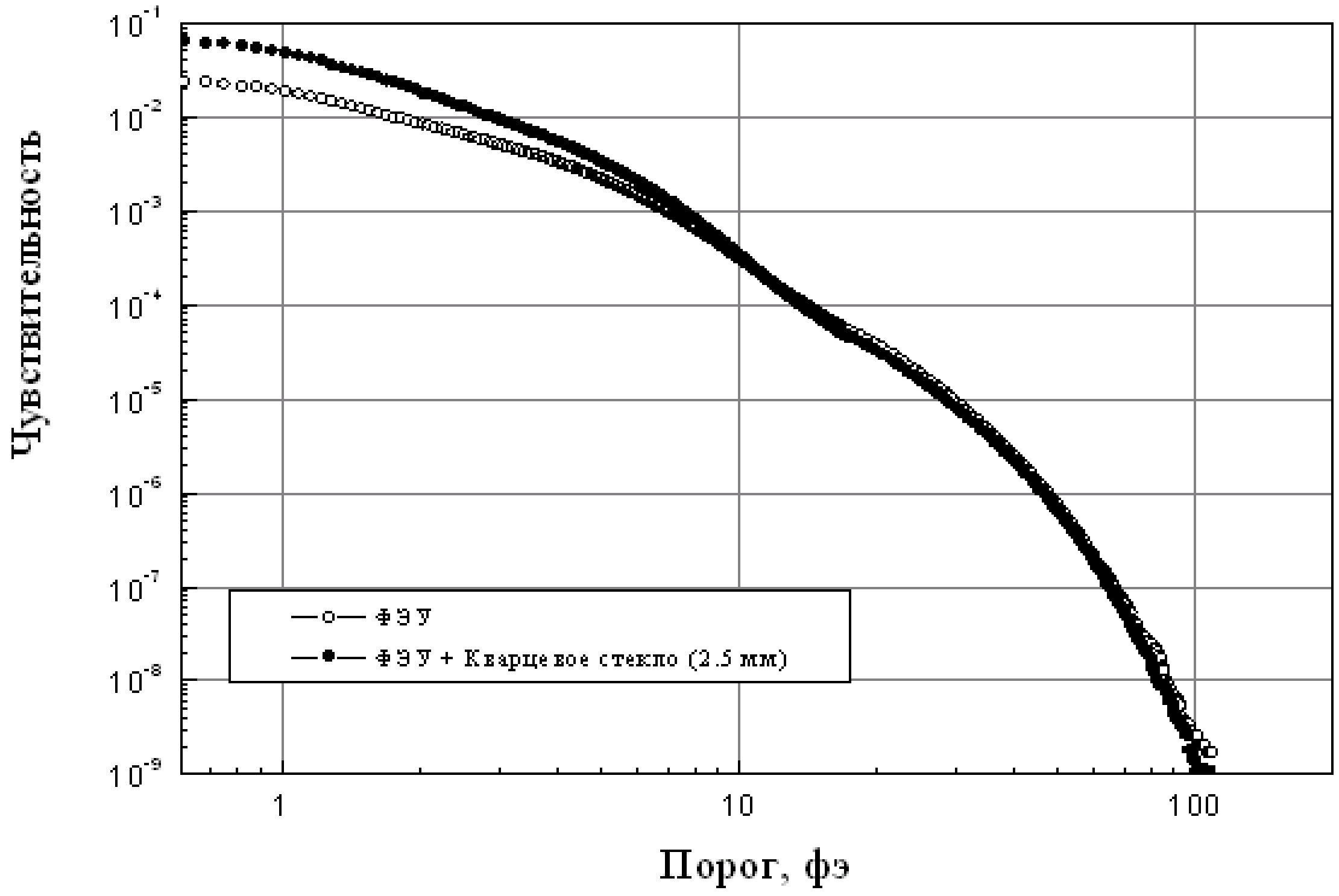

Рис. 3.9- Зависимость интегральной чувствительности ФЭУ к фоновому гамма излучению ($^{60}Co$) от порога при увеличении толщины входного окна.

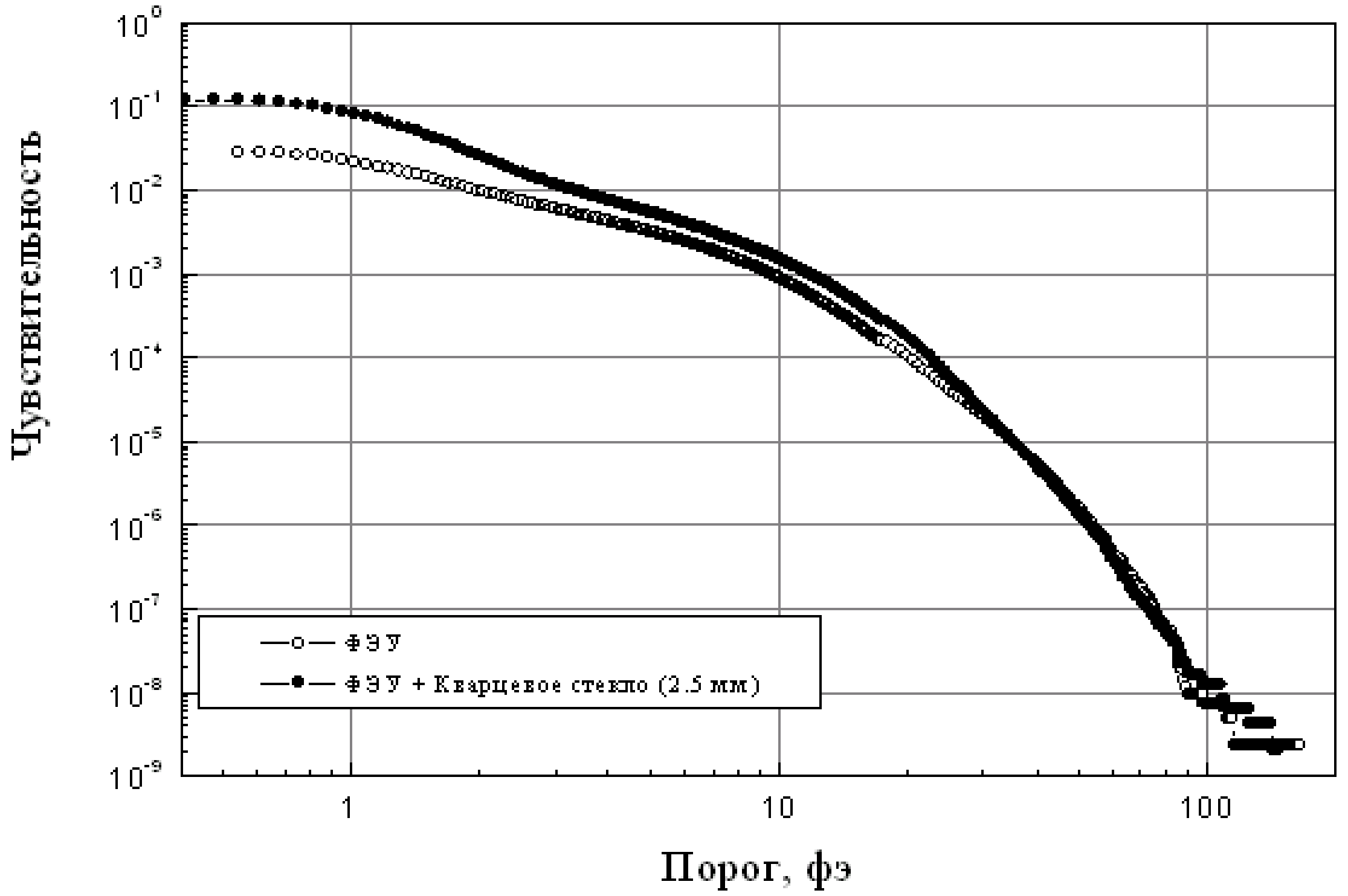

Рис. 3.10- Зависимости интегральной чувствительности ФЭУ к фоновому нейтронному и гамма излучению ($^{252}Cf$) от порога регистрации при увеличении толщины входного окна.

Из рис. 3.9 и 3.10 видно, что с ростом толщины стекла, установленного перед фотокатодом растет и полная чувствительность ФЭУ к фоновому излучению, причем этот прирост обусловлен в основном регистрацией импульсов малых и средних амплитуд в диапазоне от 0.1 фэ до 10 фэ при облучении источником $^{60}Co$ и от 0.1 фэ до 30 фэ при облучении источником $^{252}Cf$. Прирост чувствительности, очевидно, связан с увеличением доли гамма квантов взаимодействующих со стеклом. Т.е. ФЭУ, у которых толщина входного окна больше, будут иметь и более высокую чувствительность к фоновому гамма излучению. Из таблицы 3.1 видно, что миниатюрные ФЭУ (R5600U и R5600U-06), которые имеют меньшую, чем у других толщину входного окна, обладают самой низкой чувствительностью к фоновому излучению.

Входное окно ФЭУ обычно изготавливается из боросиликатного стекла / 67 /. Известно, что примесь $^{10}B$, обладающего аномально высоким сечением реакции (n,α), в боросиликатном стекле составляет 9 %. Поэтому с целью учета вклада взаимодействия нейтронов с входным окном были проведены измерения амплитудного распределения импульсов при облучении его альфа частицами. Результаты измерения дифференциального спектра импульсов при облучении входного окна ФЭУ альфа частицами от источника $^{238}Pu$ ($E_\alpha$=5.5 МэВ) представлены на рисунке 3.11.

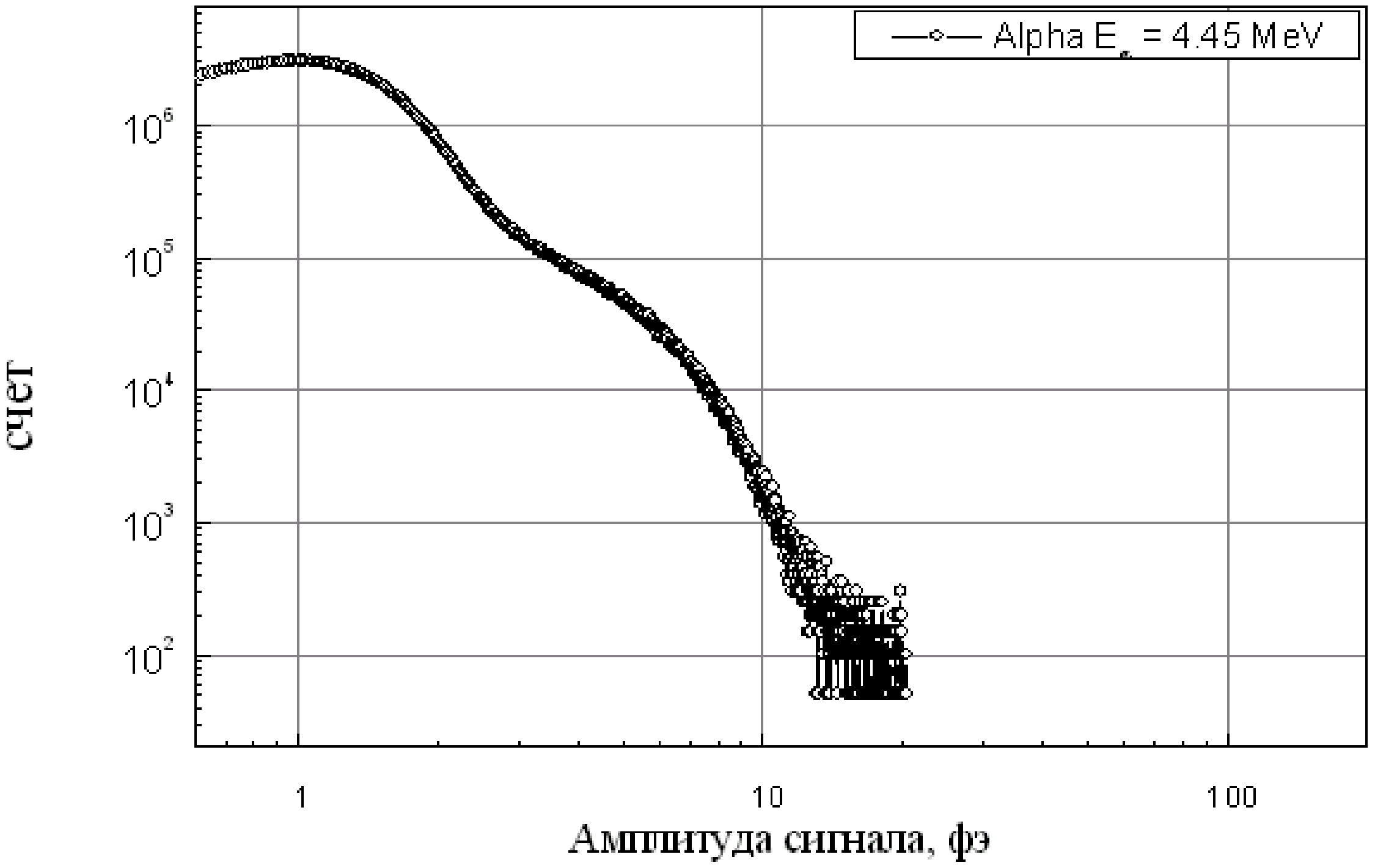

Рис. 3.11- Дифференциальный спектр альфа частиц с энергией $E_\alpha$= 4.45 МэВ

Из рисунка 3.11 отчетливо видно, что при поглощении альфа частиц с энергией $E_\alpha$= 4.45 МэВ, максимальная амплитуда импульсов достигает 20 фэ, при этом среднее число отчетов, возникающих при поглощении одной альфа частицы указанной энергии составляет ~ 10, что, по всей видимости, может быть, объяснено учетом послесвечения стекла. Как следует из этих результатов сигналы больших амплитуд (> 100 фэ) нельзя объяснить генерацией альфа частиц, образующихся в боросиликатном стекле под действием нейтронов.

Таким образом, исходя из приведенных выше, экспериментальных и расчетных данных, мы можем утверждать, что фотокатодная камера дает определяющий вклад в полную чувствительность ФЭУ к регистрации фонового излучения. При этом формирование амплитудного распределения фоновых сигналов, так или иначе, связано со стеклом входного окна детектора. Следует также отметить, что с ростом толщины входного окна растет и полная чувствительность ФЭУ к фоновому излучению.

Рассмотрим более подробно основные процессы, отвечающие за формирование амплитудного распределения сигналов на выходе ФЭУ при облучении нейтронами и гамма квантами от источника $^{252}$Cf.

## 5. Анализ основных процессов, определяющих амплитудное распределение фоновых сигналов на выходе ФЭУ под действием излучения.

Представим результаты измерений дифференциальной чувствительности к фоновому нейтронному и гамма излучению в зависимости от амплитуды сигналов на выходе ФЭУ. Отметим, что выбор ФЭУ (R3998-02) был продиктован тем, что этот детектор обладает хорошо выраженным одноэлектронным пиком, что позволяет легко выразить амплитудное распределение в количестве фотоэлектронов. На рис. 3.13 приведены зависимости дифференциальной чувствительности ФЭУ от амплитуды сигнала, измеренные для трех источников излучения ($^{252}$Cf , $^{137}$Cs и $^{60}$Co). По оси абсцисс отложены амплитуды сигналов на выходе ФЭУ, выраженные в фотоэлектронах, а по оси ординат дифференциальная чувствительность ФЭУ.

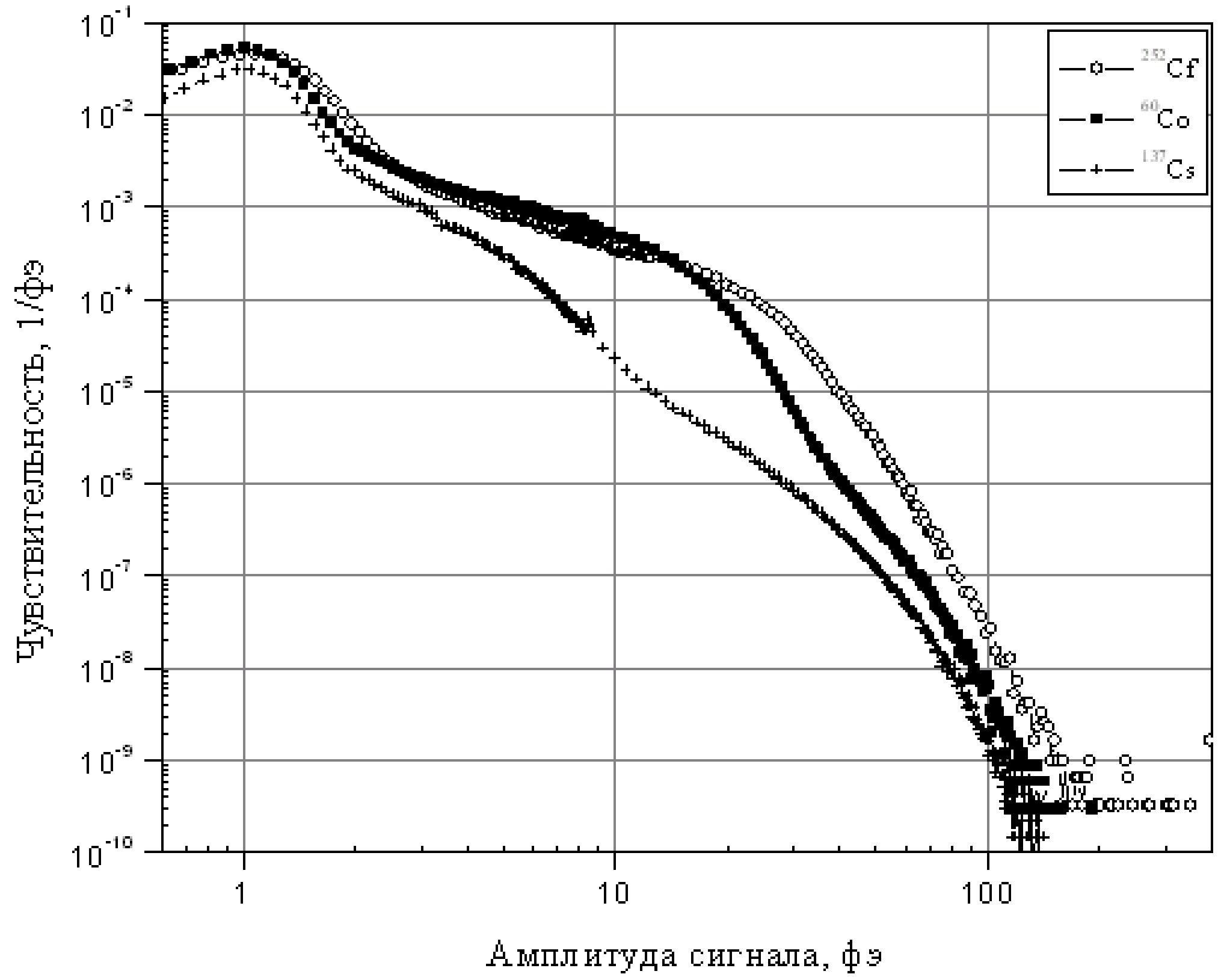

Рис. 3.13- Зависимость дифференциальной чувствительности ФЭУ к фоновому излучению от амплитуды сигнала для различных источников излучения.

Из рисунка 3.13 видно, что все амплитудные распределения имеют одинаковый характер в области одноэлектронного пика. Также все они быстро спадают с увеличением амплитуды сигналов. Однако, чем больше средняя энергия падающего излучения на входное окно ФЭУ, тем больше средняя амплитуда сигналов. Наибольшие амплитуды фоновых сигналов наблюдаются в случае облучения $^{252}Cf$.

Для оценки вклада различных процессов в чувствительность ФЭУ к фоновому нейтронному и гамма излучению была использована компьютерная программа, разработанная на кафедре Экспериментальной Ядерной Физики при Санкт-Петербургском государственном техническом университете. Программа была основана на методе Монте-Карло. Алгоритм моделирования учитывал взаимодействие гамма квантов и генерируемых ими электронов, как с элементами конструкции камеры, окружающей детектор, так и непосредственно с элементами детектора. Вероятность взаимодействия гамма квантов с веществом определялась исходя из сечений для фотоэффекта, комптоновского рассеяния и рождения пар. Значения этих сечений были получены из базы данных, представленной в интернете / 92 – 94 /.

При этом угловое и энергетическое распределения гамма квантов при комптоновском рассеянии определялось в соответствии с известным выражением, Клейна-Нишины-Тамма / 95 /:

$$\frac{d\sigma}{d\Omega} \approx \left\{ \frac{1}{\left(1+k\left(1-\cos\theta_\gamma\right)\right)^2} \left[ 1+\cos^2\theta_\gamma + \frac{k^2\left(1-\cos\theta_\gamma\right)^2}{1+k\left(1-\cos\theta_\gamma\right)} \right] \right\}$$

где $k = E_\gamma / mc^2$, а $E'_\gamma$ определяется следующей формулой:

$$E'_\gamma = E_\gamma / (1 + k(1-\cos\theta_\gamma))$$

Угловое и энергетическое распределения электронов, генерируемых при взаимодействии гамма квантов с веществом определялось, следующим образом:

- Для фотоэффекта была использована формула углового распределения Заутера:

$$\frac{d\sigma}{d\Omega} \approx \frac{\sin^2\theta_e}{\left(1-\beta\cos\theta_e\right)^3} \left( \frac{1}{1-\beta\cos\theta_e} + \frac{3-3\sqrt{1-\beta^2}-2\beta^2}{\left(1-\beta^2\right)^{3/2}} \right)$$

$$\text{где } \beta = \frac{\sqrt{1+2mc^2/E_e}}{1+mc^2/E_e}$$

- Для комптоновского рассеяния энергия электрона вычислялось как:

$$E_e = E_\gamma - E`_\gamma$$

При этом угол вылета электрона определялся по известной формуле:

$$ctg\theta_e = (1+k)tg\frac{\theta_\gamma}{2}, \text{ а } \varphi_e = \pi + \varphi_\gamma$$

- В случае рождения пары использовалась формула Хьюфа для энергетического распределения элементов пары:

$$\frac{d\sigma}{dx} \approx u\{1 + 0.135[Q - 0.52]u(1-u^2)\}, \quad k \geq 4.2$$

$$\frac{d\sigma}{dx} \approx uQ, \text{ при } 2 < k < 4.2$$

$$Q = (1-\gamma)\left[\frac{1}{3}(4-\gamma^2)(L-1) - \gamma^2 F(F-1) - \gamma^4(L-F)F\right]$$

где $\gamma = 2/k$, $L = 2(1-\gamma^2)^{-1}ln(\gamma^{-1})$, $F = (1-\gamma^2)^{-1/2}ln(\gamma^{-1}+(\gamma^{-2}-1)^{1/2})$

$$u = 2[x(1-x)]^{\frac{1}{2}}, \qquad x = \frac{E_{e-}}{E_\gamma - 2mc^2}, \qquad E_{e+-} = E_\gamma - 2mc^2 - E_{e-}$$

Угловое распределение электронов в этом случае определялось по формуле Зоммерфельда:

$$\frac{d\sigma_\chi}{d\Omega} \approx \frac{1-\beta^2}{4(1-\beta\cos\theta_{e-})^2}$$

Угловое распределение позитрона $\theta_{e+}$ определяется из закона сохранения импульса

$$\mathbf{p}_\gamma = \mathbf{p}_{e-} + \mathbf{p}_{e+} \qquad p_\gamma = E_\gamma/c\ ,\ p_e = E_e/c\ (1 + 2mc^2/E)^{1/2}$$

Прохождение электронов, генерируемых гамма квантами через вещество моделировалось следующим образом. При заданной энергии и угле вылета электрона определялась вероятность выхода электрона за предел элемента в соответствии с аналитическим приближением, полученным из анализа экспериментальных данных по прохождению электронов через вещество / 96, 97 /.

$$P(\chi) = e^{-(1.73\chi)^3}$$

где: $\chi = \frac{L}{R}$ - пройденный путь в длинах свободного пробега. L – расстояние от точки рождения электрона до границы элемента. R – Средняя длина пробега электрона с заданной энергией в данном веществе.

В случае, если электрон не выходил за границу элемента тогда рассматривалась вероятность генерации кванта тормозного излучения. Разница начальной энергии электрона и энергии тормозного кванта полагалась равной энергии, выделенной электроном в веществе. В противном случае (если электрон выходит за границу элемента) по модифицированной формуле Ландау вычислялся выход электрона с энергией $E_1$ (энергия вылетевшего электрона) при этом средняя потерянная энергия $\overline{\Delta E}$, используемая в формуле Ландау определялась по методу остаточного пробега.

Таким образом, данная модель достаточно точно учитывала процесс диссипации энергии электрона при прохождении через вещество. Однако она не учитывала процесс рассеяния электронов в веществе. Это приближение в нашем случае достаточно корректно, так как первичным источником электронов является гамма излучение или нейтронный поток, имеющий изотропное распределение.

На рис. 3.14 представлены результаты моделирования дифференциальной чувствительности ФЭУ к излучению $^{252}Cf$ полученные с помощью описанной выше программы.

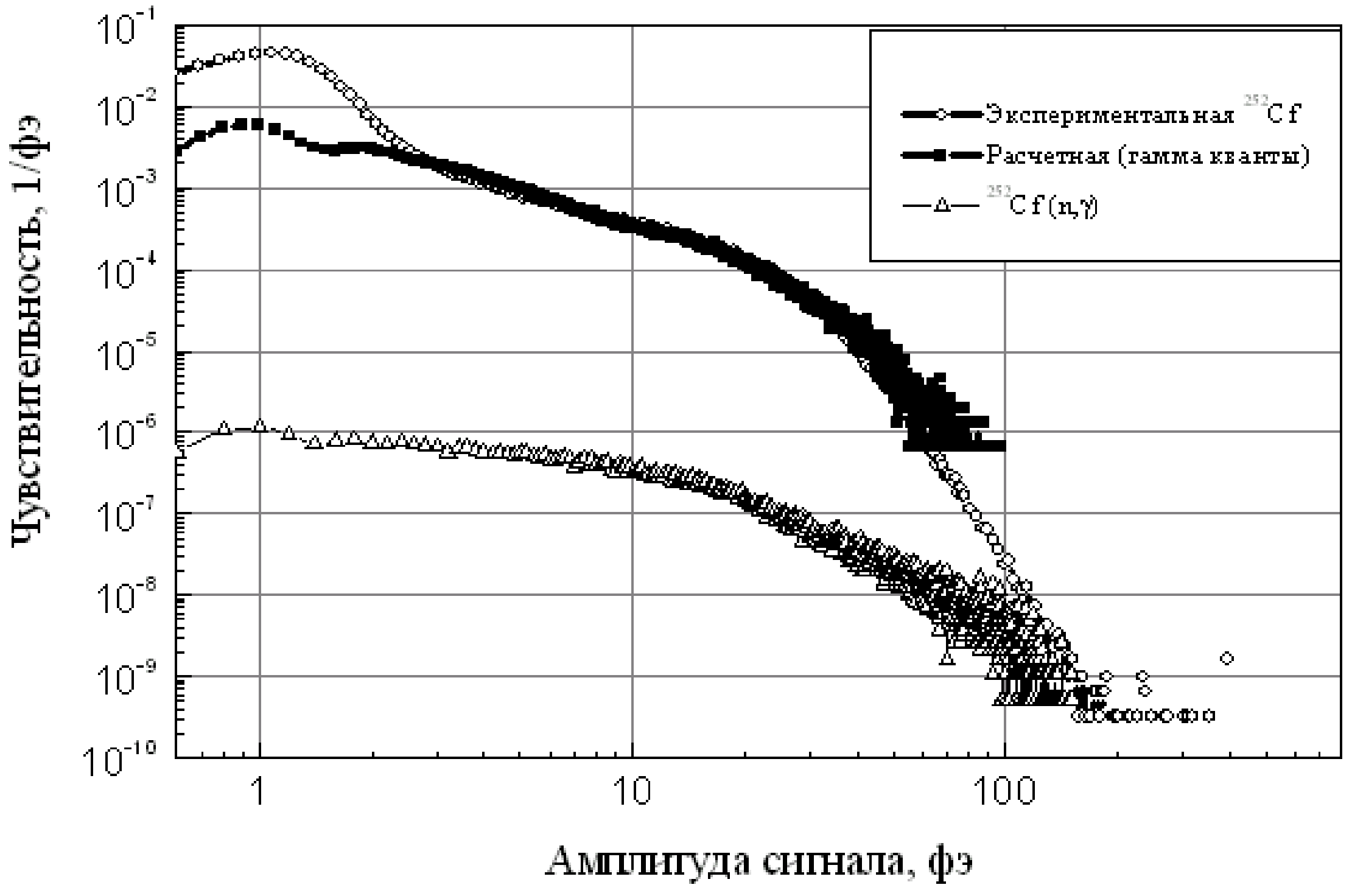

Рис. 3.14- Сравнение зависимостей дифференциальных чувствительностей ФЭУ к фоновому излучению от амплитуды сигнала при облучении $^{252}$Cf и моделирования.

На этом рисунке представлены три кривые, первая из них – экспериментальная. Вторая кривая представляет амплитудное распределение сигналов, обусловленное сцинтилляциями, возникающими при взаимодействии гамма квантов $^{252}$Cf с входным окном ФЭУ. Третья кривая отвечает моделированию взаимодействия гамма квантов, генерируемых в элементах конструкции окружающей детектор за счет реакции (n,γ) и также вызывающих сцинтилляции во входном стекле ФЭУ. Надо отметить, что при моделировании значение средней энергии, затрачиваемой на образование одного фотоэлектрона в боросиликатном стекле, составило ~ 70 КэВ. Из этого рисунка видно, что суммарный учет процессов, описываемых второй и третьей кривой, дает хорошее совпадение в области амплитуд превышающих 2 фэ. Однако расчетные значения дифференциальной чувствительности заметно расходятся в области амплитуд вблизи одноэлектронного пика. По всей видимости, это может быть связанно с тем, что в наших расчетах не учитывалось наличие медленной компоненты свечения входного стекла под действием излучения. Наличие медленной компоненты приводит к тому, что часть поглощенной энергии высвечивается вне времени формирования сигнала в основном усилителе (~ 1 мксек) и приводит к регистрации избыточного числа одноэлектронных импульсов. Данный факт проверялся экспериментально при облучении входного стекла ФЭУ источником альфа частиц (см. Рис. 3.11). В этом эксперименте было получено, что система регистрация давала в среднем ~ 10 отчетов с амплитудами, лежащими в области одноэлектронного пика на каждую поглощенную альфа частицу. В области же больших амплитуд число, зарегистрированных событий совпадало с числом альфа частиц, падающих на входное окно ФЭУ.

Таким образом, можно утверждать, что выбранная нами модель адекватно описывает амплитудное распределение сигналов на выходе ФЭУ при облучении источником $^{252}Cf$ в области амплитуд, превышающих одноэлектронные сигналы.

Дополнительные аргументы в пользу описанной выше модели были получены при реализации следующих экспериментов. Между источником и камерой детектора устанавливался блок свинца толщиной 50 мм, который эффективно поглощает гамма квантов низких энергий (см. Рис. 3.17). Для иллюстрации выше сказанного предположения на рисунке 3.16. приведены два спектра гамма квантов $^{252}Cf$ до, и после прохождения свинцовой защиты.

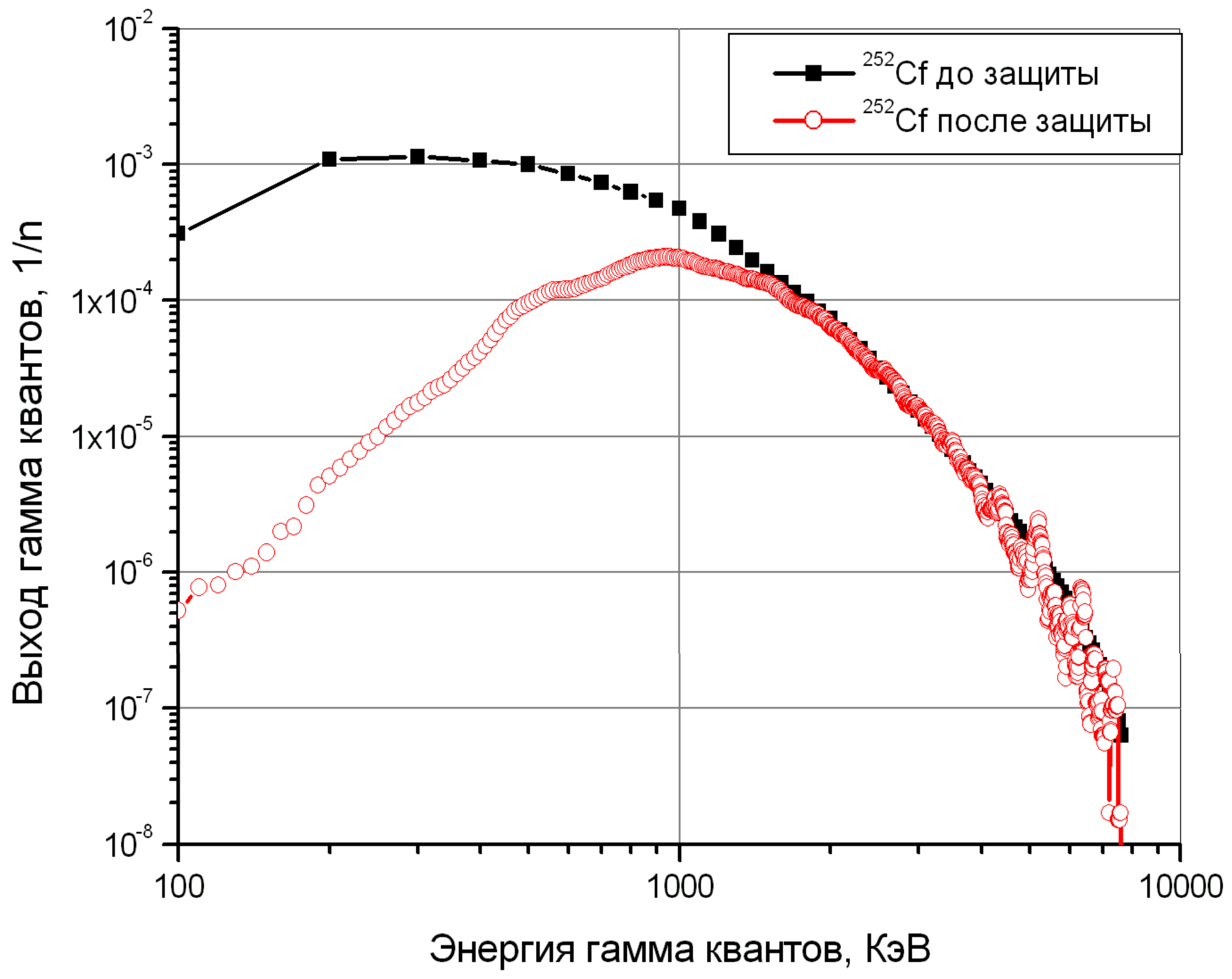

Рис. 3.16- Спектры гамма квантов до и после прохождения свинцовой защиты толщиной 50 мм.

Как видно из рисунка такая свинцовая защита эффективно поглощает гамма кванты до энергии в сотни КэВ.

Результаты измерения чувствительностей ФЭУ к фоновому излучению при облучении детектора с защитой в виде свинцового блока и без нее представлены на рисунке 3.17

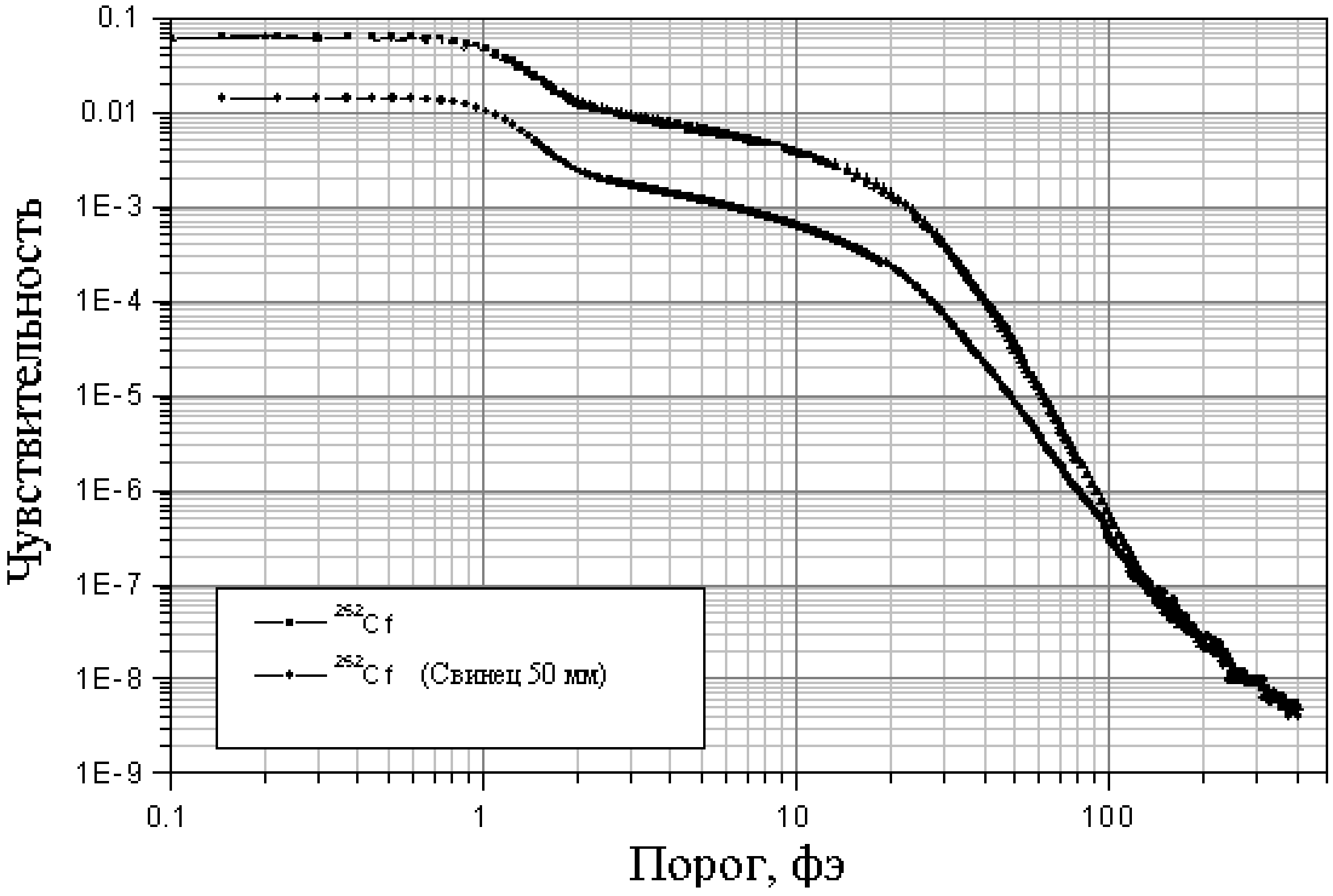

Рис. 3.17- Зависимость интегральной чувствительности ФЭУ к нейтронному фоновому излучению при облучении с защитой и без нее.

Из рис. 3.17 видно, что именно в области амплитуд сигналов > 100 фэ, как и ожидалось чувствительность ФЭУ к фоновому излучению осталось прежней. Это связанно с тем, что ослабление потока гамма квантов высокой энергии, при прохождении через такую защиту практически не происходит. Напротив заметное уменьшение в чувствительности ФЭУ к фоновому излучению произошло в области амплитуд сигналов от одного до десятков фотоэлектронов. Тем самым полностью подтверждается гипотеза о том, что именно вклад гамма квантов относительной небольшой энергии является определяющим чувствительность ФЭУ к фоновому нейтронному и гамма излучению.

Второй эксперимент заключался в том, что был установлен блок полиэтилена толщиной 80 мм между источником и детектором. При этом спектр гамма излучения, испускаемого от источника $^{252}$Cf, практически не изменился. Однако замедление нейтронов в полиэтиленовом блоке привело к увеличению вероятности генерации гамма квантов по каналу реакции (n,γ) в стенках железной камеры. При этом в железе происходит генерация гамма квантов, в том числе и очень высокой энергий (до 10 МэВ). На рисунке 3.18 представлено сравнение результатов измерения дифференциальных чувствительностей ФЭУ при облучении нейтронами от $^{252}Cf$ с замедлителем и без него.

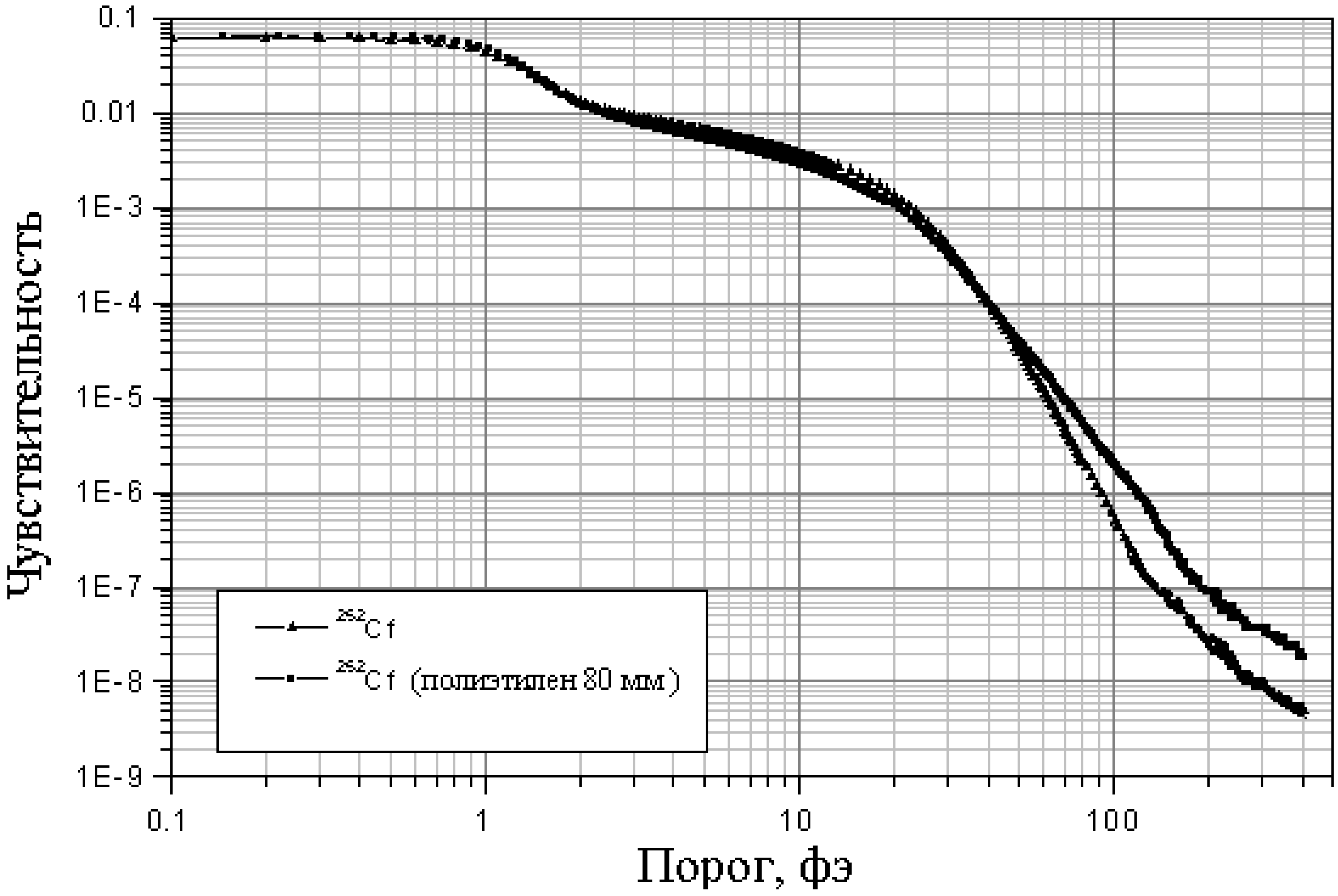

Рис. 3.18- Зависимость интегральной чувствительности ФЭУ к нейтронному фоновому излучению при облучении $^{252}Cf$ с замедлителем и без него.

Из рис. 3.18 видно, что в области порогов > 70 фэ произошло заметное увеличение чувствительности ФЭУ к фоновому излучению, что соответствует области амплитуд сигналов, возникающих при взаимодействии гамма квантов высокой энергии. В нашем случае сечение реакции (n,γ) в железе усредненное по спектру $^{252}Cf$ составило ~ 18 мб, что при толщине железной крышки ~ 20 мм дает вероятность генерации гамма квантов ~ $10^{-3}$. Учитывая, что вероятность поглощения во входном окне таких гамма квантов составляет ~ $10^{-2}$ в лучшем случае то мы получаем, что заметный рост чувствительности ФЭУ в этом случае должен ожидаться на уровне $10^{-5}$, что мы и наблюдаем на рис. 3.18.

Представим результаты моделирования взаимодействия гамма квантов, испускаемых источниками $^{137}Cs$ и $^{60}Co$ с входным окном ФЭУ.

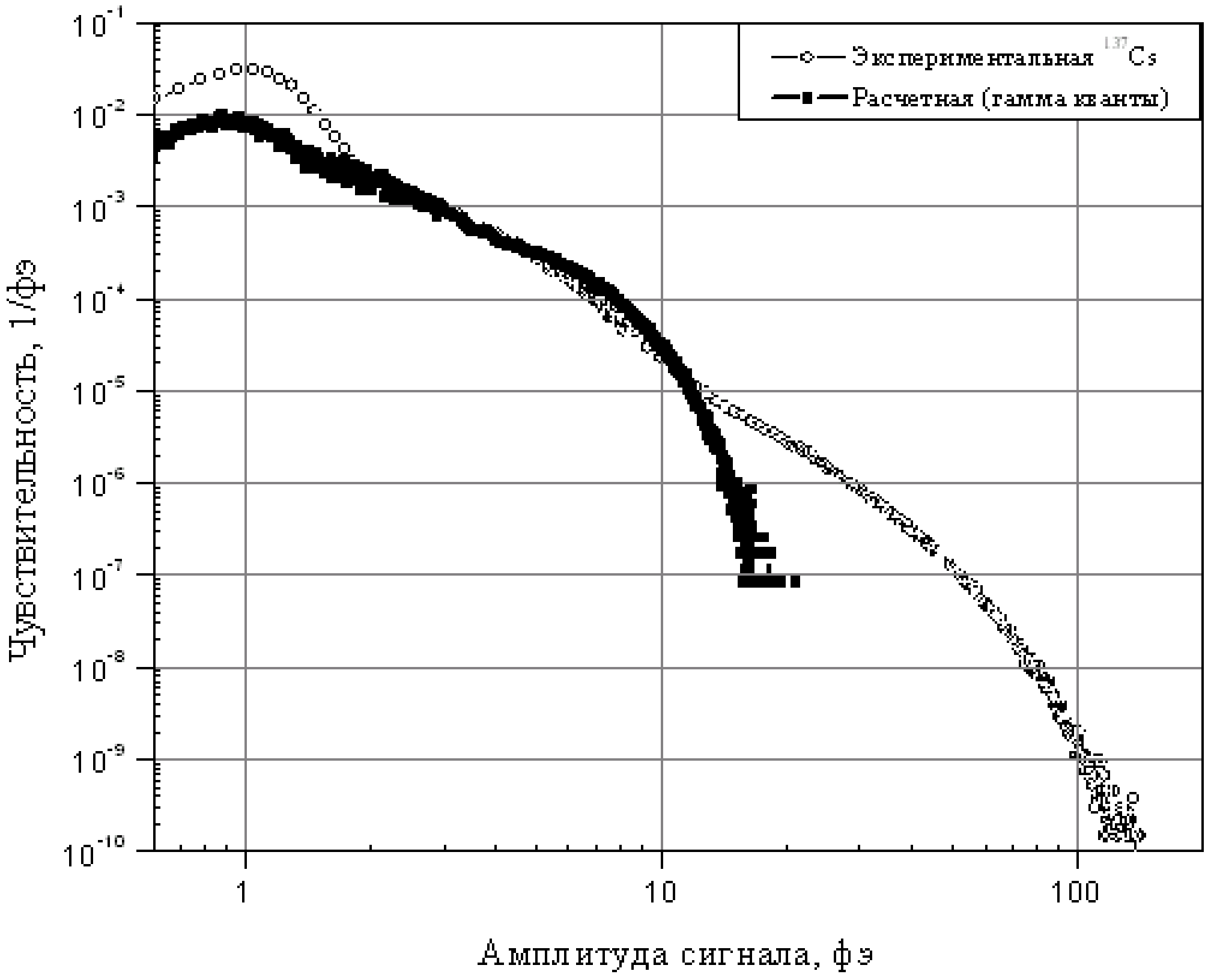

Рис. 3.17- Сравнение зависимостей дифференциальных чувствительностей ФЭУ к фоновому излучению от амплитуды сигнала при облучении $^{137}Cs$.

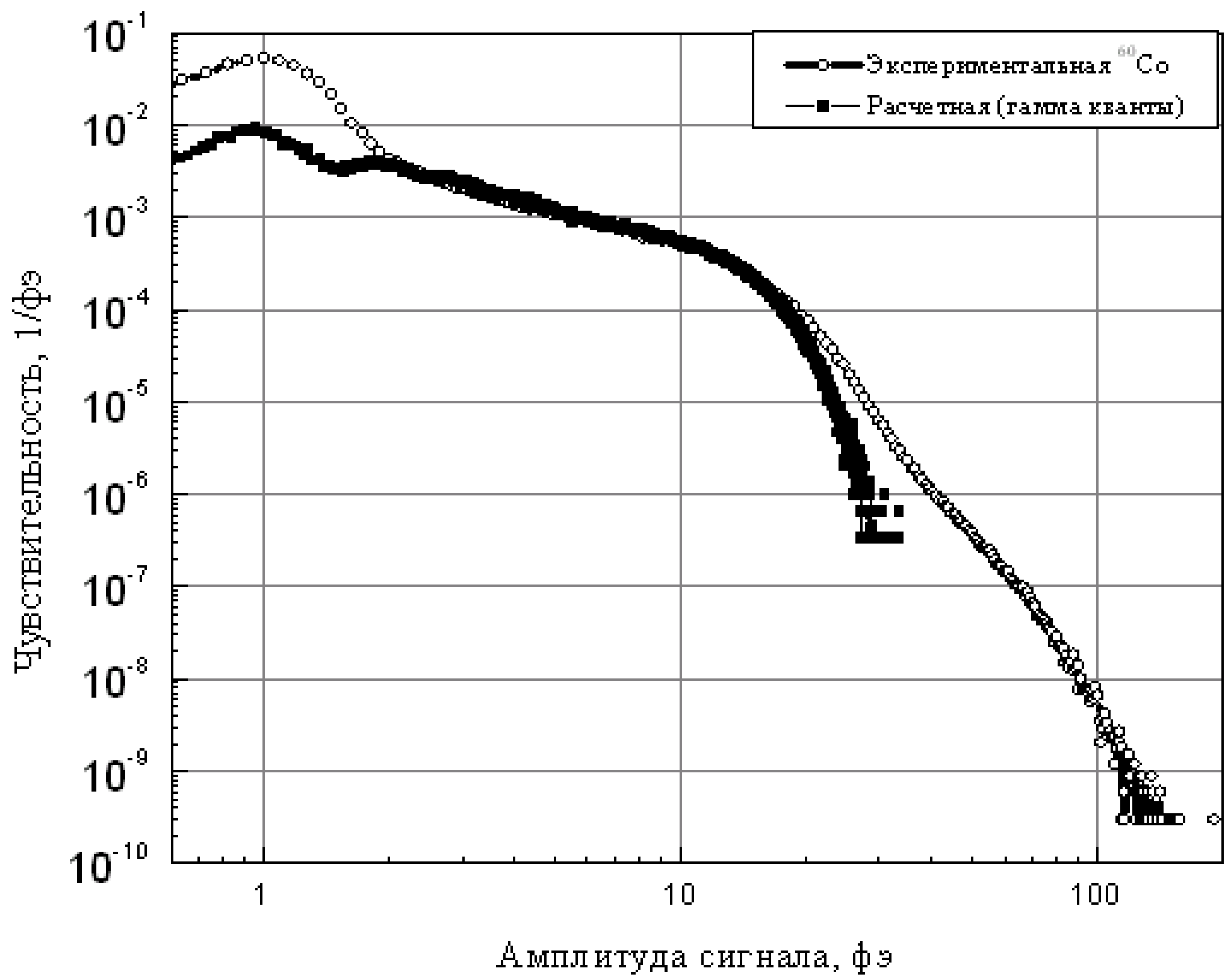

Рис. 3.18- Сравнение зависимостей дифференциальных чувствительностей ФЭУ к фоновому излучению от амплитуды сигнала при облучении $^{60}$Co.

На рисунках 3.17 и 3.18 представлены две кривые, первая из них соответствует экспериментальным данным, а вторая – расчетным. Вторая – была получена при моделировании взаимодействия гамма квантов, с входным окном ФЭУ. Из этих рисунков видно, что заметное расхождение наблюдается в области одноэлектронного пика, как и для $^{252}$Cf и, по всей видимости, природа их возникновения аналогична. В области амплитуд сигналов от 2 до 12 фэ. (для $^{137}$Cs) и от 2 до 20 фэ. (для $^{60}$Co) наблюдается хорошее совпадение расчетных кривых с экспериментом. Однако надо отметить, что расчетные данные практически не дают сигналов с амплитудами превышающие десятки фотоэлектронов, которые наблюдаются в эксперименте. По всей видимости, это может быть объяснено тем, что в программу не был заложен процесс взаимодействия электронов, эмитированных входным окном, с фотокатодом ФЭУ.

Оценим вклад этого процесса в генерацию сигналов больших амплитуд. Для этого, используя описанную выше программу, мы рассчитали, долю и энергетический спектр электронов, вышедших из входного окна ФЭУ под действием облучения $^{137}Cs$ и $^{60}Co$. В результате было получено, что полное число электронов выходящих из стекла, составляет величину порядка $10^{-4}$. Так как прохождение этих электронов через фотокатод вызывает вторичную эмиссию с вероятностью близкой к 100 %, то это число и будет определять вероятность эмиссии электронов из фотокатода под действием гамма квантов облучающих камеру детектора. Для оценки средней энергии оставленной электронами ΔE, при прохождении через фотокатод мы использовали следующее выражение:

$$\Delta \mathrm{E} = \frac{dE}{dx} \cdot t , \qquad (5)$$

где: ΔE – среднее количество энергии выделенной при прохождении электронов через фотокатод, $\frac{dE}{dx}$ - тормозная способность вещества фотокатода для данной энергии электронов, t – толщина фотокатода.

Можно отметить, что средняя энергия ΔE выделенная в фотокатоде слабо зависит от величины энергии проходящих через него электронов. Это является следствием того, что тормозная способность вещества, из которого изготовлен фотокатод, мало меняется в рассматриваемом диапазоне энергий. Величина ΔE составила ~ 50 эВ (при толщине фотокатода ~ 1000 Å), причем для широкого диапазона энергии электронов ($E_e$ = 200 – 1000 КэВ)/ 98 – 100 /, проходящих через фотокатод. Так как для образования одного электрона вторичной эмиссии затрачивается примерно 1 эВ, то можно ожидать что, такой процесс может приводить к генерации сигналов больших амплитуд при облучении источниками $^{137}Cs$ и $^{60}Co$. При облучении $^{252}Cf$ генерация сигналов больших амплитуд, как было установлено выше, связана с взаимодействием гамма квантов больших энергией с входным окном ФЭУ и этот процесс имеет более высокую вероятность реализации, чем эмиссия вторичных электронов из фотокатода.

Надо отметить, что наблюдаемое расхождение расчетных значений чувствительностей с экспериментальными данными связано с несовершенством модели, используемой для расчета. Так как при моделировании не был учтен вклад медленной компоненты высвечивания входного стекла ФЭУ, кроме того, в модели ФЭУ были использованы следующие приближения:

- Во-первых, чувствительная область ФЭУ, при моделировании на компьютере была представлена только входным окном. Тогда как в реальных условиях фотокатодная камера включает и боковые стенки. Это привело к тому, что не был учтен вклад в амплитудное распределение сигналов, возникающих за счет взаимодействия фонового излучения с боковыми стенками фотокатодной камеры. Учет боковой стенки фотокатодной камеры затруднем в связи с тем, что конверсионная эффективность фотокатода покрывающего боковые стенки камеры не паспортизуется.

- Во-вторых, в качестве средней энергии, затрачиваемой, на образование одного фотоэлектрона в боросиликатном стекле мы брали значение ~ 70 КэВ. Необходимо подчеркнуть, что эта величина является приближенной, так как она получена при анализе амплитудного распределения при облучении бета частицами сплошного спектра со средней энергией 99 КэВ. Надо отметить, что в литературе отсутствуют данные о конверсионной эффективности боросиликатного стекла.

Таким образом, можно утверждать, что использованная нами модель позволяет предсказать амплитудное распределение сигналов, возникающих на выходе ФЭУ под действием фонового излучения и тем самым оценить вклад различных процессов в полную чувствительность. Кроме того, на основе представленных расчетных и экспериментальных данных для источников $^{137}$Cs, $^{60}$Co и $^{252}$Cf, также можно утверждать, что основным процессом, определяющим чувствительность ФЭУ к фоновому нейтронному и гамма излучению является взаимодействие гамма квантов с входным окном ФЭУ и поглощение в нем электронов, генерируемых в элементах конструкции окружающей детектор.

6. Определение чувствительности сцинтилляционных детекторов к фоновому нейтронному и гамма излучению.

Следующий раздел посвящен анализу чувствительности сцинтилляционного детектора к фоновому излучению. Как известно, сцинтилляционный детектор состоит из сцинтилляционного кристалла, находящегося в оптическом контакте с входным окном ФЭУ (см. рис. 3.19). Так как нас интересуют сцинтилляционные кристаллы с толщинами, сравнимыми с пробегом, регистрируемых частиц, то далее мы рассмотрим чувствительность к фоновому нейтронному и гамма излучению для сцинтилляционных детекторов CsI(Tl) с толщинами кристаллов от одного до десяти микронов.

На рис. 3.19 представлена схема сцинтилляционного детектора. Такая конфигурация была использована для исследования фоновых характеристик сцинтилляционных детекторов CsI(Tl) при облучении источниками $^{252}$Cf и $^{60}$Co.

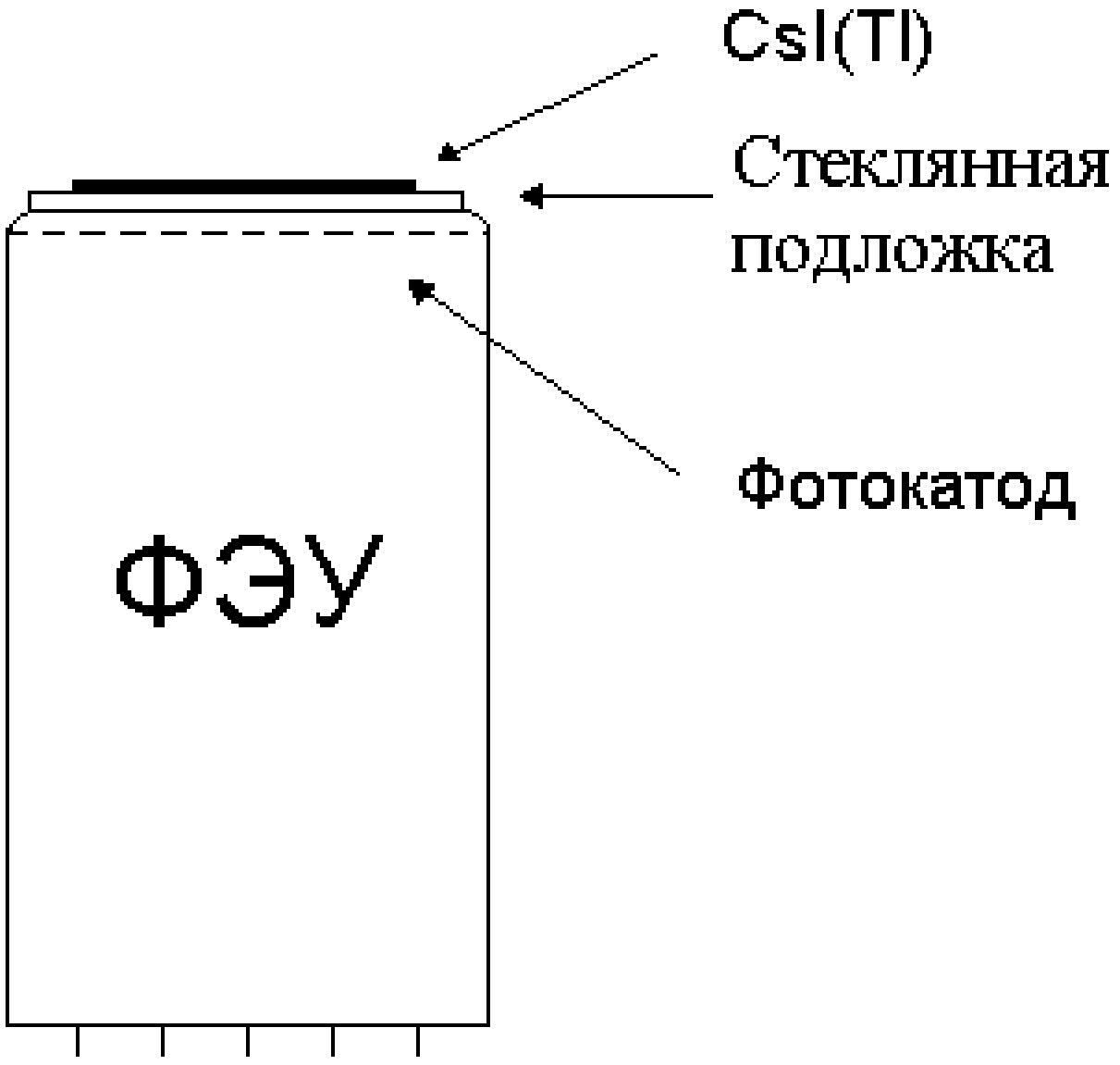

Рис. 3.19 Схема сцинтилляционного детектора.

Результаты измерения интегральной чувствительности сцинтилляционного детектора CsI(Tl) к фоновому излучению в зависимости от величины порога регистрации и толщины сцинтиллятора приведены на рисунке 3.20. Приведенные на этом рисунке зависимости были измерены при облучении источником $^{252}$Cf. Толщина, исследуемых сцинтилляционных кристаллов менялась от одного до восьми микронов.

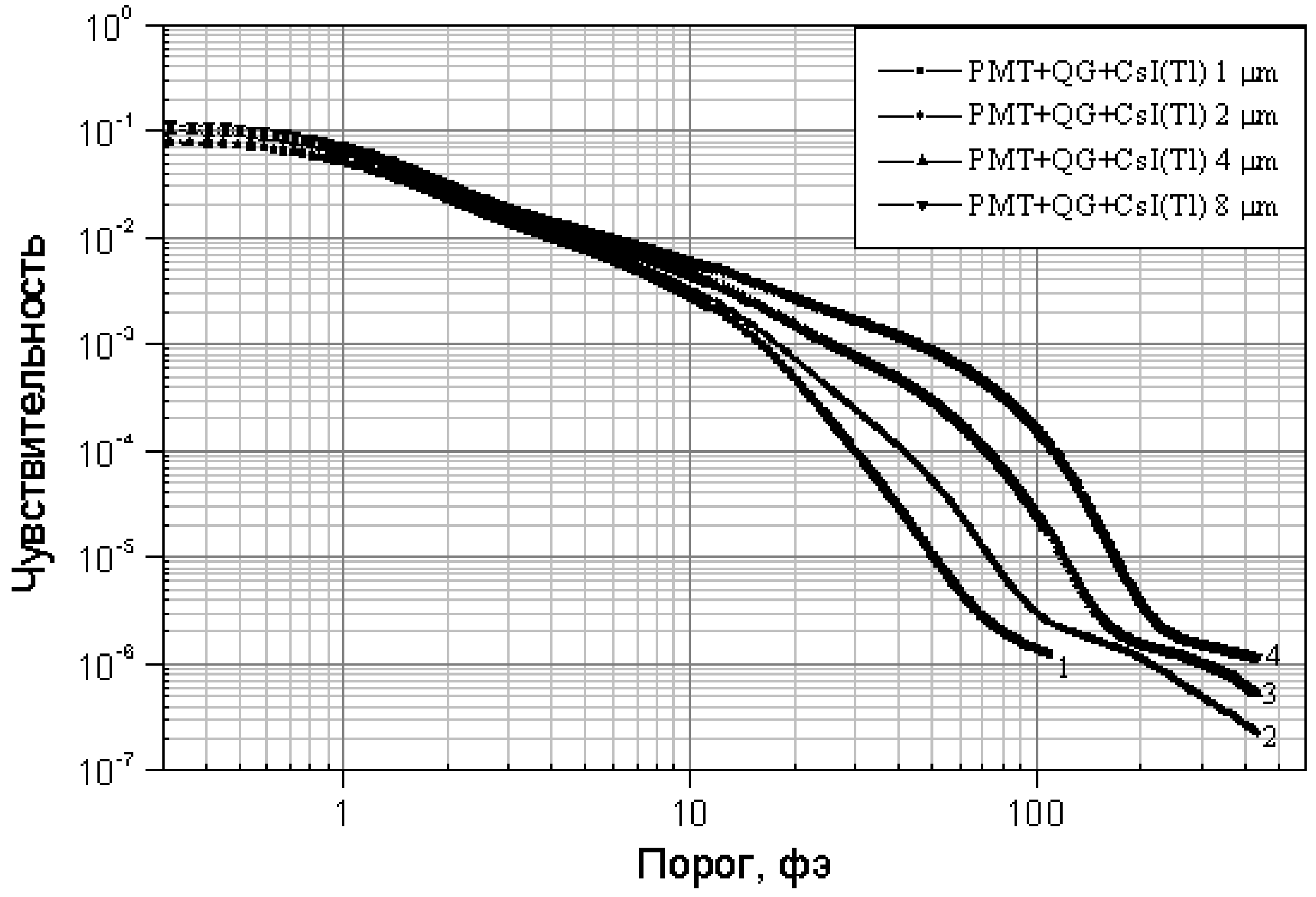

Рис. 3.20 – Зависимости интегральных чувствительностей к фоновому излучению от порога регистрации для сцинтилляторов CsI(Tl) разных толщин: 1- 1, 2- 2, 3- 4, 4- 8 мкм

Из рисунка 3.20 видно, что во всех случаях при увеличении толщины напыленного сцинтиллятора заметно изменяется чувствительность сцинтилляционных детекторов к фоновому излучению. Однако в области больших порогов регистрации увеличение толщины кристалла приводило к заметному увеличению чувствительности детектора к фоновому излучению. По всей видимости, прирост чувствительности в этой области обусловлен взаимодействием быстрых электронов, генерируемых гамма квантами, со сцинтилляционным кристаллом / 62 – 65, 69, 70 /. Известно, что эти электроны взаимодействуют со сцинтилляционным кристаллом с вероятностью близкой к 100%. Так как электроны могут проходить через кристалл, в том числе, и под скользящими углами, то в этом случае они могут оставлять значительное количество энергии в кристалле, и таким образом генерировать сигналы больших амплитуд. При этом, чем больше толщина кристалла, тем выше среднее энерговыделение при прохождении через него электронов. Для оценки приведем средние значения энергии, потерянной электронами при прохождении через сцинтилляционные кристаллы CsI(Tl). Эти величины были получены с помощью описанной выше программы. В таблице 3.2 представлены величины средней энергии, оставленной электронами в кристалле CsI(Tl). Эти электроны генерируются при прохождении гамма квантов через материалы конструкции окружающей детектор.

Таблица 3.2

| Толщина кристалла t, мкм | Средняя энергия, выделенная в кристалле $<\Delta E>$, КэВ | Среднее число фотоэлектронов | |
|---|---|---|---|
| | | $<N_ф>_{расч}$ | $<N_ф>_{эксп}$ |
| 1 | 2.75 | 5.5 | 6.0 |
| 2 | 4.80 | 9.6 | 9.0 |
| 4 | 9.10 | 18.2 | 18.0 |
| 8 | 16.75 | 33.5 | 30.0 |

Из таблицы видно, что увеличение толщины сцинтилляционного кристалла CsI(Tl) приводит к линейному росту средней выделенной в нем энергии. Зная среднее количество выделенной в кристалле энергии можно вычислить среднюю амплитуду сигнала на выходе детектора. Эти результаты приведены в третьей колонке с учетом средней энергии, затрачиваемой на образование одного фотоэлектрона ~ 500 эВ.

Кривая чувствительности в диапазоне порогов от 10 до 200 фэ была аппроксимирована в соответствии с выражением:

$$S(n_{\phi}) = S_0 \cdot e^{-\frac{n_{\phi}}{\overline{N_{\phi}}}}$$

где: $S(n_{\phi})$ - интегральная чувствительность сцинтилляционного детектора CsI(Tl) к фоновому излучению выше заданного фотоэлектронного порога, $S_o$- вероятность генерации сигнала под действием быстрых электронов, возникающих при прохождении гамма квантов через материалы конструкции камеры, окружающей детектор, $\overline{N_{\phi}}$- среднее число фотоэлектронов, образующих в этом процессе (значения были взятые из таблицы 3.2).

Из таблицы видно, что полученные значения $<N_ф>_{эксп}$ достаточные близки к расчетным $<N_ф>_{расч}$. Поэтому мы можем утверждать, что именно процесс взаимодействия электронов со сцинтилляционным кристаллом CsI(Tl) дает вклад в чувствительность, наблюдаемую в диапазоне фотоэлектронных порогов выше 10 фэ.

В области порогов до 10 фэ процесс генерации сигналов как было отмечено выше, в основном обусловлен взаимодействием гамма квантов с стеклом, установленным между сцинтилляционным кристаллом и фотокатодом.

Таким образом, можно утверждать, что основной вклад в полную чувствительность сцинтилляционных детекторов CsI(Tl) микронной толщины обусловлен взаимодействием гамма квантов с входным стеклом ФЭУ. Вклад же сцинтиллятора проявляется в увеличении чувствительности в области относительно больших порогов регистрации. Причем чувствительность в этой области резко растет с увеличением толщины сцинтилляционного кристалла CsI(Tl), что и требует выбрать минимально возможной толщины кристалла при регистрации частиц заданной энергии.

# ГЛАВА IV. АНАЛИЗАТОР НЕЙТРАЛЬНЫХ ЧАСТИЦ МЭВ-ДИАПАЗОНА ЭНЕРГИЙ GEMMA-2M И ИЗОТОПНЫЙ СЕПАРАТОР ISEP

В Физико-техническом институте им А.Ф. Иоффе РАН был разработан и испытан атомный анализатор GEMMA-2M, который в дальнейшем составил основу комплексов корпускулярной диагностики в МэВ-диапазоне энергий. Анализаторы этого типа были установлены на крупнейших термоядерных установках мира JT-60, TFTR и JET. Там же был разработан прототип прибора для анализа содержания водорода, дейтерия и трития в высокотемпературной плазме, названный ISEP. Системы детектирования всех этих приборов были разработаны и изготовлены на кафедре экспериментальной ядерной физики Санкт-Петербургского Государственного Технического университета.

Рассмотрим более подробно результаты оптимизации детектирующих систем для анализаторов нейтральных частиц GEMMA-2M и ISEP.

## 1.- Анализатор нейтральных частиц МэВ-диапазона энергий GEMMA-2M

### 1.1.- Конструкция атомного анализатора GEMMA-2M.

Атомный анализатор частиц GEMMA-2M / 101 – 104 / предназначен для измерения абсолютного потока и энергетического спектра изотопов водорода (H, D, T) в диапазоне энергий 0.2-2 МэВ и атомов гелия ($^{3}$He и $^{4}$He ) в диапазоне энергий 0.4-4 МэВ. Принцип работы анализатора основан на стандартной схеме приборов подобного типа. Обдирка атомов происходит в тонкой пленке, а энергетический и массовый анализ вторичных ионов осуществляется в параллельных магнитном и электрическом полях.

Отличительными чертами данного прибора является использование в качестве приемников – специально разработанных сцинтилляционных детекторов оптимизированных по чувствительности к фоновому нейтронному и гамма излучению плазмы.

На рис. 4.1 представлено схематическое изображение основных элементов анализатора GEMMA-2M (рис. 4.2 фотография анализатора). В камере коллиматоров расположены вертикальная 1 и горизонтальная 2 щели, перемещающиеся в двух взаимно перпендикулярных направлениях и позволяющие дискретно изменять площадь входной апертуры анализатора в пределах от 1 до 50 мм$^2$. В камере электромагнита установлены основные диспергирующие элементы – полюсные наконечники (полюсники) электромагнита 5 и высоковольтный конденсатор 6. На входе в магнит в межполюсном зазоре размещен подвижный механизм 3 с обдирочными углеродными пленками, юстировочными отверстиями и радиоактивным альфа источником, использованным для тестирования. На оптической оси прибора расположен патрубок прямого канала с детектором 4, используемым в калибровочном эксперименте для измерения полного потока атомов, входящих в анализатор. Этот же канал предназначен для точной юстировки прибора на плазменной установке. В камере детекторов 7 установлена панель из восьми детекторов, позволяющих одновременно проводить измерения потока атомов для восьми разных значений энергии.

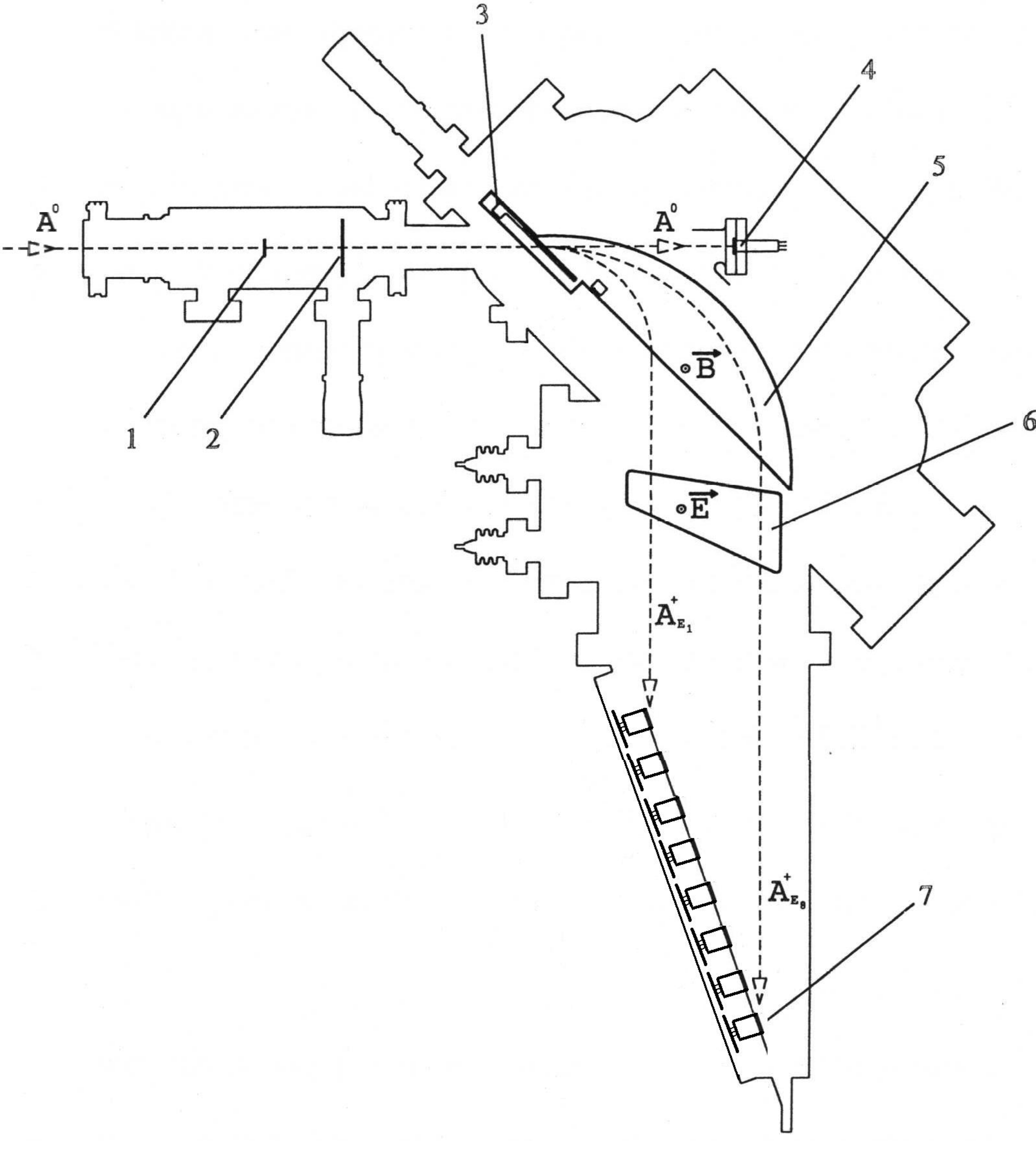

Рис. 4.1.- Схематическое изображение основных элементов атомного анализатора GEMMA-2M: 1, 2 – вертикальная и горизонтальная входные щели, 3- механизм смены обдирочных пленок, 4- детектор прямого канала, 5- полюсники анализирующего магнита, 6- пластины анализирующего конденсатора, 7- детекторный фланец, в котором расположена линейка детекторов.

Рис. 4.2.

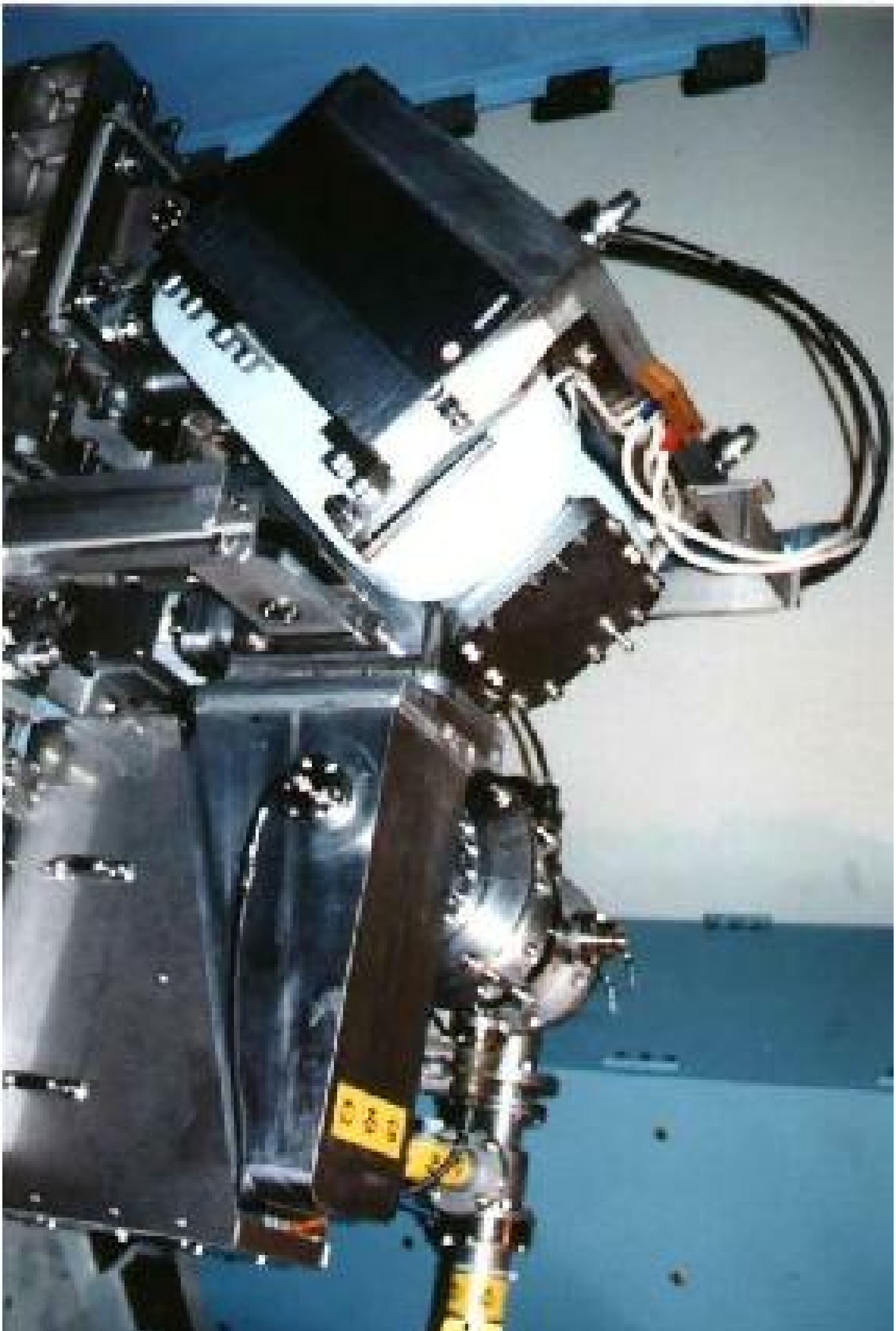

## 1.2.- Конструкция детектирующей системы анализатора

В настоящее время на этом анализаторе установлена новая линейка детекторов, разработанная с использованием описанной выше методики на кафедре экспериментальной ядерной физики Санкт-Петербургского Государственного Технического университета / 62 /. Детектирующая система состоит из восьми сцинтилляционных детекторов, расположенных эквидистантно друг относительно друга (см. рис. 4.1). Детекторы, использованные в данном приборе, отличаются друг от друга тем, что толщина сцинтилляционных кристаллов CsI(Tl) варьируется от 1 до 6.3 мкм.

На рисунках 4.3 и 4.4 представлены для сравнения измеренные зависимости чувствительностей к фоновому излучению $^{252}$Cf и $^{60}$Co для старой и новой детектирующих систем. Видно, что оптимизация детектирующей системы позволила существенно понизить чувствительность детектора к фоновому нейтронному и гамма излучению. Фактически, было достигнуто уменьшение полной чувствительности в 7 раз, а для больших порогов регистрации, уменьшение составило несколько порядков. Этот результат является следствием оптимизации детектирующей системы в соответствии с приведенной выше методикой. Как было показано выше, чем больше толщина стекла между сцинтилляционным кристаллом и фотокатодом ФЭУ, тем выше полная чувствительность детектора к фоновому излучению. Поэтому для уменьшения чувствительности системы детектирования было решено располагать детекторы внутри вакуумной камеры и тем самым отказаться от использования вакуумного стекла, а кроме того, убрать световод и использовать тонкую стеклянную подложку толщиной 0.5 мм для напыления сцинтилляционного кристалла. Таким образом, удалось существенно уменьшить толщину стекла с 7.5 мм (вакуумное стекло 2.5 мм и световод 5 мм) до 0.5 мм. Кроме того, сцинтилляционные кристаллы были оптимизированы по толщине. Подбор толщины сцинтилляционных кристаллов был проведен так, чтобы его толщина соответствовала среднему пробегу регистрируемых частиц в данном канале.

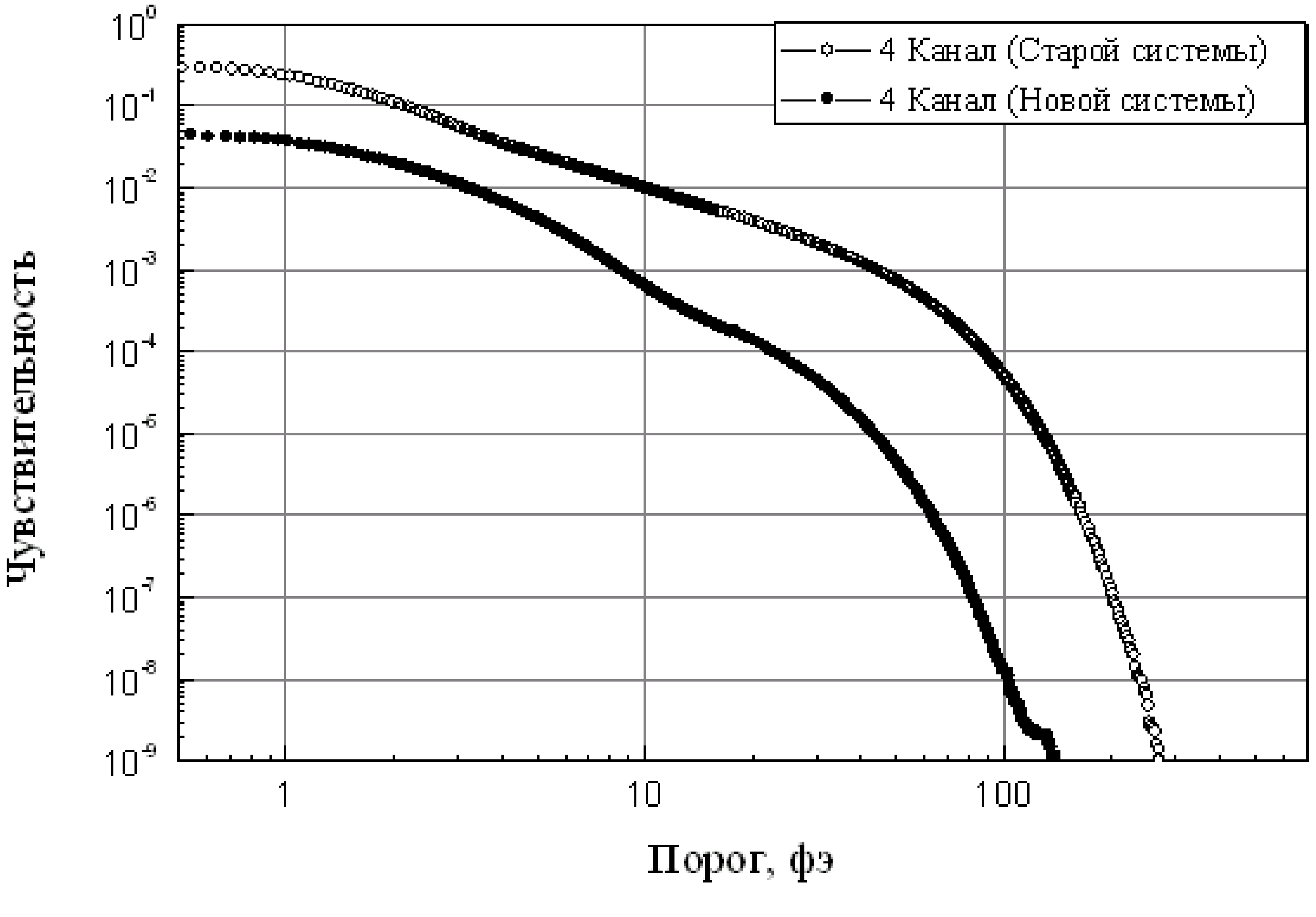

Рис. 4.3.- Сравнительные кривые чувствительности старого и нового детекторов к гамма излучению от источника $^{60}Co$ для четвертого канала анализатора нейтральных частиц GEMMA-2M.

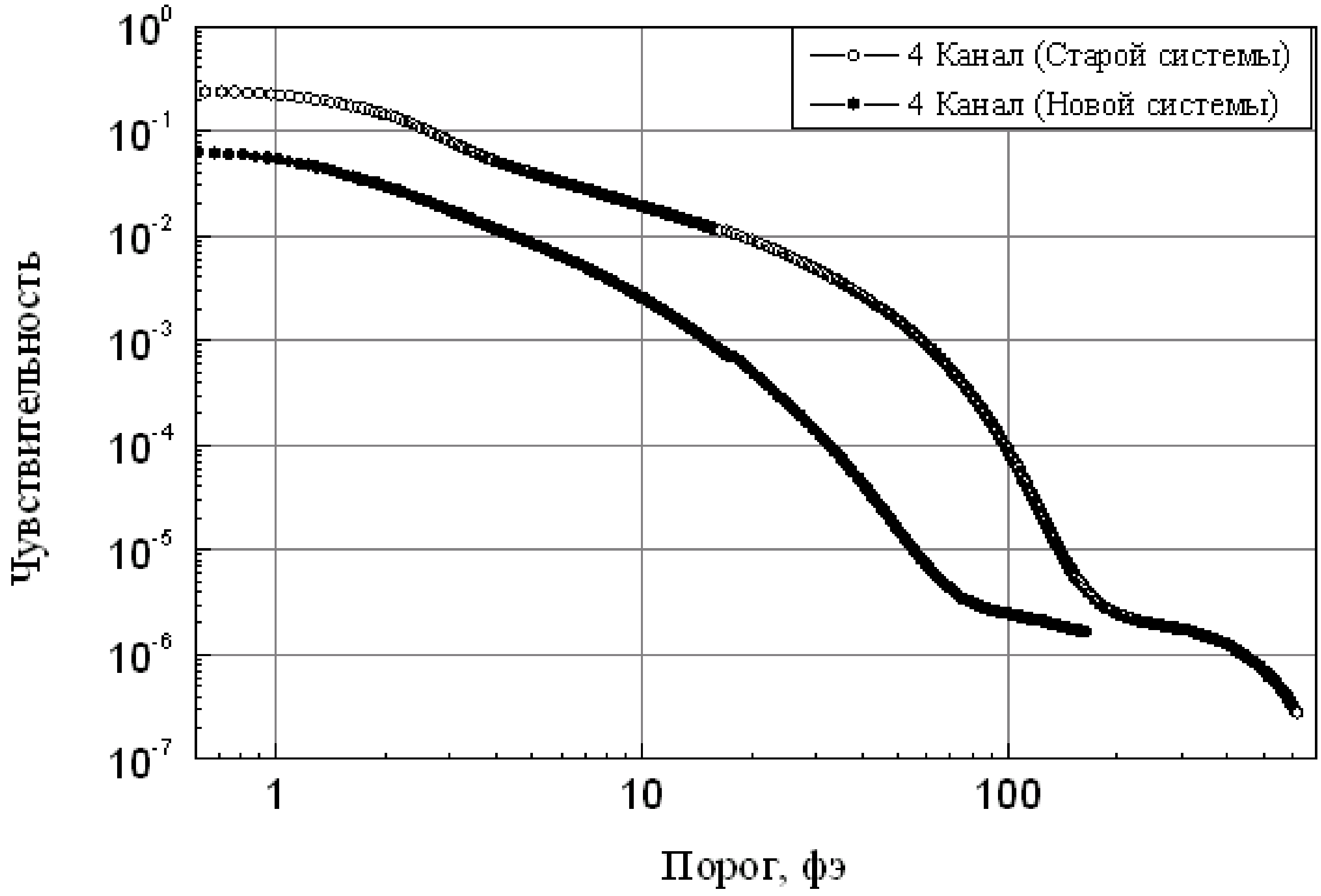

Рис. 4.4.- Сравнительные кривые чувствительности старого и нового детекторов к нейтронному и гамма излучению от источника $^{252}Cf$ для четвертого канала анализатора нейтральных частиц GEMMA-2M.

Таким образом, в результате проделанных изменений в новой системе детектирования было достигнуто заметное уменьшение чувствительности детектирующей системы к фоновому излучению.

# 2.- Изотопный Сепаратор ISEP.

## 2.1.- Конструкция изотопного сепаратора ISEP

Изотопный сепаратор представляет собой анализатор нейтральных потоков водорода: дейтерия и трития, излучаемых высокотемпературной плазмой. Диапазон регистрации различных изотопов водорода лежит в интервале энергий для водорода H (от 5 до 740 КэВ), для дейтерия D (от 5 до 370 КэВ) и для трития T (от 5 до 250 КэВ). Изотопный сепаратор по принципу действия аналогичен анализатору нейтральных частиц рассмотренному выше. Его схематическое изображение представлено на рис. 4.6.

Его принципиальное отличие от разработанных ранее анализаторов заключается в том, что используется доускорение изотопов после обдирки. Процесс доускорения изотопов после обдирки примерно на 100 КВ был реализован с целью сообщить дополнительную энергию, регистрируемым частицам, и тем самым улучшить соотношение сигнал/фон, что особенно важно при регистрации частиц низкой энергии (< 100 КэВ). В этом случае можно установить более высокий порог дискриминации и получить более низкую чувствительность детектора к фоновому излучению (см. рис. 4.9).

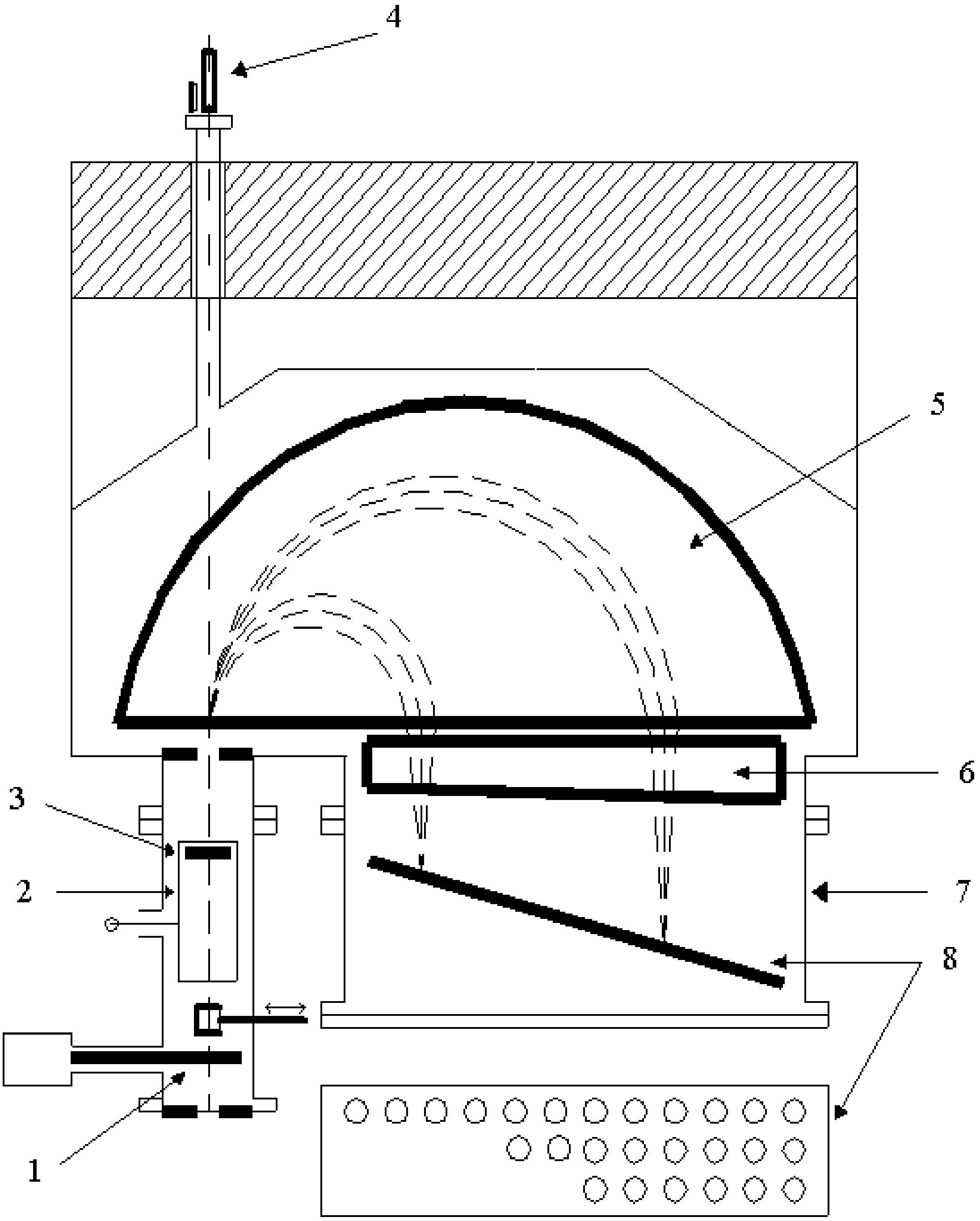

Рис. 4.6.- Схематическое изображение изотопного сепаратора ISEP: 1- механизм коллимации, 2- высоковольтный ускоритель, 3- обдирочная пленка, 4- детектор прямого канала, 5- полюсники анализирующего магнита, 6- пластины конденсатора, 7- камера детектирования, 8- линейки детекторов.

Кроме того, Форма магнитного поля была подобрана так, что пучок частиц на выходе из магнита фокусировался на детектор. Это позволило заметно уменьшить размеры чувствительной поверхности детектора, использовать детекторов меньших размеров и тем самым уменьшить регистрацию фоновых сигналов при той же эффективности регистрации частиц.

## 2.2.- Конструкция детектирующей системы анализатора

Детектирующая система изотопного сепаратора размещена на фланце в камере детектирования и состоит из тридцати двух сцинтилляционных детекторов, расположенных в трех рядах, для одновременной регистрации соответственно трех типов изотопов H, D, T (см. рис. 4.7).

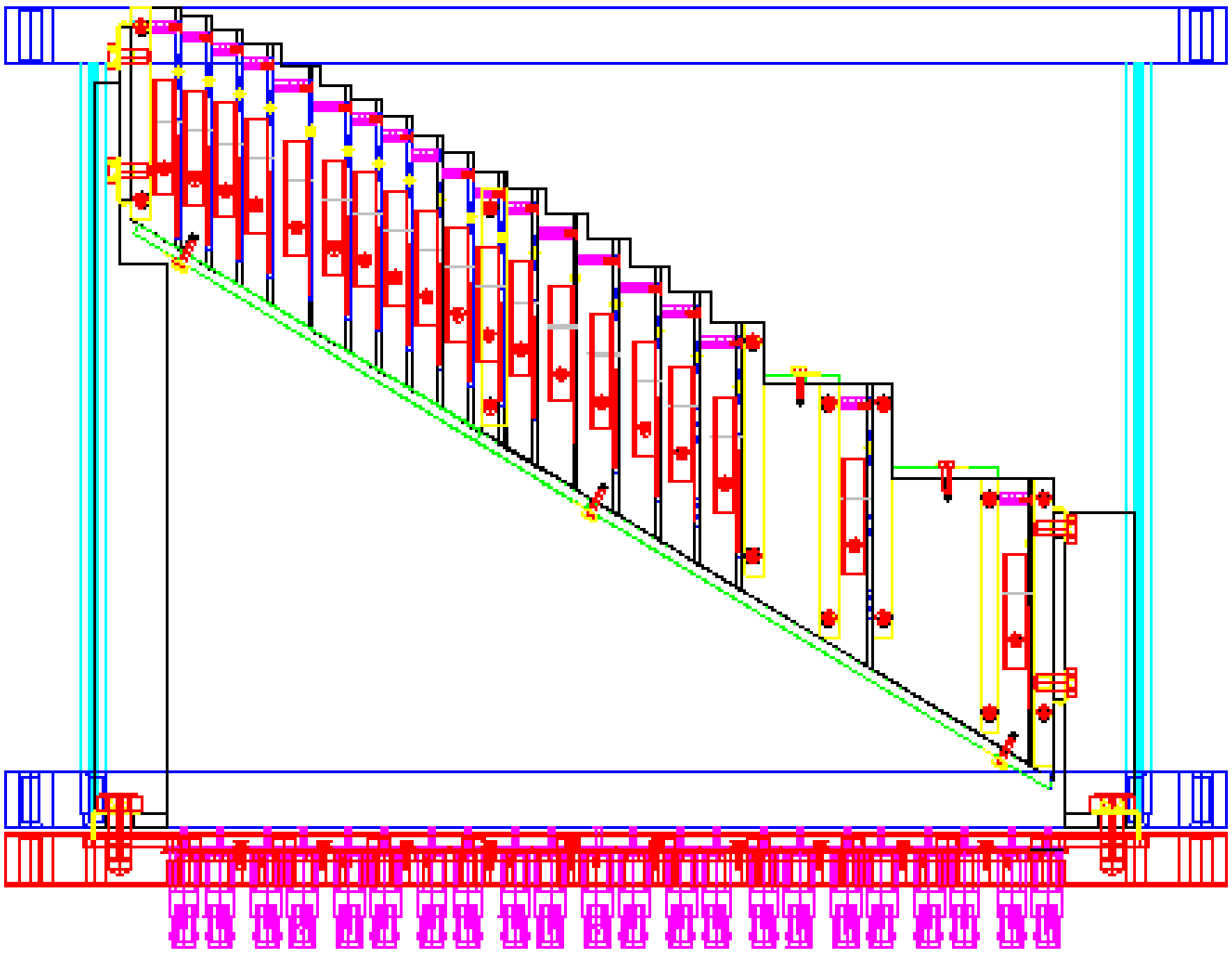

Рис. 4.7.- Камера детектирования изотопного сепаратора (ISEP)

Детекторы, специально разработанные для использования в данном приборе, также были изготовлены на кафедре экспериментальной ядерной физики Санкт-Петербургского Государственного Технического университета. Эти детекторы представляют собой спектрометрические сцинтилляционные счетчики, на основе тонкого вакуумного сцинтиллятора CsI(Tl), напыленного в вакууме на стеклянную подложку. Толщина сцинтилляторов варьировалась от 1 до 7 мкм.

На рисунке 4.8 представлена схема конструкции детекторов, используемых в изотопном сепараторе ISEP.

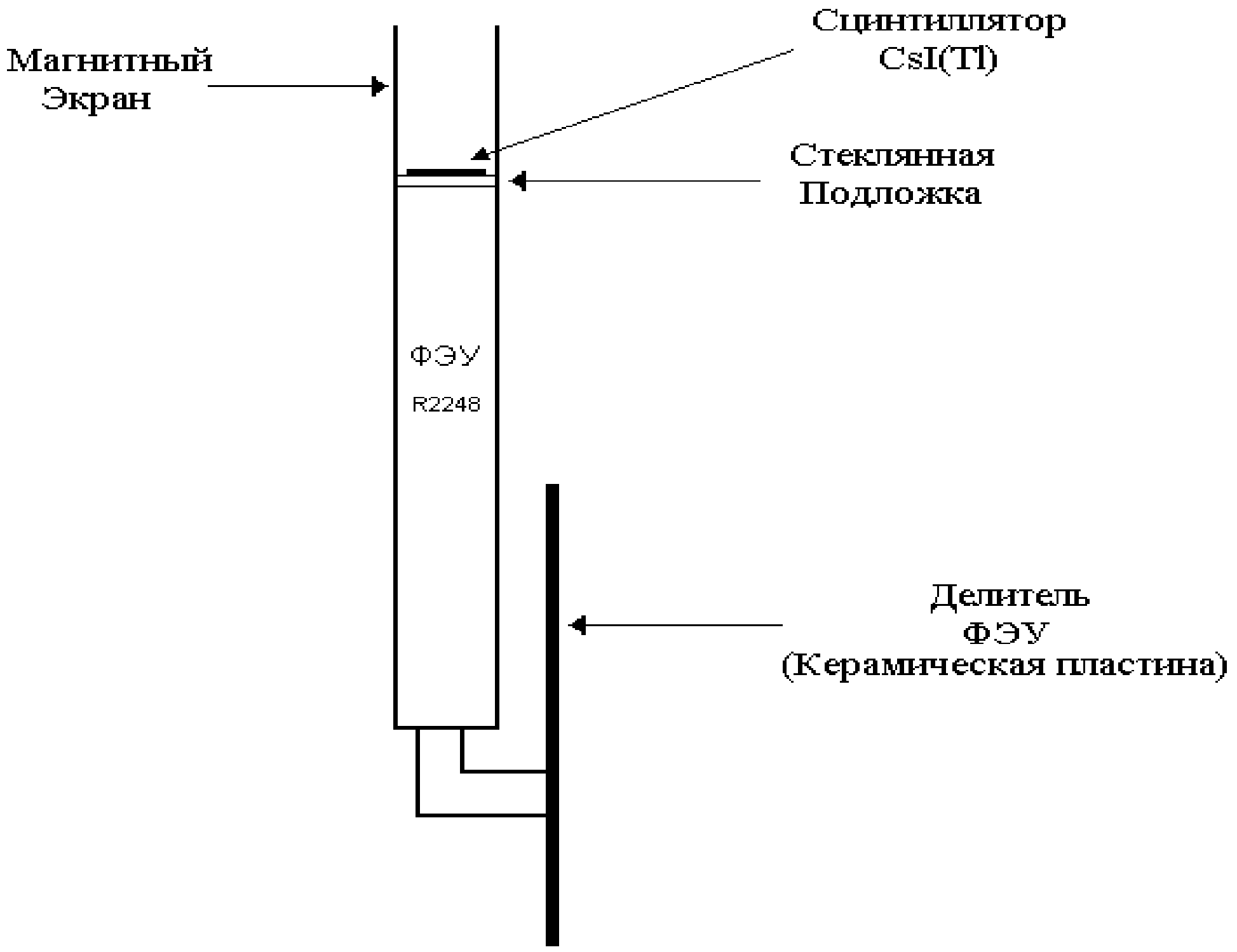

Рис. 4.8 Схема конструкции детектора для изотопного сепаратора ISEP

На фланце камеры детектирования изотопного сепаратора ISEP были размещены 32 таких детектора. При этом в линейке водорода установлено 14 детекторов, в линейке дейтерия 10 и в линейке трития 8. Расположение этих детекторов на фланце и толщины сцинтилляционных кристаллов CsI(Tl) приведены в таблице 4.1

Таблица 4.1

| Номер Канала | T(t, мкм) | D(t, мкм) | H(t, мкм) |
|---|---|---|---|
| 1 | 1.2 | 1.0 | 1.2 |
| 2 | 1.2 | 1.2 | 1.2 |
| 3 | 1.5 | 1.2 | 1.2 |
| 4 | 1.5 | 1.5 | 1.5 |
| 5 | 1.5 | 1.5 | 1.5 |
| 6 | 2.0 | 1.5 | 2.0 |
| 7 | 2.0 | 2.0 | 2.0 |
| 8 | 2.0 | 2.0 | 2.0 |
| 9 | | 3.0 | 3.0 |
| 10 | | 3.0 | 3.0 |
| 11 | | | 4.0 |
| 12 | | | 4.0 |
| 13 | | | 5.0 |
| 14 | | | 7.0 |

Из таблицы видно, что с увеличением номера канала увеличивалась и толщина сцинтилляционного кристалла CsI(Tl). Это связано с тем, что при увеличении номера канала увеличивается и энергия регистрируемых частиц.

На рис. 4.9 представлены зависимости дифференциальных чувствительностей сцинтилляционных детекторов к фоновому нейтронному и гамма излучению от амплитуды сигнала на выходе детектора. На этом же рисунке представлены амплитудные распределения сигналов при облучении альфа частицами. Измерения были выполнены для всех используемых толщин сцинтилляционных кристаллов CsI(Tl).

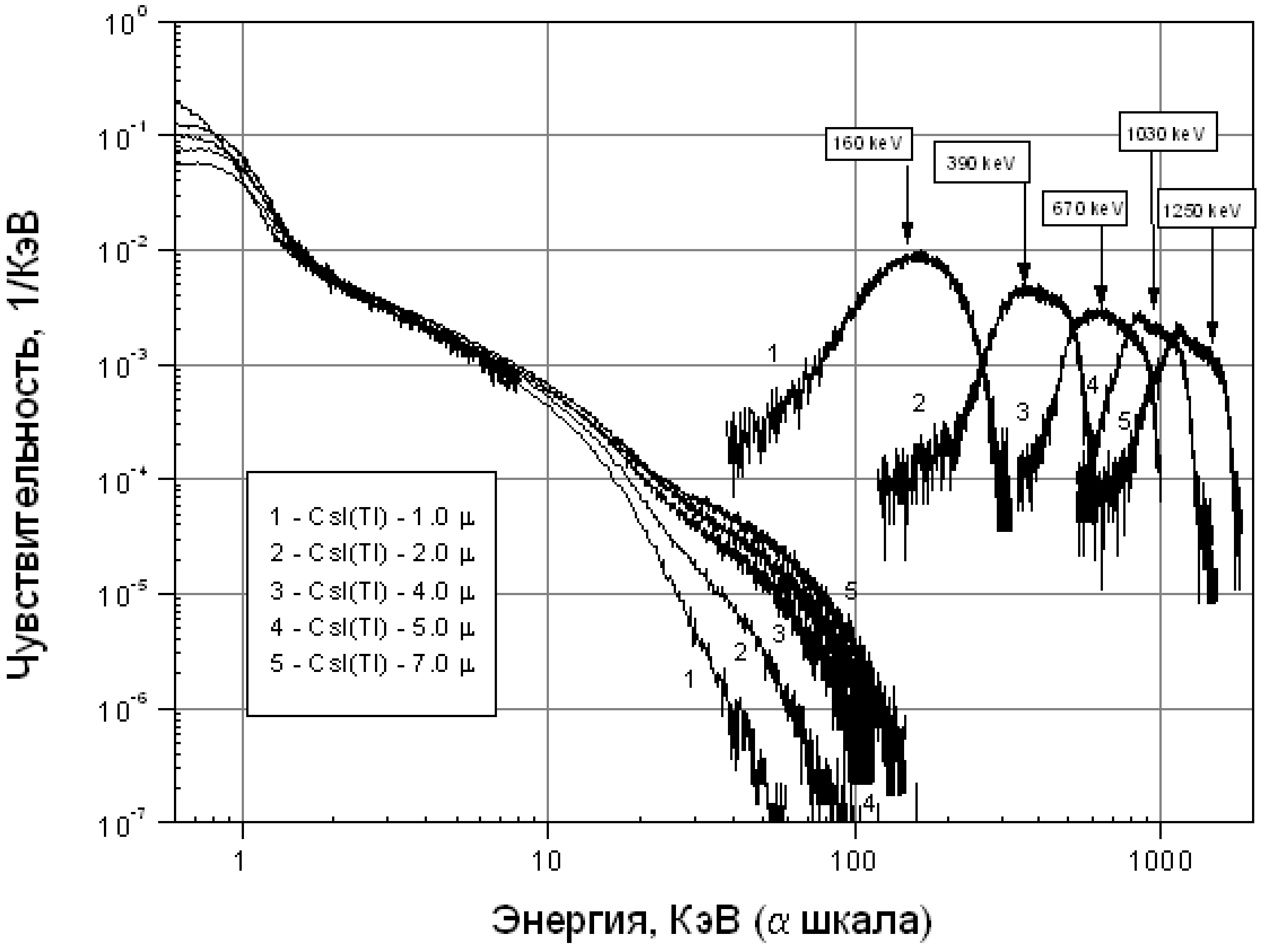

Рис 4.9- Дифференциальная чувствительность сцинтилляционных детекторов к фоновому нейтронному и гамма излучению для разных толщин: (1- 1.0, 2- 2.0, 3- 4.0,4- 5.0, 5- 7.0 мкм) и амплитудные распределения альфа частиц с энергией отвечающей данному каналу

Изотопный сепаратор ISEP недавно был установлен и в настоящее время испытывается на термоядерной установке JET (Англия). Были получены первые результаты измерения корпускулярных потоков в реальных условиях фонового излучения плазмы. На рис. 4.9 представлено амплитудное распределение сигналов, возникающих при регистрации дейтронов (D) с энергией 10 КэВ (Канал 7). Эти данные были получены в разряде 50694 с интегральным потоком фонового излучения ~ $6.0 \cdot 10^{16}$ n/сек. При этом ускоряющее напряжение в ускоряющим модуле составило ~ 10 КВ. На этом же рисунке, приведено амплитудное распределение сигналов, возникающих под действием только фонового излучения в этом же детектора. Эти данные были получены в разряде (50890) с параметрами аналогичному разряду (50694) при закрытом шибере (в этом случае поток корпускулярного излучения не попадал на вход прибора).

Из рис. 4.10 видно, что число отчетов, возникающих при регистрации дейтронов (кривая B) на несколько порядков превышает уровень фона (кривая D). Отсюда следует, что система работоспособна при существующим уровне защиты по крайне мере в дейтериевых разрядах.

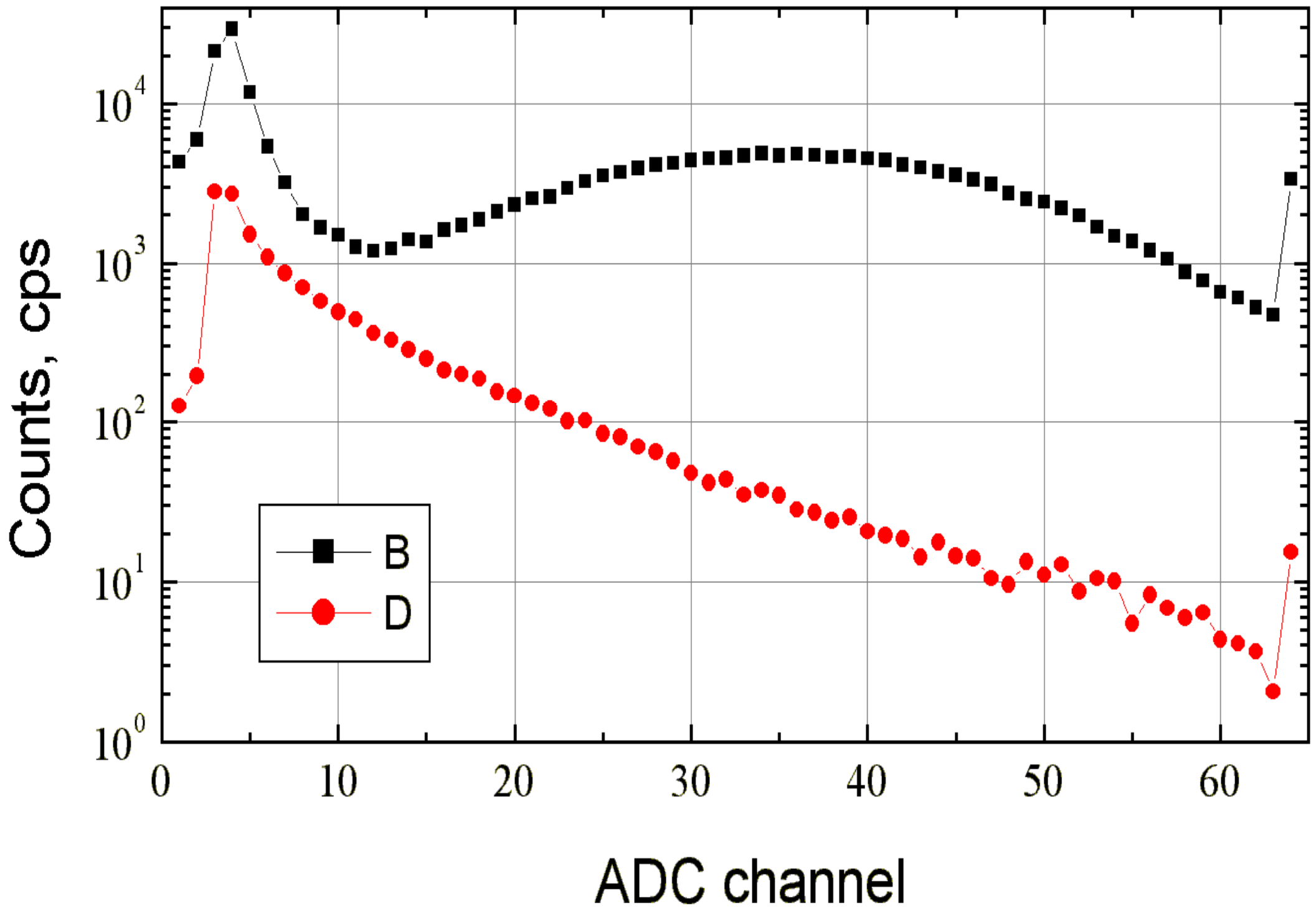

Рис 4.10- Амплитудное распределение сигналов, возникающих в 7 канале дейтериевой линейки разрядных 50694 и 50890 в JET: в- при открытом шибере, d – при закрытом шибере.

В заключение можно утверждать, что предложенная выше методика оптимизации системы регистрации корпускулярного излучения плазмы позволяет разрабатывать детектирующие системы для анализаторов нейтральных частиц работоспособные в условиях интенсивного нейтронного и гамма излучения плазмы.

# ЗАКЛЮЧЕНИЕ

В настоящее время увеличение мощности плазменных разрядов в крупных термоядерных установках привело к тому, что уровень фонового излучения плазмы, облучающего детектирующую систему анализаторов корпускулярной диагностики резко увеличивался. В связи с этим возникла проблема разработки новых систем детектирования с пониженной чувствительностью к фоновому нейтронному и гамма излучению. Нам удалось создать системы детектирования корпускулярных потоков анализаторов нейтральных частиц GEMMA-2M и ISEP, для работы в условиях в условиях интенсивного нейтронного и гамма излучения. Опыт использования данных анализаторов на современных крупных токамаках показал, что они способны обеспечить высокую информативность и надежность результатов измерения.

Достигнутые результаты позволяют проектировать системы детектирования корпускулярных потоков, работоспособны в условиях интенсивного фонового нейтронного и гамма излучения плазмы, как для существующих крупных токамаков, так и для проектируемого в настоящее время термоядерного реактора нового поколения ITER. Повышение уровня нейтронного и гамма фона, ожидаемое в проектируемом термоядерном реакторе ITER, способствует росту интереса к дальнейшему развитию данной методики. Укажем основные направления ее развития:

– проектирование более эффективной защиты элементов конструкции детектора от проникающего излучения,
– использование неорганических сцинтилляционных кристаллов со временем высвечивания меньшим, чем у сцинтилляторов CsI(Tl).

Данная диссертация представляет собой часть этой работы и отражает все этапы ее развития: разработку, изготовление систем детектирования и измерение их параметров по отношению к нейтронному и гамма излучению.

На защиту выносится следующие результаты:

1. Исследование поведения интегральной чувствительности детекторов, используемых в корпускулярной и оптической диагностике плазмы к фоновому нейтронному и гамма излучению, позволило установить, что основным фактором, определяющим чувствительность детекторов типа вторичный электронный умножитель (ВЭУ), канальный электронный умножитель (КЭУ), микроканальная пластина (МКП) и фотоэлектронный умножитель (ФЭУ) к фоновому нейтронному и гамма излучению является взаимодействие γ-квантов с чувствительной поверхностью детектора и окружающими его элементами конструкции. При этом ФЭУ обладает наибольшей чувствительностью к фоновому излучению, которая составляет ~ $10^{-2}$, КЭУ и МКП ~ $10^{-3}$, а наименьшей чувствительностью к фоновому излучению обладает ВЭУ и составляет ~ $10^{-4}$.

2. Детальное исследование природы возникновения сигналов, вызываемых облучением нейтронами и γ-квантами позволило установить, что для ФЭУ основним элементом, определяющим полную чувствительность детектора к фоновому нейтронному и гамма излучению, является входное окно. Обнаружена сильная корреляция полной чувствительности ФЭУ к фоновому излучению от толщины стекла входного окна. Найдено, что при облучении нейтронами не наблюдается заметного вклада продуктов реакции (n,α) в полную чувствительность детектора к фоновому излучению для ФЭУ с входным окном из боросиликатного стекла

3. Анализ амплитудных распределений, полученных при облучении нейтронами и γ-квантами позволил установить, что для сцинтилляционных детекторов на основе тонких кристаллов CsI(Tl) основной вклад в чувствительность детектора к фоновому излучению в области больших амплитуд дают электроны, генерируемые взаимодействием γ-квантов со стенками камеры, окружающей детектор. Обнаружена квазилинейная зависимость средней оставленной энергией электронами от толщины сцинтилляционного кристалла CsI(Tl).

4. Результаты измерения корпускулярных потоков из плазмы в реальных условиях термоядерной установки JET (Англия) свидетельствуют о том, что разработанная методика позволяет создавать системы детектирования, работоспособные в условиях интенсивного фонового излучения.

В заключение хочу поблагодарить моего научного руководителя С.С. Козловского за большую помощь на всех этапах данной работы.

Благодарю всем сотрудникам кафедры ядерной физики и особенно В.В. Гребенщикову, А.В. Дечу за  плодотворную совместную работу.

Хочу также поблагодарить мою жену Риикку Елину Вайнио-Гарсия и всю мою семью за помощь в написании этой работы. А также семью Яновских: Валерия Витальевича и Дмитрия Валериевича за оказанную помощь в переводе рукописи из русского на русский и всех тех, кто способствовал ее успешному завершению.

# Список использованной литературы